\documentclass[longbibliography,reprint, amsmath,amssymb, aps,prx,superscriptaddress]{revtex4-2}
\usepackage{amssymb}
\usepackage{amsmath}
\usepackage{amsfonts}
\usepackage[colorlinks,linkcolor=red,urlcolor=blue,citecolor=blue]{hyperref}
\usepackage{graphicx}
\usepackage{txfonts}
\usepackage{braket}
\usepackage{color}
\usepackage{bm,bbm}
\usepackage{multirow}

\newcommand{\EE}{{\rm E}}
\newcommand{\Var}{{\rm Var}}
\newcommand{\Cov}{{\rm Cov}}
\newcommand{\dd}{{\rm d}}
\renewcommand{\ss}{{\rm ss}}
\newcommand{\jss}{{\rm jss}}
\newcommand{\ee}{{\rm e}}
\newcommand{\ii}{{\rm i}}
\newcommand{\T}{{\sf T}}
\newcommand{\Texp}{\overleftarrow{\mathrm{T}} \exp }

\graphicspath{{figs/}}

\begin{document}

\title{Optimal time estimation and the clock uncertainty relation for stochastic processes}

\author{Kacper Prech}
\email{kacper.prech@unibas.ch}
\affiliation{Department of Physics and Swiss Nanoscience Institute, University of Basel, Klingelbergstrasse 82, 4056 Basel, Switzerland}

\author{Gabriel T. Landi}
\email{glandi@ur.rochester.edu}
\affiliation{Department of Physics and Astronomy, University of Rochester, Rochester, New York 14627, USA}

\author{Florian Meier}
\affiliation{Atominstitut, TU Wien, 1020 Vienna, Austria}

\author{Nuriya Nurgalieva}
\affiliation{Department of Physics, University of Zurich, Winterthurerstrasse 190, 8057 Z\"urich, Switzerland}

\author{Patrick P. Potts}
\affiliation{Department of Physics and Swiss Nanoscience Institute, University of Basel, Klingelbergstrasse 82, 4056 Basel, Switzerland}

\author{Ralph Silva}
\affiliation{Institute for Theoretical Physics, ETH Zurich, Wolfgang-Pauli-Strasse 27, 8093 Z\"urich, Switzerland}

\author{Mark T. Mitchison}
\email{mark.mitchison@kcl.ac.uk}
\affiliation{School of Physics, Trinity College Dublin, College Green, Dublin 2, D02 K8N4, Ireland}
\affiliation{Department of Physics, King’s College London, Strand, London, WC2R 2LS, United Kingdom}

\begin{abstract}
Time estimation is a fundamental task that underpins precision measurement, global navigation systems, financial markets, and the organisation of everyday life. Many biological processes also depend on time estimation by nanoscale clocks, whose performance can be significantly impacted by random fluctuations. In this work, we formulate the problem of optimal time estimation for Markovian stochastic processes, and present its general solution in the asymptotic (long-time) limit. Specifically, we obtain a tight upper bound on the precision of any time estimate constructed from sustained observations of a classical, Markovian jump process. This bound is controlled by the mean residual time, i.e.~the expected wait before the first jump is observed. As a consequence, we obtain a universal bound on the signal-to-noise ratio of arbitrary currents and counting observables in the steady state. This bound is similar in spirit to the kinetic uncertainty relation but provably tighter, and we explicitly construct the counting observables that saturate it. Our results establish ultimate precision limits for an important class of observables in non-equilibrium systems, and demonstrate that the mean residual time, not the dynamical activity, is the measure of freneticity that tightly constrains fluctuations far from equilibrium.
\end{abstract}

\maketitle

\section{Introduction}
\label{sec:intro}

\begin{figure*}
\includegraphics[width=0.93\linewidth]{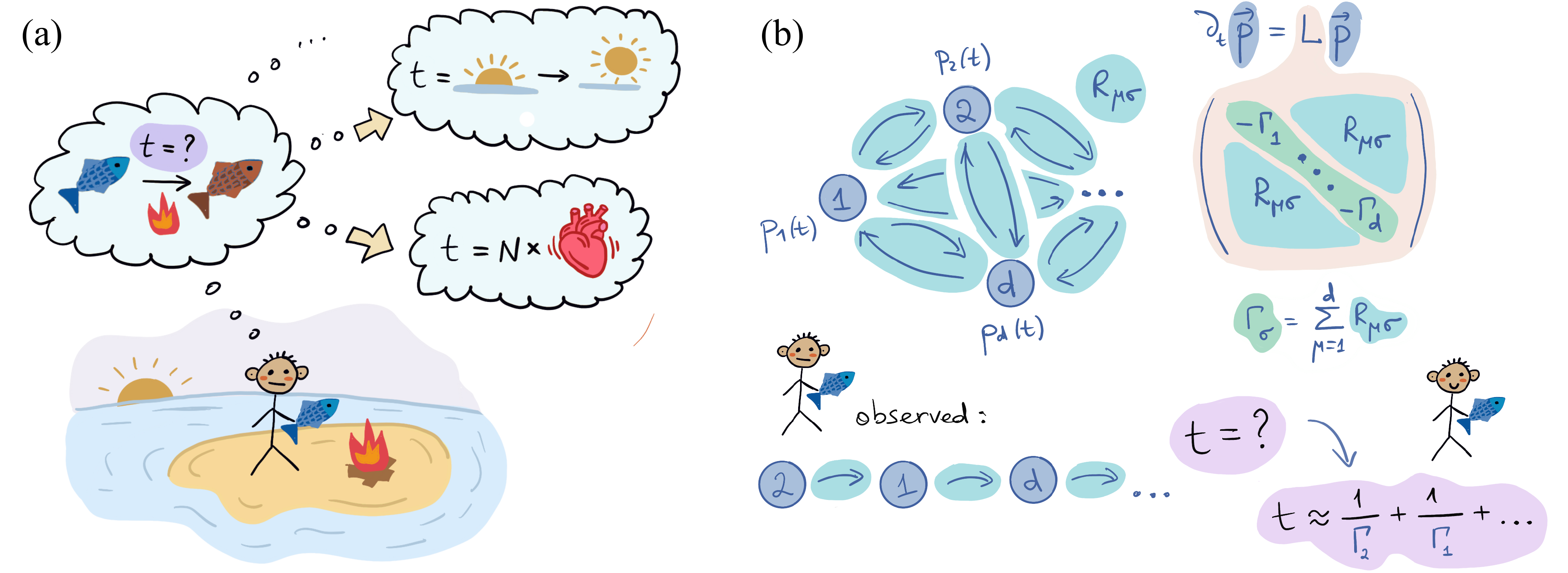}
\caption{Optimal time estimation for classical stochastic processes. (a) Successful cooking (among other important tasks) requires an accurate and precise time estimator to be constructed from an observer's record of events. Stochastic phenomena can be used for this purpose but have varying degrees of frequency, regularity, and inter-dependency. (b) We focus on Markovian jump processes described by a master equation on a discrete network of $d$ states, where $p_\sigma$ is the probability to find the system in state $\sigma$, $R_{\mu\sigma}$ is the rate for a jump $\sigma\to \mu$, and $\Gamma_\sigma$ is the escape rate from $\sigma$. An observer lacking a clock may record the stochastic sequence of states visited by the system, $\bm{\sigma} = \sigma_0\to \sigma_1 \to \cdots$. An asymptotically optimal time estimator is the sum of mean dwell times $\Gamma_\sigma^{-1}$ for each state observed in the sequence. 
Timing precision is inversely proportional to the mean residual time, which is illustrated in Fig.~\ref{fig:residual_time} and defined by $\mathcal{T}=\sum_\sigma p_\sigma^\ss \Gamma_\sigma^{-1}$, where $p_\sigma^\ss$ is the steady state.
\label{fig:fishing}}
   \end{figure*}

Imagine a castaway stranded on a desert island, who has painstakingly succeeded in catching a fish. In order to cook this fish safely and reproducibly on the campfire (without watching it continuously), the castaway must be able to estimate how much time has elapsed. Fortunately, even in the absence of modern technology, the environment provides several approximately periodic phenomena that could be exploited to measure time, e.g.~the motion of the sun in the sky, waves lapping on the shore, and the beating of the human heart  (see Fig.~\ref{fig:fishing}). These processes have widely varying timescales and regularity, and they are also correlated with each other to some extent. Using this data, how can the castaway optimally estimate the cooking time (i.e.~with high accuracy and precision) and thus survive another day?

While we sincerely hope that our readers never find themselves in this situation, the task of timekeeping using imperfectly periodic events is central to current research aiming to uncover the physical limits on nanoscale clocks in quantum~\cite{Erker2017, Woods2019,Milburn2020,Schwarzhans2021,Woods2021,Woods2022,Meier2023}, electronic~\cite{Pearson2021,He2023,Culhane2024,Gopal2024}, and biomolecular~\cite{Barato2016,Barato2017,Marsland2019} settings. Quantifying the precision of a ticking clock can be seen as an instance of the more general problem of counting statistics in non-equilibrium systems~\cite{Silva2023}. A major breakthrough in this field was the discovery of the thermodynamic (TUR)~\cite{Barato2015a,Gingrich2016} and kinetic (KUR)~\cite{Garrahan2017,Terlizzi2018} uncertainty relations, which impose universal bounds on fluctuations far from equilibrium. These relations not only limit the accuracy of clocks but also constrain the fluctuations of generic currents and other observables in systems undergoing Markovian stochastic dynamics.  Important applications include performance bounds for nanoscale heat engines~\cite{Shiraishi2016,Pietzonka2018,Miller2021} and methods to infer entropy production from experimentally accessible data~\cite{Li2019,Manikandan2020,Vu2020}, while generalisations encompass transient fluctuations~\cite{Pietzonka2017,Horowitz2017}, first-passage times~\cite{Garrahan2017,Gingrich2017,Hiura2021}, driven systems~\cite{Proesmans2017,Koyuk2020}, ballistic transport~\cite{Brandner2018}, quantum dynamics~\cite{Macieszczak2018,Guarnieri2019,Hasegawa2020,Vu2022,Salazar2024, Prech2025}, feedback control \cite{Potts2019}, general counting observables~\cite{Pietzonka2024}, and relations to fluctuation theorems~\cite{Timpanaro2019,Hasegawa2019b,Ray2024}; see Refs.~\cite{Seifert2019,Horowitz2020} for reviews. 

Loosely speaking, stochastic uncertainty relations imply that higher precision comes at a greater dissipative cost. More precisely, they are inequalities of the general form $\mathcal{S}\leq \mathcal{C},$ where $\mathcal{S}$ is the signal-to-noise ratio (SNR), e.g.~$\mathcal{S}=J^2/D$ for a current with steady-state mean $J$ and diffusion coefficient $D$~\cite{Landi2024}, while the cost function $\mathcal{C}$ depends on the bound in question. The TUR quantifies this cost via the entropy production rate~\cite{Landi2021}, which measures irreversibility and provides a tight bound on current fluctuations near equilibrium, where dissipation is weak. By contrast, the KUR involves the dynamical activity~\cite{Maes2020}, $\mathcal{C}=\mathcal{A}$, which quantifies how frequently the system jumps from one state to another and remains bounded even for strongly irreversible dynamics, where the entropy production rate diverges. The KUR is thus tighter than the TUR in the far-from-equilibrium regime, but it is nonetheless a loose bound that cannot be saturated in general.

\begin{figure}
    \centering
    \includegraphics[width=\linewidth]{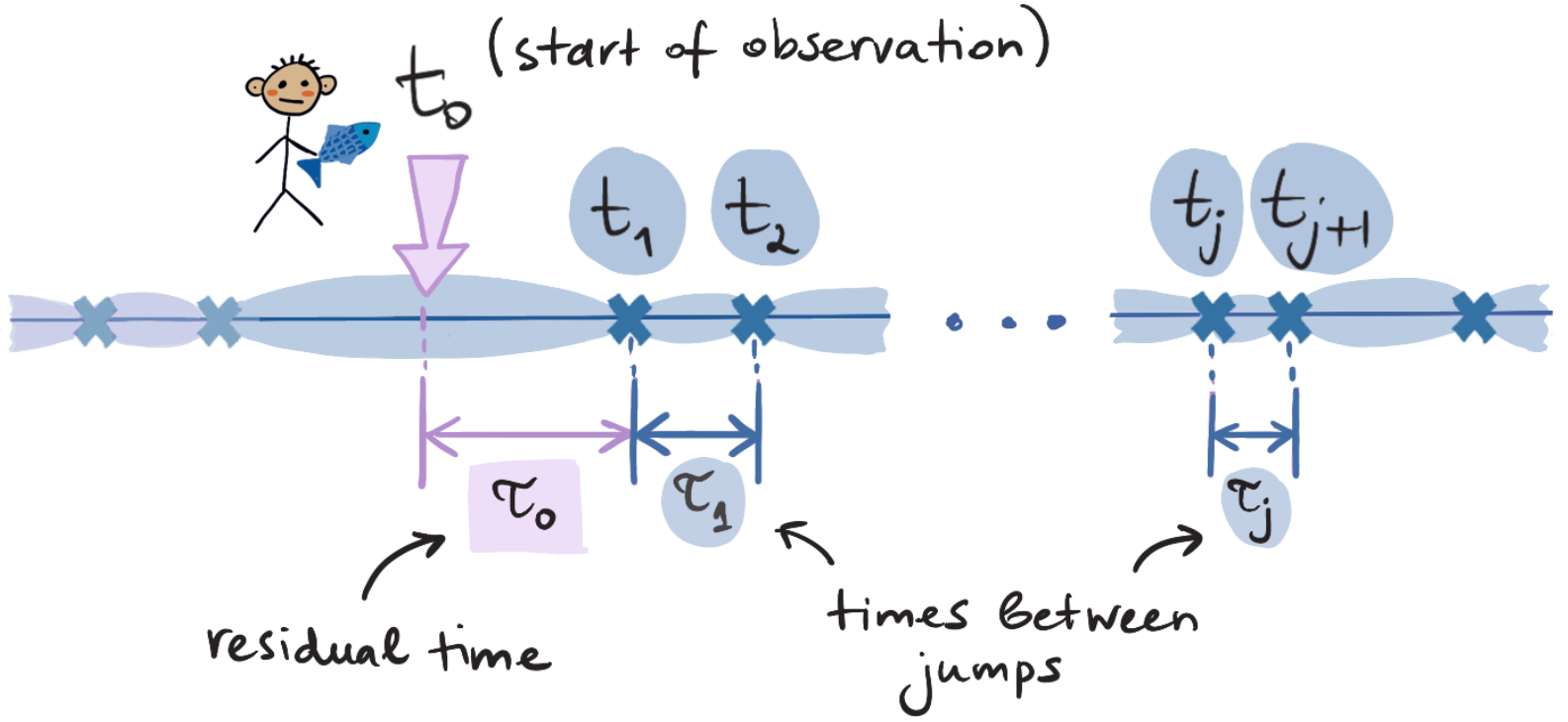}
    \caption{Illustration of the residual and waiting times, with jumps depicted as blue crosses on the time axis. The mean residual time $\mathcal{T}$ is the expected value of $\tau_0$, the time interval until the first jump after observations begin at $t_0$. Paradoxically, this exceeds the mean of the waiting time between jumps, $\tau_{j}$, because the jumps tend to cluster in bunches and the system spends more time in long-lived states.}
    \label{fig:residual_time}
\end{figure}

In this work, we uncover a tight, universal bound on steady-state fluctuations in systems described by Markovian jump dynamics on a discrete set of states. Our bound takes the form
\begin{equation}
	\label{CUR}
	\mathcal{S} \leq \mathcal{T}^{-1},
\end{equation}
where $\mathcal{T}$ is the mean residual time, i.e.~the interval expected before the first jump observed in a sequence, when observations start from an arbitrary (random) instant in time. This differs from the mean waiting time between jumps, which is given by the inverse dynamical activity, $\mathcal{A}^{-1}$. Surprisingly, $\mathcal{T}\geq \mathcal{A}^{-1}$, a manifestation of the notorious inspection paradox~\cite{Stein1985, Pal2022}, which implies that Ineq.~\eqref{CUR} is tighter than the KUR. See Sec.~\ref{sec:optimal_estimator} for precise definitions of the residual and waiting times and Fig.~\ref{fig:residual_time} for an illustration.

As explained below, Ineq.~\eqref{CUR} emerges naturally from the problem of time estimation, so we term it the clock uncertainty relation (CUR). Like the KUR, the CUR applies to general counting observables, it holds arbitrarily far from equilibrium, and it does not require local detailed balance. Physically, both the KUR and the CUR state that the signal-to-noise ratio is limited by the activity of the process, as measured by $\mathcal{A}$ and $\mathcal{T}^{-1}$, respectively. Unlike the KUR, however, Ineq.~\eqref{CUR} can always be saturated and we explicitly construct the counting observables that do so. This distinguishes the inverse mean residual time, $\mathcal{T}^{-1}$, as the proper measure of activity controlling fluctuations in non-equilibrium steady states. Moreover, since $\mathcal{T}$ is directly observable and the CUR bound is tight, it can be used as a witness to verify or falsify that the observed dynamics is autonomous, classical, and memoryless.

To obtain this result, we formulate and solve the problem of optimal time estimation for classical stochastic processes described by a Markovian master equation, as depicted schematically in Fig.~\ref{fig:fishing}. We work within the framework of local parameter estimation theory, where the Fisher information constrains the variance of any time estimate via the Cram\'{e}r-Rao bound (CRB)~\cite{Kay2013}. We note that the Fisher information of the instantaneous statistical state has recently appeared in speed limits derived from the information geometry of stochastic thermodynamics~\cite{Ito2018,Ito2020,Nicholson2020}. Here, we consider a completely different object: the Fisher information of the distribution of states visited along the entire trajectory, which accounts for all data available to an observer who lacks an external clock. 

We first  prove that this Fisher information, $\mathcal{F}_t$, scales inversely to the mean residual time, $\mathcal{F}_t \sim 1/(\mathcal{T}t)$, when the time $t$ is large (Sec.~\ref{sec:optimal_estimator}). In our setting, this result yields the ultimate upper bound on time precision when all events and possible correlations between them are accounted for. We also construct an unbiased time estimator that asymptotically saturates the CRB, which is far simpler than the maximum-likelihood estimator because it only involves counting of events. 

We then formulate the CUR~\eqref{CUR} for more general observables in non-equilibrium steady states (Sec.~\ref{sec:CUR}). We prove that this bound can be saturated by an explicit class of counting observables: those that correspond to efficient time estimators. Quantitative comparisons of different precision bounds are presented for random networks and unicyclic clock models, elucidating the conditions under which the TUR, KUR, and CUR differ. We also prove that the KUR is not generally saturable on basic information-theoretic grounds, and explain why the CRB for time estimation gives the tightest bound on fluctuations of counting observables. As an application of the CUR, we show that its experimental violation can be used to detect the presence of unobservable states or transitions, a common scenario in many biological settings~\cite{Roldan2010,Hilfinger2016,Martinez2019}. This provides a novel tool for characterising stationary stochastic systems, complementing recent proposals to detect hidden states from transient behaviour~\cite{Ito2020} or infer entropy production from partially observed dynamics~\cite{Skinner2021, Harunari2022,vanderMeer2022}.

Finally, we discuss how other metrics of clock performance fit into our framework (Sec.~\ref{sec:beyond_precision}). We find that the CUR imposes a bound on any autonomous classical clock with discrete states and Markovian dynamics:
\begin{equation}
\label{accuracy_resolution_CUR}
    \nu\mathcal{N} \mathcal{T} \leq 1,
\end{equation}
where the resolution $\nu$ is the rate at which the clock ticks and the accuracy $\mathcal{N}$ is the mean number of ticks before the clock is off by one tick  (precise definitions are given in Sec.~\ref{sec:beyond_precision}). This proves a tighter special case of the general accuracy-resolution tradeoff  for quantum clocks derived recently in Ref.~\cite{Meier2023}.

Our work not only reveals a hitherto unknown physical principle constraining fluctuations in non-equilibrium systems, but it also embeds emerging research on stochastic and quantum clocks within the established statistical framework of metrology. This deepens the connection between estimation theory and stochastic thermodynamics, while unlocking the mathematical tools of both fields to study the foundations of timekeeping. We postpone further discussion of our results, their significance, and their potential extensions to Sec.~\ref{sec:discussion}.

\section{Optimal time estimation}
\label{sec:optimal_estimator}

Our framework is motivated by recent research aiming to understand the minimal resources needed to build a clock~\cite{Erker2017,Schwarzhans2021,Woods2021,Woods2022, Silva2023}. In that context, fair bookkeeping of resources forces one to consider autonomous systems, i.e.~those that are not driven by some externally prescribed, time-dependent protocol, which would be tantamount to the prior existence of an ideal clock that could be used to measure time instead.  

Following this logic, here we formulate the problem of time estimation for autonomous classical processes described by a Markovian master equation~\cite{Gardiner2009}. This standard model in stochastic thermodynamics encompasses a range of natural and artificial non-equilibrium systems, including chemical reaction networks~\cite{Schmiedl2007}, biological clocks and oscillators~\cite{Goldbeter1996}, and steady-state open quantum systems in the secular approximation~\cite{Breuer2007}. In Sec.~\ref{sec:setup} we describe our general setup and introduce elementary notions of estimation theory, then in Sec.~\ref{sec:residual_vs_waiting} we explain the difference between residual and waiting times. In Sec.~\ref{sec:sufficient_statistics} we prove that the order of events is irrelevant for time estimation in this context, before deriving an asymptotically efficient estimator and the corresponding long-time Fisher information in Sec.~\ref{sec:asymptotic_fisher}.

\subsection{General setup}
\label{sec:setup}

Consider a Markovian, time-continuous, and stationary stochastic process over a discrete set of $d$ states labelled by the integer $\sigma = 1,2,\ldots,d$. Let $p_\sigma(t)$ be the probability for the system to be in state $\sigma$ at time $t$, and let the probability vector be $\mathbf{p} = (p_1,\ldots,p_d)^\T$, where the superscript $\T$ denotes the matrix transpose. The ensemble dynamics is then described by the master equation $\partial_t\mathbf{p} = \mathsf{L}\mathbf{p}$, where the generator $\mathsf{L}$ has matrix elements
\begin{equation}
    \label{rate_matrix}
    L_{\mu\sigma} = \begin{cases}
        R_{\mu\sigma} &( \mu \neq \sigma) \\
        -\Gamma_\sigma  & (\mu = \sigma)
    \end{cases},\qquad \Gamma_{\sigma} := \sum_{\mu=1}^d R_{\mu\sigma}.
\end{equation}
Above, $R_{\mu\sigma}$ is the transition rate for a jump $\sigma \to \mu$, $\Gamma_\sigma$ is the total escape rate from state $\sigma$, and we set $R_{\sigma\sigma} = 0$ by convention. The steady-state distribution $p_\sigma^{\ss}$ is given by the solution of $\mathsf{L}\mathbf{p}^{\ss} = 0$, and we assume this to be unique. 

Suppose that an observer monitors the transitions between these states as they randomly occur (see Fig.~\ref{fig:fishing}). To find the ultimate limits on time estimation, we assume that all such transitions can be detected and distinguished with perfect efficiency. The observer's record of events thus comprises a stochastic sequence of transitions between states of the form $\bm{\sigma} = \sigma_0 \to \sigma_1 \to \cdots \to \sigma_{N}$. The total number of transitions occurring in a given trajectory is denoted by $N$, which is itself a random variable. 

The probability of observing such a sequence of jumps $\sigma_{j-1} \to \sigma_{j}$ at times $t_j$ within a total observation time $t$ is given by (see Appendix~\ref{SM:t-ensemble})
\begin{align}
	\label{path_probability_with_times}
	&  P(\bm{\sigma},\mathbf{t}|t) = \frac{W(\tau_N|\sigma_N)}{\Gamma_{\sigma_N}} \left( \prod_{j=0}^{N-1}   \pi(\sigma_{j+1}|\sigma_{j})W(\tau_j|\sigma_j) \right)
    p^{\ss}_{\sigma_0},
\end{align}
where $\mathbf{t} = (t_1,\ldots,t_N)$, and we define the waiting time after the $j^{\rm th}$ jump as $\tau_j = t_{j+1} - t_j$, with $t_0 \equiv 0$ and $t_{N+1}\equiv t$. We also introduce the transition probability for a jump $\sigma\to \mu$, $\pi(\mu|\sigma) = R_{\mu\sigma}/\Gamma_\sigma$, and the distribution of waiting times between jumps, $W(\tau|\sigma) = \Gamma_\sigma \ee^{-\Gamma_\sigma \tau}$. Equation \eqref{path_probability_with_times} holds for $t_{j+1}\geq t_j$, while $P(\bm{\sigma},\mathbf{t}|t)=0$ otherwise. Since the observer does not know the time \textit{a priori}, they can record the sequence of jumps $\bm{\sigma}$ but not the times $\mathbf{t}$ at which jumps occur. The probability of recording such a sequence is obtained by marginalizing Eq.~\eqref{path_probability_with_times} over the jump times:
\begin{equation}
	\label{path_probability}
	P(\bm{\sigma}|t) = \int_0^t \dd t_N\int_0^{t_N} \dd t_{N-1}\cdots \int_0^{t_2}\dd t_1  \,P(\bm{\sigma},\mathbf{t}|t) .
\end{equation}

The problem is now to estimate the total elapsed time, $t$, using the observation record $\bm{\sigma}$. A time estimator $\Theta(\bm{\sigma})$ is a function of the sequence $\bm{\sigma}$, which ideally should produce an accurate and precise estimate of $t$. An accurate estimator should be unbiased,
\begin{equation}
	\label{unbiased_estimator}
	\EE[\Theta] = \sum_{\bm{\sigma}}  P(\bm{\sigma}|t) \Theta(\bm{\sigma}) = t,
\end{equation}
where $\EE[\bullet]$ denotes an expectation value and we use the shorthand notation $\sum_{\bm{\sigma}} = \sum_{N=0}^\infty \sum_{\sigma_0=1}^d\cdots \sum_{\sigma_N=1}^d$. A precise estimator has a variance, $\Var[\Theta] = \EE[\Theta^2] - \EE[\Theta]^2$, which should be as small as possible. In the following, we will mostly restrict to unbiased estimators, for which the variance is equivalent to the mean-square error, $\Var[\Theta] = \EE[(\Theta-t)^2]$. 

The variance of any time estimator obeys the Cram\'er-Rao bound (CRB)~\cite{Kay2013}
\begin{equation}
    \label{CRB}
    \Var[\Theta] \mathcal{F}_t \geq \left(\frac{\partial \EE[\Theta]}{\partial t}\right)^2,
\end{equation}
where the right-hand side is $(\partial_t \EE[\Theta])^2=1$ for unbiased estimators, while the Fisher information is defined as
\begin{equation}
	\label{Fisher_info}
	\mathcal{F}_t = \EE\left [\left (\frac{\partial\ln P(\bm{\sigma}|t)}{\partial t}\right )^2\right ].
\end{equation}
Unbiased estimators that saturate the CRB are known as efficient. We emphasise that here we are considering the Fisher information of the sequence probability $P(\bm{\sigma}|t)$ given in Eq.~\eqref{path_probability}. This is distinct from the Fisher information of the statistical state $p_\sigma(t)$  considered in previous works~\cite{Ito2018,Nicholson2020,Ito2020}. 

\subsection{Residual time versus waiting time and the jump steady state}
\label{sec:residual_vs_waiting}

Before presenting our results, we introduce a few quantities that are central to the rest of our work. First, we define the mean residual time
\begin{align}
    \label{mean_residual_time}
    & \mathcal{T} \equiv \sum_{\sigma=1}^d \frac{p_\sigma^\ss}{\Gamma_\sigma },
\end{align}
also known as the forward recurrence time~\cite{McFadden1962} or the persistence time~\cite{Jung2005}. To understand its meaning, note that if the system is found in state $\sigma$, the mean waiting time until the next jump is $\Gamma_\sigma^{-1} = \int_0^\infty\dd \tau \,W(\tau|\sigma)\tau$~\footnote{Technically, this assumes that the total time $t$ is large enough to avoid boundary effects due to the constraint $\sum_{j=0}^{N}\tau_j=t$.}. (The waiting time preceding any jump $\sigma \to \mu$ is statistically independent of $\mu$; see Appendix~\ref{SM:t-ensemble} for discussion of this point.) The mean residual time~\eqref{mean_residual_time} is thus $\mathcal{T} = \EE[\tau_0]$, i.e.~the expected waiting time until the first jump after observation begins at $t_0=0$, where the initial state $\sigma$ is sampled from the steady-state distribution, $p_\sigma^\ss$ (see Fig.~\ref{fig:residual_time}).

This differs from the mean waiting time between jumps in the steady state, which is dictated by the dynamical activity
\begin{equation}
    \label{dynamical_activity}
    \mathcal{A}\equiv \sum_{\sigma=1}^d p_\sigma^\ss \Gamma_\sigma.
\end{equation}
To show this, we need the probability that any given jump originates from state $\sigma$. Given that a jump is observed, the probability that the specific transition $\sigma\to \mu$ has occurred is $R_{\mu\sigma} p_\sigma^\ss/\mathcal{A}$ (see Appendix~\ref{SM:jump_ss}). Marginalising this over the final state $\mu$ yields the so-called jump steady state~\cite{Landi2023a}
\begin{equation}
\label{jump_steady_state}
    p_\sigma^\jss \equiv \frac{\Gamma_\sigma p^\ss_\sigma}{\mathcal{A}},
\end{equation} 
which can be interpreted as the expected frequency of the state $\sigma$ within a long sequence $\bm{\sigma}$. Conversely, the steady-state distribution $p_\sigma^\ss$ represents the expected fraction of the total time $t$ spent in state $\sigma$. The mean waiting time after a given jump $j$ in the bulk of the sequence, with $1\leq j< N$, is therefore $\EE[\tau_j] =\sum_\sigma p_\sigma^\jss \Gamma_\sigma^{-1} = \mathcal{A}^{-1}$. 

Note that $\mathcal{T} = \EE[\Gamma^{-1}_\sigma]$ is the \textit{arithmetic} mean of the conditional waiting times $\{\Gamma^{-1}_\sigma\}$ with respect to the steady-state distribution $\mathbf{p}^\ss$, while $\mathcal{A}^{-1}=\EE[\Gamma_\sigma]^{-1}$ is the \textit{harmonic} mean. We therefore have $\mathcal{T} \geq \mathcal{A}^{-1}$, with equality if and only if all escape rates $\Gamma_\sigma$ are equal~\footnote{This can be proved, for example, using the Cauchy-Schwarz inequality $\mathbf{a}^\T\mathbf{b} \leq \lVert \mathbf{a}\rVert \cdot \lVert \mathbf{b}\rVert$, with $a_\sigma = \sqrt{p^\ss_\sigma \Gamma_\sigma}$ and $b_\sigma = \sqrt{p^\ss_\sigma/\Gamma_\sigma}$. The necessary and sufficient condition for equality is $a_\sigma = \mathrm{const.} \times b_\sigma$, which implies $\Gamma_\sigma = \mathrm{const.}$ whenever $p^\ss_\sigma\neq 0$.}, i.e.~the residual time is greater than the waiting time between jumps on average. This surprising result arises because the unconditional waiting-time distribution, $W(\tau) = \sum_{\sigma} p_\sigma^\jss W(\tau|\sigma)$, is hyper-exponential and thus leads to bunching of jumps, as explained in Appendix~\ref{SM:inspection_paradox} and illustrated in Fig.~\ref{fig:residual_time}. This is an example of the inspection paradox~\cite{Stein1985, Pal2022}: the system is likely to be encountered in long-lived states where it spends more time, even if those states appear infrequently in a typical sequence~$\bm{\sigma}$.

\subsection{Counting states is sufficient for time estimation}
\label{sec:sufficient_statistics}

As a first step, we observe that the sequence probability~\eqref{path_probability} 
can be factorised as 
\begin{equation}
	\label{Fisher_Neyman_factorise}
	P(\bm{\sigma}|t) = \Pi(\bm{\sigma})g(\mathbf{n},t),
\end{equation}
where $\Pi(\bm{\sigma}) = \Gamma_{\sigma_N}^{-1} \prod_{j=0}^{N-1}\pi(\sigma_{j+1}|\sigma_j) p^{\ss}_{\sigma_0}$ and $g(\mathbf{n},t)$ is a convolution of $N+1$ waiting-time distributions. This can be written in the Laplace domain as a product, so that (see Appendix~\ref{SM:t-ensemble})
\begin{equation}
	\label{Fisher_Neyman_factor}
	g(\mathbf{n},t) = \int_{-\infty}^\infty \frac{\dd \chi}{2\pi}\, \ee^{-\ii \chi t} \prod_{\mu = 1}^d \left( \frac{\Gamma_\mu}{\Gamma_\mu - \ii \chi}\right )^{n_\mu}.
\end{equation}
Above, $n_\mu$ is the number of times that state $\mu$ is visited along the trajectory $\bm{\sigma}$, i.e.
\begin{equation}
    \label{empirical_distribution}
    n_\mu(\bm{\sigma}) = \sum_{j=0}^N \delta_{\mu\sigma_j}.
\end{equation}
The collection of these numbers, $\mathbf{n}(\bm{\sigma})= (n_1,\ldots,n_d)$, is known as the empirical distribution~\cite{Smiga2023} and is equivalent to the histogram of the states appearing in the sequence $\bm{\sigma}$.
Note that here it is convenient to use the unnormalised empirical distribution as defined in Eq.~\eqref{empirical_distribution}, in contrast to the standard definition~\cite{Touchette2011} $n'_\mu= n_\mu/(N+1)$ which obeys $\sum_\mu n'_\mu = 1$~\footnote{This should be distinguished from the empirical measure considered in large deviation theory~\cite{Garrahan2017}, which gives the fraction of total time spent in state $\sigma$. The empirical measure tends to the steady-state distribution $p_\sigma^\ss$ when averaged over a long observation time, while the empirical distribution tends to the jump steady state; see Appendix~\ref{SM:jump_ss} for an illustration}. 

Crucially, we see from Eq.~\eqref{Fisher_Neyman_factorise} that $P(\bm{\sigma}|t)$ depends on $t$ only through the factor $g(\mathbf{n},t)$. Therefore, by the Fisher-Neymann factorisation theorem, the empirical distribution $\mathbf{n}$ constitutes jointly sufficient statistics for estimating $t$~\cite{Kay2013}. In other words, an efficient time estimator (if it exists) can be constructed as a function of the empirical distribution alone, $\Theta(\bm{\sigma}) = \Theta(\mathbf{n})$. This is generally a significant data compression, since $\mathbf{n}$ comprises $d$ numbers whereas the trajectory $\bm{\sigma}$ describes $N$ jumps (or $N+1$ states), with $N$ arbitrarily large. Remarkably, therefore, the order of events is unimportant for time estimation. By contrast, for a generic parameter encoded in a Markov process, the Fisher information of the ordered sequence $\bm{\sigma}$ may greatly exceed that of the empirical distribution~\cite{Smiga2023}.

\subsection{Asymptotic Fisher information and efficient time estimator} 
\label{sec:asymptotic_fisher}

In general, evaluating the Fisher information is complicated, but the problem simplifies in the asymptotic limit of large $t$, where many jumps have occurred. In this limit, the integral in Eq.~\eqref{Fisher_Neyman_factor} is dominated by the contribution from the saddle point and we can write (see Appendix~\ref{SM:Fisher_limits})
\begin{equation}
	\label{score_saddle_point}
	\partial_t \ln P(\bm{\sigma}|t) = \left (\sum_{\sigma=1}^d \frac{\EE[n_\sigma]}{\Gamma_\sigma^2}\right )^{-1}\left (\sum_{\mu=1}^d\frac{n_\mu}{\Gamma_\mu} - t \right ) + \mathcal{O}(t^{-1}).
\end{equation}
Here, we have assumed that the empirical distribution scales with time, $n_\mu \sim t$, and that its fluctuations away from the mean, $\delta n_\mu = n_\mu - \EE[n_\mu],$ are subleading in the sense that $\delta n_\mu= \mathcal{O}(t^{1/2})$~\footnote{Note that we use the asymptotic (big-$\mathcal O$) notation following Knuth~\cite{Knuth1976} and by $X(t)\sim Y(t)$ we mean that $X(t)=\mathcal O(Y(t))$ and $Y(t)=\mathcal O(X(t))$, i.e.~$X$ is proportional to $Y$ asymptotically for large $t$.}. 

\begin{figure}
\includegraphics[width=\linewidth]{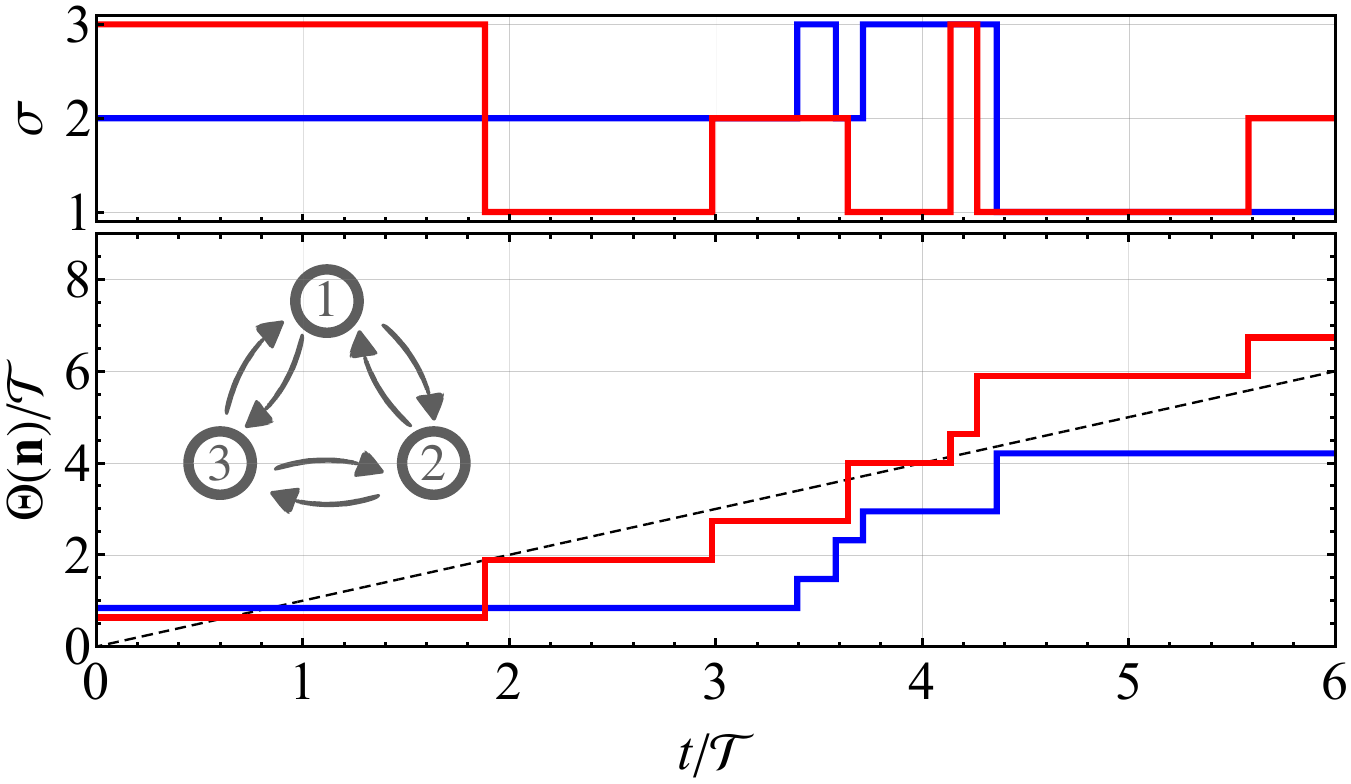}
\caption{Stochastic time estimation in a three-state system (sketched in the inset). The upper panel shows the state along two randomly sampled trajectories and the lower panel shows the associated time estimates given by Eq.~\eqref{efficient_estimator}, with the exact time $t$ indicated by the dashed line. We choose transition rates $\nolinebreak{R_{\mu\sigma} = \sigma \delta_{\mu,\sigma+1} + \delta_{\mu,\sigma-1}}$, with periodic boundary conditions implied ($p_{\sigma+3} \equiv p_\sigma$). The escape rates obey $\Gamma_1 < \Gamma_2 < \Gamma_3$. All times are expressed in units of the mean residual time $\mathcal{T}$.\label{fig:trajectories}}
\end{figure}

Equation \eqref{score_saddle_point} is of the form $\partial_t \ln P(\bm{\sigma}|t) = f(t)\left [\Theta(\mathbf{n}) -t\right ]$, where $f(t)$ is independent of the random sequence $\bm{\sigma}$. This is the necessary and sufficient condition for the estimator 
\begin{equation}
\label{efficient_estimator}
	\Theta(\mathbf{n}) = \sum_{\sigma=1}^d\frac{n_\sigma}{\Gamma_\sigma}
\end{equation}
to be unbiased and efficient, with $f(t)$ equal to the Fisher information~\cite{Kay2013}. The efficient estimator in Eq.~\eqref{efficient_estimator} has a simple physical interpretation: whenever the state $\sigma$ is observed, one increments the estimate by the expected waiting time until the next jump, $\Theta\to \Theta + \Gamma^{-1}_\sigma$; see Fig.~\ref{fig:trajectories} for an illustration. Despite its simplicity, this estimator achieves the CRB~\eqref{CRB} asymptotically, with the corresponding Fisher information given by the prefactor of Eq.~\eqref{score_saddle_point}:
\begin{equation}
	\label{asymptotic_Fisher}
	\mathcal{F}_t = \left (\sum_{\sigma=1}^d\frac{\EE[n_\sigma]}{\Gamma_\sigma^2}\right )^{-1} + \mathcal{O}(t^{-3/2})= \frac{1}{\mathcal{T}t} + \mathcal{O}(t^{-3/2}).
\end{equation}
Here, we have used $\EE[n_\sigma] =(1+\Gamma_\sigma t)p_\sigma^\ss$ (see Appendix~\ref{SM:small_n_estimator}). This is consistent with the definition of the jump steady state~\eqref{jump_steady_state}, as it implies $\EE[n_\sigma/(N+1)] = p_\sigma^\jss +\mathcal{O}(t^{-1/2}).$

For any unbiased time estimator, the CRB with Eq.~\eqref{asymptotic_Fisher} dictates that $\Var[\Theta] \geq \mathcal{T}t$ asymptotically. This implies that the relative error decreases no faster than $\Var[\Theta]/t^2 \sim t^{-1}$ for large times. This is the typical ``shot-noise'' scaling of the error with time in a continuous measurement, expected whenever temporal correlations are finite-ranged. Here, it appears as a consequence of the scaling $\mathcal{F}_t\sim t^{-1}$ embodied by Eq.~\eqref{asymptotic_Fisher}, whereas for estimation of a static parameter, $\lambda$, the long-time Fisher information scales as $\mathcal{F}_\lambda \sim t$~\cite{Radaelli2023}. The scaled Fisher information
\begin{equation}\label{Fisher_info_long_time}
	\lim_{t\to \infty} t\mathcal{F}_t = \mathcal{T}^{-1}
\end{equation} 
thus quantifies the asymptotic rate at which time information is acquired. 

We can also obtain a simple approximation for the scaled Fisher information in the short-time limit as $t\to 0$. Here, only trajectories with $N=0$ and $N=1$ jumps need be taken into account, and we find (see Appendix~\ref{SM:Fisher_limits})
\begin{equation}\label{Fisher_info_short_time}
\lim_{t\to 0} t\mathcal{F}_t = \mathcal{A} .
\end{equation}
In fact, we show in Sec.~\ref{sec:CUR_vs_KUR} [Ineq.~\eqref{Fisher_upper_bound}] that the upper bound $t \mathcal{F}_t \leq \mathcal{A}$ holds for all $t\geq 0$, implying that $t\mathcal{F}_t$ is maximal at short times. The corresponding short-time CRB is saturated by the unbiased estimator $\Theta(N) = N/\mathcal{A}$ (see Appendix~\ref{SM:Fisher_limits}). However, this estimator is generally not efficient away from the $t\to 0$ limit, i.e.~its variance grows more quickly than $\mathcal{T}t$, in general.
By contrast, the estimator defined in Eq.~\eqref{efficient_estimator} is only unbiased and efficient in the limit of large $t$, where the saddle-point approximation holds. In particular, as $t\to 0$, it has a bias $\EE[\Theta] = \EE[\Gamma_\sigma^{-1}]=\mathcal{T}$ and variance $\Var[\Theta] = \Var[\Gamma_\sigma^{-1}]$ (here averages are taken with respect to $\mathbf{p}^\ss$), as shown in Appendix~\ref{SM:small_n_estimator} and illustrated in Fig.~\ref{fig:trajectories}. The constant bias can be removed by the trivial shift $\Theta(\mathbf{n}) \to \Theta(\mathbf{n})  - \mathcal{T}$, although the resulting shifted estimator can take negative values when $t$ is small. However, there exist positive, unbiased, and asymptotically efficient estimators with smaller variance in the short-time limit, as we show in Sec.~\ref{sec:counting_observables}.

\begin{figure}
	\includegraphics[width = \linewidth]{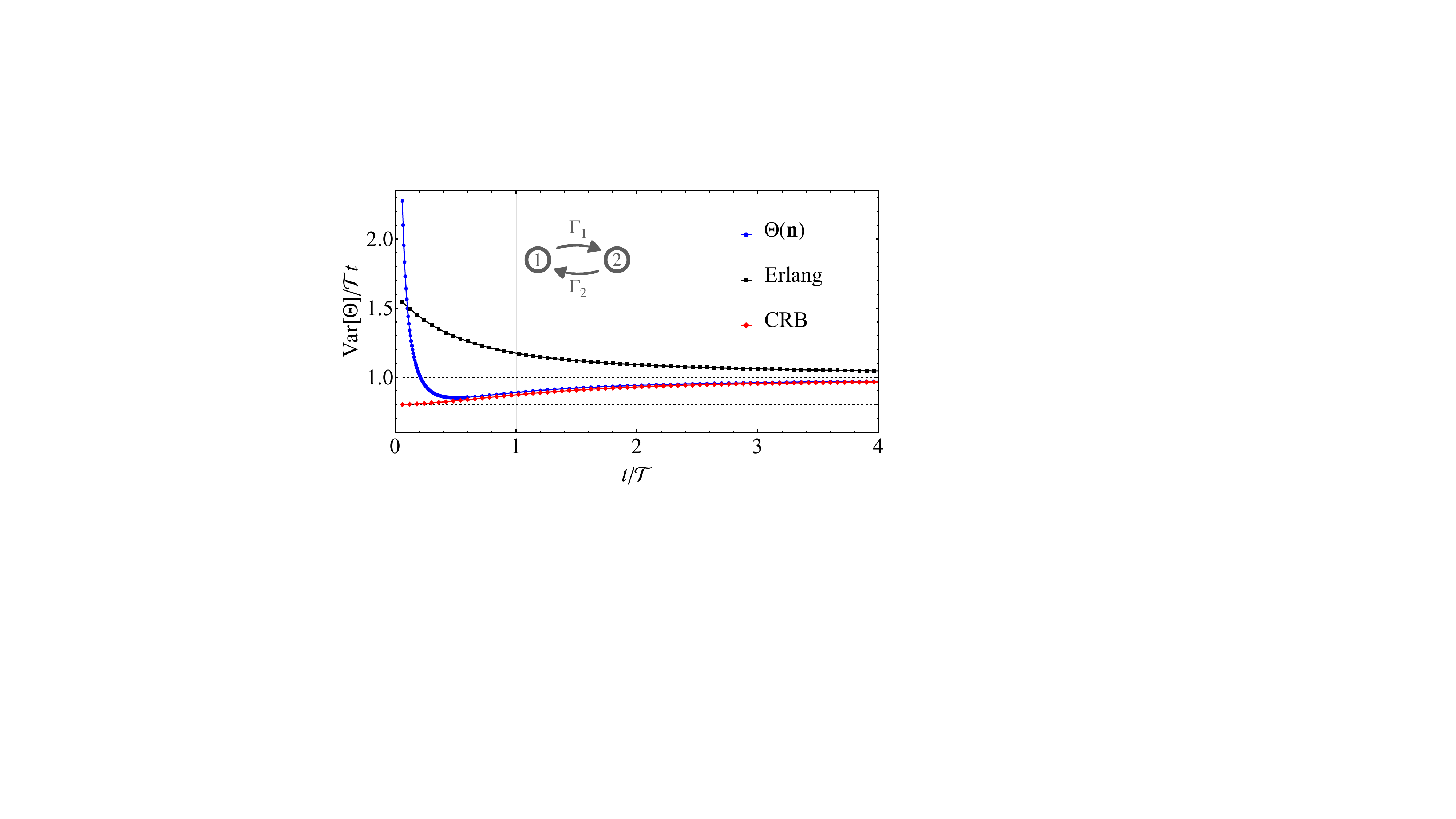}
	\caption{Time estimation error in the exactly solved two-state system (see inset) with imbalanced transition rates such that $\Gamma_1 = 3\Gamma_2$. Markers show the scaled variances of the asymptotically efficient estimator defined in Eq.~\eqref{efficient_estimator} (blue circles) and the Erlang estimator discussed in Sec.~\ref{sec:resolution} (black squares), as well as the Cram\'er-Rao bound set by the exact Fisher information, $(t\mathcal{F}_t)^{-1}$ (red diamonds). Horizontal dashed lines indicate the long- and short-time limits given by the mean residual time $\mathcal{T}$ (upper line) and the mean waiting time $\mathcal{A}^{-1}$ (lower line), respectively. 	\label{fig:estimators}}
\end{figure}

To assess the efficiency of the estimator~\eqref{efficient_estimator} and the validity of our approximations for the Fisher information, we analyse the example of a two-state system ($d=2$), which can be solved exactly (see Appendix~\ref{SM:exact_solution_d2} for details). Figure~\ref{fig:estimators} shows the scaled variance, $\text{Var}[\Theta]/t$, of the estimator in Eq.~\eqref{efficient_estimator} as a function of time for the two-state system (blue circles in Fig.~\ref{fig:estimators}). For comparison, we also plot the inverse scaled Fisher information $1/(t \mathcal{F}_t)$, computed without any approximation, which represents the lower bound on the scaled variance of any unbiased estimator (red diamonds in Fig.~\ref{fig:estimators}). We observe that the scaled variance of the estimator is significantly larger than $1/(t\mathcal{F}_t )$ for short times. However, after a relatively short time --- less than the mean residual time --- the variance drops to a value close to the CRB. The inverse scaled Fisher information itself is initially close to $\mathcal{A}^{-1}$ (lower dashed line in Fig.~\ref{fig:estimators}) as expected from Eq.~\eqref{Fisher_info_short_time}. Eventually, $1/(t\mathcal{F}_t )$ tends to the mean residual time $\mathcal{T}$ (upper dashed line in Fig.~\ref{fig:estimators}), in agreement with Eq.~\eqref{Fisher_info_long_time} and confirming the validity of the saddle-point approximation at long times.

\section{Clock uncertainty relation}
\label{sec:CUR}

Thus far, we have focussed on observables that are estimators for time. However, the generalised CRB~\eqref{CRB} applies to arbitrary functions $\Theta(\bm{\sigma})$. Therefore, using Eq.~\eqref{Fisher_info_long_time}, we can write down a general bound on the signal-to-noise ratio (SNR)
\begin{equation}
	\label{general_bound}
	\mathcal{S} = \frac{\left (\partial_t \EE[\Theta]\right )^2}{\Var[\Theta]/t} \leq \mathcal{T}^{-1},
\end{equation}
which holds in the long-time limit, with the mean residual time $\mathcal{T}$ given by Eq.~\eqref{mean_residual_time}. This is the main result of our work. 

Since Ineq.~\eqref{general_bound} constrains the precision of time estimation, we call it the clock uncertainty relation (CUR). Yet the CUR also bounds the precision of any observable $\Theta(\bm{\sigma})$ that depends on the jump sequence $\bm{\sigma}$ but not the jump times $\mathbf{t}$. The bound is non-trivial for time-extensive observables, such that $\EE[\Theta]\sim t$ and $\Var[\Theta]\sim t$. This includes the important class of counting observables, defined in Sec.~\ref{sec:counting_observables}, where we explicitly construct a counting observable that saturates the CUR. We then compare the CUR to other precision bounds in Sec.~\ref{sec:comparison} and explain its relation to the kinetic uncertainty relation in Sec.~\ref{sec:CUR_vs_KUR}. Sec.~\ref{sec:testing_CUR} concludes with an application of the CUR: the detection of hidden transitions from partial observations of steady-state jump and waiting-time statistics.

\subsection{Counting observables and the BLUE}
\label{sec:counting_observables}

A counting observable has the general form
\begin{equation}
	\label{counting_estimator}
	\Theta(\bm{\sigma}) = \sum_{\mu,\sigma = 1}^d w_{\mu\sigma} N_{\mu\sigma}= \int_0^t\dd t'\, I(t'),
\end{equation}
where $w_{\mu\sigma}$ are real constants, $N_{\mu\sigma}(t)$ is the total number of observed transitions $\sigma\to\mu$ within a time $t$, and the final equality defines the generalised current $I(t) = \dd\Theta/\dd t$. Here, a note on terminology is in order: we use the term ``current'' to describe the derivative of any counting observable such as Eq.~\eqref{counting_estimator}. This should be distinguished from ``thermodynamic currents'', which have antisymmetric weights, $w_{\mu\sigma} = -w_{\sigma \mu}$, and are therefore odd under time reversal. 

It is convenient to introduce the elementary probability currents $I_{\mu\sigma}(t) = \dd N_{\mu\sigma}/\dd t$, so that $N_{\mu\sigma}(t) = \int_0^t  \dd t' I_{\mu\sigma}(t')$ and $I(t) = \sum_{\mu,\sigma} w_{\mu\sigma} I_{\mu\sigma}(t)$.
At long times, the mean and covariance of the $\{N_{\mu\sigma}\}$ grow linearly in time, so that
\begin{align}
	\label{current_noise_def}
	&   \EE[N_{\mu\sigma}(t)] = J_{\mu\sigma} t,\notag \\
	&  \Cov[N_{\alpha\beta}(t),N_{\mu\sigma}(t)]  \to D_{\alpha\beta\mu\sigma}t,
\end{align}
where $\Cov[A,B] = \EE[AB]-\EE[A]\EE[B]$. Here, $J_{\mu\sigma} = \EE[I_{\mu\sigma}]$ is the mean steady-state current (i.e.~the mean number of transitions per unit time from $\sigma\to\mu$), while $D_{\alpha\beta\mu\sigma}$ is the diffusion tensor. Using the techniques of Ref.~\cite{Landi2024}, we show in Appendix~\ref{SM:current_statistics} that
\begin{align}
	\label{average_currents}
	&	J_{\mu\sigma} = R_{\mu\sigma} p_\sigma^{\ss},\\
	\label{diffusion_tensor}
	&	D_{\alpha\beta\mu\sigma} = \delta_{\alpha\mu}\delta_{\beta\sigma} J_{\mu\sigma} - R_{\alpha\beta}L^{+}_{\beta\mu}J_{\mu\sigma} -   R_{\mu\sigma}L^{+}_{\sigma\alpha}J_{\alpha\beta}.
\end{align}
Above, $L^{+}_{\mu\sigma} = [\mathsf{L}^{+}]_{\mu\sigma}$ are matrix elements of the Drazin inverse of $\mathsf{L}$: the unique matrix obeying $\mathsf{L}^{+}\mathsf{L} = \mathsf{L}\mathsf{L}^{+} = \mathsf{1} - \mathsf{\Pi}_0$, where $\mathsf{1}$ is the identity matrix and $\mathsf{\Pi}_0$ is the projector onto the null eigenspace of $\mathsf{L}$. It follows that the asymptotic mean and variance of the counting observable~\eqref{counting_estimator} also grow linearly in time with rates $\partial_t \EE[\Theta] = J$ and $\partial_t \Var[\Theta] \to D$, where the mean current $J$ and diffusion coefficient $D$ are given by
\begin{align}
	\label{linear_estimator_mean}
	& J = \sum_{\mu,\sigma=1}^d w_{\mu\sigma} J_{\mu\sigma} \equiv \vec{w}\cdot \vec{J},\\ 
	\label{linear_estimator_var}
	& D =  \sum_{\alpha,\beta,\mu,\sigma=1}^dw_{\alpha\beta} D_{\alpha\beta\mu\sigma}  w_{\mu\sigma} \equiv \vec{w}\cdot \mathbb{D}\cdot \vec{w}.
\end{align}
Here, $\vec{w}$ and $\vec{J}$ denote ``vectors'' with $d^2$ components, while the tensor $\mathbb{D}$ maps one such vector to another. 

To find counting observables that saturate the bound~\eqref{general_bound}, we choose weights that maximise the SNR
\begin{equation}
	\label{SNR_counting}
	\mathcal{S} = \frac{J^2}{D} = \frac{\left (\vec{w}\cdot \vec{J}\;\right )^2}{\vec{w}\cdot\mathbb{D}\cdot\vec{w}}.
\end{equation}
A current that maximises the SNR is known as a hyperaccurate current~\cite{Busiello2019,Timpanaro2023}; here, we extend this notion to general counting observables. Since $\mathcal{S}$ is invariant under uniform rescaling of the weights, $\vec{w}\to r\vec{w}$,  we are free to rescale $\vec{w}$ to satisfy $\vec{w}\cdot\vec{J}=1$. This guarantees an unbiased estimator [cf.~Eqs.~\eqref{unbiased_estimator} and~\eqref{linear_estimator_mean}]. Maximising $\mathcal{S}$ is then equivalent to minimising the denominator of Eq.~\eqref{SNR_counting} under the constraint $\vec{w}\cdot\vec{J}=1$, which yields a condition that the optimal weights must satisfy (see Appendix~\ref{SM:current_statistics}):
\begin{equation}
	\label{BLUE}
	\frac{\mathbb{D}\cdot \vec{w}}{\vec{w}\cdot\mathbb{D}\cdot\vec{w}}= \vec{J}.
\end{equation}
If $\mathbb{D}$ were invertible, the solution of this equation would be of the form $\vec{w}\propto \mathbb{D}^{-1}\cdot \vec{J}$, which gives more weight to elementary currents with low variance relative to their mean, while also accounting for current-current correlations. In general, however, the diffusion tensor is singular: it possesses at least $d-1$ linearly independent, null eigenvectors $\vec{w}$ such that $\mathbb{D}\cdot\vec{w} = 0$ (see Appendix~\ref{SM:current_statistics} for their explicit construction). Therefore, the solution of Eq.~\eqref{BLUE} is not unique, and there exist several inequivalent counting observables with maximal SNR. 

A particularly elegant solution of Eq.~\eqref{BLUE} is $w_{\mu\sigma} = \Gamma_\sigma^{-1}$, as can be verified directly (see Appendix~\ref{SM:current_statistics} for details). The resulting time estimator is
\begin{equation}
	\label{BLUE_estimator}
	\Theta_{\rm BLUE}(\bm{\sigma}) =  \vec{w}\cdot \vec{N} = \sum_{\mu,\sigma=1}^d \frac{N_{\mu\sigma}}{\Gamma_\sigma},
\end{equation}
which we refer to as the best linear unbiased estimator (BLUE), following Ref.~\cite{Kay2013}. The corresponding SNR in the long-time limit is $\mathcal{S} = \mathcal{T}^{-1}$, which shows that the BLUE saturates the CUR~\eqref{general_bound}. Strictly speaking, any solution of Eq.~\eqref{BLUE} qualifies as a BLUE because it yields an unbiased estimator with the same SNR, but for clarity we reserve this terminology for the estimator defined by Eq.~\eqref{BLUE_estimator}. 

The meaning of Eq.~\eqref{BLUE_estimator} is that, whenever a transition $\sigma\to\mu$ is observed, the BLUE increments by the expected waiting time preceding the jump, $\Gamma_\sigma^{-1}$. It is therefore closely analogous, but not strictly equivalent, to the estimator $\Theta(\mathbf{n})$ defined by Eq.~\eqref{efficient_estimator}. More precisely, we have $\Theta_{\rm BLUE}(\bm{\sigma}) = \Theta(\mathbf{n}) - \Gamma_{\sigma_N}^{-1}$, because $n_\sigma = \sum_\mu N_{\mu\sigma} + \delta_{\sigma\sigma_N}$. While $\Theta(\mathbf{n})$ is increased by $\Gamma_\sigma^{-1}$ whenever a jump \textit{into} state $\sigma$ is observed (including the first observation $\sigma_0$), $\Theta_{\rm BLUE}(\bm{\sigma})$ is increased by $\Gamma_\sigma^{-1}$ whenever a jump \textit{out of} state $\sigma$ is observed. The final state $\sigma_N$ therefore does not contribute to the BLUE. This corrects for the bias and excess variance of $\Theta(\mathbf{n})$ at short times, where $N = 0$ with high probability. At long times, however, the difference between the two estimators becomes negligible.  

In fact, the BLUE saturates the CUR for all $t$, not only in the long-time limit. To show this, we compute the current autocorrelation function $F(\tau) = \Cov[I(t+\tau), I(t)]$, finding
\begin{equation}
	\label{autocorrelation_function}
	F_{\rm BLUE}(\tau) = \mathcal{T}\delta(\tau),
\end{equation}
as shown in Appendix~\ref{SM:current_statistics}. The variance then follows as
\begin{align}
	\label{variance_autocorrelation}
	& \Var[\Theta_{\rm BLUE}] = \int_0^t\dd t_1\int_0^t\dd t_2\, F_{\rm BLUE}(t_1-t_2) = \mathcal{T}t,
\end{align}
and hence Ineq.~\eqref{general_bound} is satisfied for all $t$. Note that this implies that the variance of $\Theta(\mathbf{n})$ [Eq.~\eqref{efficient_estimator}] may drop below that of the BLUE at intermediate times, as illustrated in Fig.~\ref{fig:estimators} and discussed in more detail in Appendix \ref{SM:small_n_estimator}.

Surprisingly, the form of Eq.~\eqref{autocorrelation_function} is reminiscent of pure Poissonian or white noise~\cite{Landi2024}, whereas intuitively one expects a good clock to tick at regularly spaced intervals. In Sec.~\ref{sec:beyond_precision} we discuss other optimal estimators that  yield more regular ticks but at the cost of reduced time resolution.

\subsection{Comparison to other precision bounds}
\label{sec:comparison}

\begin{table}
\begin{ruledtabular}
	\begin{tabular}{ccc}
		Bound $\mathcal{S}\leq \mathcal{C}$ & Cost function $\mathcal{C}$ &  Observable $\Theta$ \\ 	 \toprule \vspace{-3mm}	\\
		CUR  &$\mathcal{T}^{-1} = \left ( \displaystyle\sum_\sigma \frac{p_\sigma^\ss}{\Gamma_\sigma } \right )^{-1} $ &  $\Theta(\bm{\sigma})$ \\
  \vspace{-3mm}	\\
        KUR &  $ \mathcal{A} = \displaystyle\sum_\sigma p_\sigma^\ss \Gamma_\sigma $   & $\Theta(\bm{\sigma})$ \\
        
		TUR & $\dfrac{\dot{\Sigma} }{2} =\displaystyle\frac{1}{2}\sum_{\mu,\sigma}J_{\mu\sigma} \ln\left (\frac{J_{\mu\sigma}}{J_{\sigma\mu}}\right ) $  & $\begin{array}{c}
			\Theta(\bm{\sigma}) = \vec{w}\cdot\vec{N} \\ w_{\mu\sigma} = -w_{\sigma\mu}
		\end{array}$
	\end{tabular}
\end{ruledtabular}
	\caption{Summary of three bounds on the signal-to-noise ratio (SNR), $\mathcal{S} = (\partial_t\EE[\Theta])^2/(\Var[\Theta]/t)$, specifying the upper bound $\mathcal{C}$ and the class of relevant observables $\Theta$. \label{tab:bounds}}
\end{table}

Recently, numerous general bounds on precision have been discovered for observables in non-equilibrium steady states of classical Markovian systems, including the thermodynamic uncertainty relation (TUR)~\cite{Barato2015a,Gingrich2016} and kinetic uncertainty relation (KUR)~\cite{Terlizzi2018,Garrahan2017}. In general, we can write such bounds as $\mathcal{S}\leq \mathcal{C}$, where the quantity $\mathcal{C}$ is a measure of dissipation or activity (see Table~\ref{tab:bounds} for precise definitions). The CUR~\eqref{general_bound} ($\mathcal{C}=\mathcal{T}^{-1}$) is most closely related to the KUR ($\mathcal{C}=\mathcal{A}$) because both $\mathcal{A}$ and $\mathcal{T}^{-1}$ can be understood as a measure of the activity or freneticity of the stochastic process. However, since $\mathcal{T}^{-1} \leq \mathcal{A}$, with equality if and only if all escape rates $\Gamma_\sigma$ are equal, the CUR is generally tighter than the KUR. 

\begin{figure}
	\includegraphics[width = \linewidth]{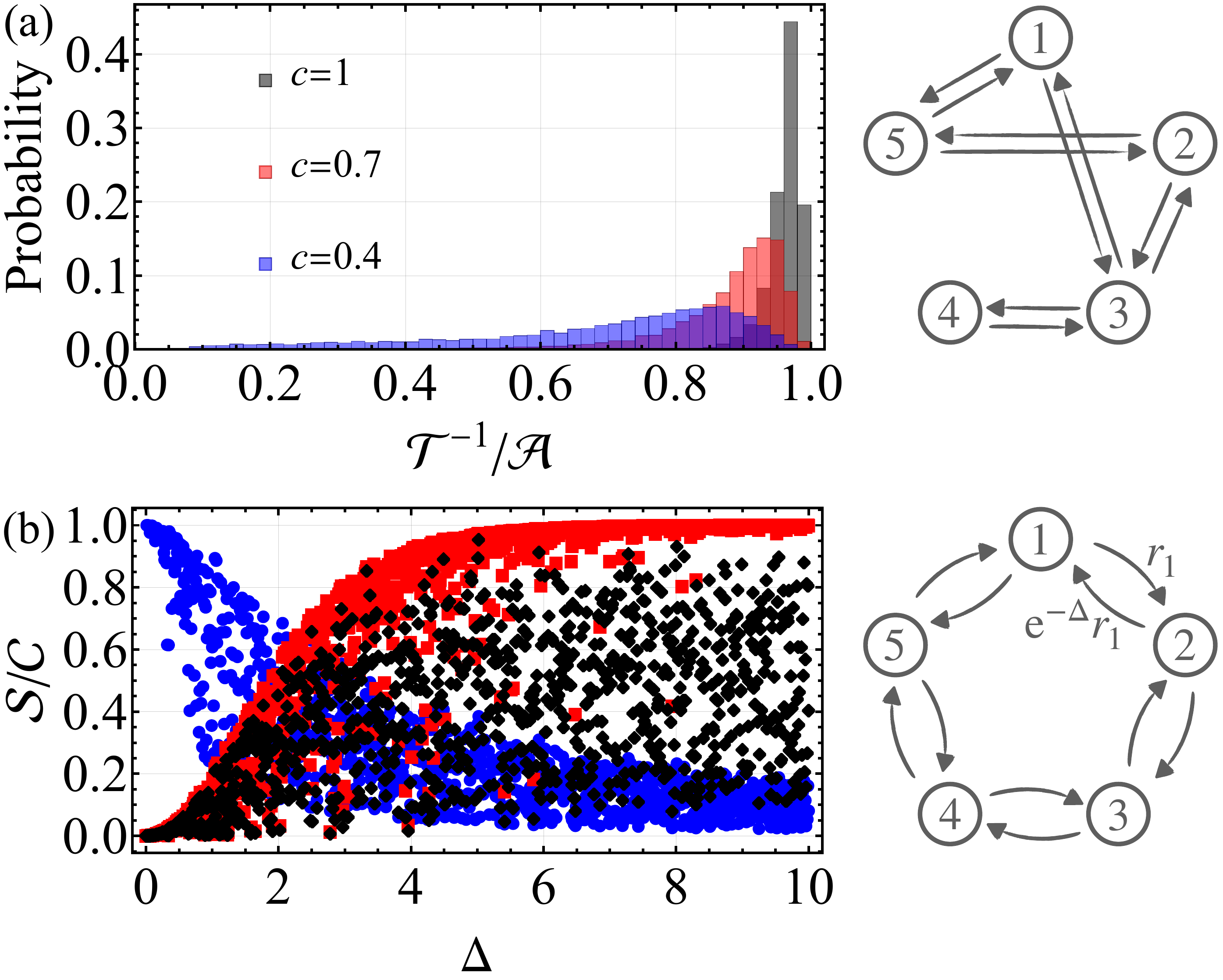}
	\caption{Comparison of precision bounds (see Table~\ref{tab:bounds}) for random sparse and unicyclic networks; diagrams on the right show example networks with $d=5$. (a)~Histogram of the ratio between the inverse mean residual time, $\mathcal{T}^{-1}$, and the dynamical activity, $\mathcal{A}$, in random networks with connectivity $c$. With probability $c$, a given entry of the rate matrix is sampled independently from a uniform distribution $R_{\mu\sigma}\in [0,1]$, and with probability $1-c$ we set $R_{\mu\sigma}=0$. Only networks with a single connected component are considered. Results are shown for $d=10$ states and $10^4$ random networks sampled for each value of $c$. The distribution of $\mathcal{T}^{-1}/\mathcal{A}$ moves to the left as $c$ decreases. 
		(b)~Precision-cost ratio $\mathcal{S}/\mathcal{C} \leq 1$ for random ring networks as a function of the bias, $\Delta $. For each value of $\Delta$, we generate a ring network with transition rates $R_{\mu\sigma} = r_\sigma \delta_{\mu,\sigma+1} + \ee^{-\Delta } r_\mu \delta_{\mu,\sigma-1} $, where $r_\sigma \in [0,1]$ is a uniformly distributed random number, the bias $\Delta$ favours jumps in the clockwise direction, $\sigma\to \sigma+1$, and periodic boundary conditions are imposed. For each network generated, we compute the ratio $\mathcal{S}/\mathcal{C} $ for the TUR (blue circles, $\mathcal{C} = \tfrac{1}{2}\dot{\Sigma}$), KUR (black diamonds, $\mathcal{C} = \mathcal{A}$), and CUR (red squares, $\mathcal{C}=\mathcal{T}^{-1}$). Results are shown for $d=10$ states and the counting observable $\Theta = \vec{w}\cdot\vec{N}$, where $w_{21}= - w_{12}= 1$ and all other $w_{\mu\sigma}=0$.
		\label{fig:bounds}}
\end{figure}

To clarify the difference between these two measures of freneticity for generic systems, we generate random networks by sampling stochastic generators $\mathsf{L}$ with random coefficients and compare the corresponding values of $\mathcal{T}^{-1}$ and $\mathcal{A}$. Results are shown in Fig.~\ref{fig:bounds}(a) for networks with $d=10$; see the caption for details of our sampling procedure. When every state is connected to every other one by a non-zero transition rate $R_{\mu\sigma}\neq 0$, we find that $\mathcal{T}^{-1}\approx \mathcal{A}$ with high probability (black bars in Fig.~\ref{fig:bounds}(a)). This effect becomes more pronounced as the dimension $d$ grows larger, which means that the KUR and CUR are approximately equivalent for high-dimensional generic networks lacking any particular structure. This is easily understood by the fact that, for a typical structureless network with many states, the steady state closely approximates a uniform distribution, $p^\ss_\sigma \approx d^{-1}$, while the escape rates concentrate around their mean value, $\Gamma_\sigma \approx d\mathcal{A} + \mathcal{O}(d^{1/2})$, as a consequence of the central limit theorem. To leading order for large $d$, therefore, all escape rates are approximately equal and their harmonic and arithmetic means are indistinguishable.

This picture changes when we add structure to the network by reducing its connectivity. Specifically, we generate sparse random networks by allowing direct transitions between pairs of states ($R_{\mu\sigma}\neq 0$) with probability $c$, while with probability $1-c$ we set $R_{\mu\sigma}=0$. We consider only networks with a single connected component, which ensures that the steady state is unique. As the connectivity $c$ is decreased, the differences between $\mathcal{T}^{-1}$ and $\mathcal{A}$ become more pronounced on average and the distribution of their ratio broadens to encompass smaller values (red and blue bars in Fig.~\ref{fig:bounds}(a)). Qualitatively similar results are found for networks with higher dimension $d$, but the connectivity $c$ needed to see substantial differences between $\mathcal{T}^{-1}$ and $\mathcal{A}$ gets progressively smaller as $d$ grows large. In summary, the CUR imposes a tighter bound on fluctuations than the KUR whenever $\mathcal{T}^{-1}<\mathcal{A}$, which occurs generically in low-dimensional or sparse stochastic networks. 

While optimal counting observables exist that saturate the CUR for any stochastic network (Sec.~\ref{sec:counting_observables}), one may also be interested in the fluctuations of one specific, sub-optimal observable that is singled out by the structure of the network. To illustrate how the TUR, KUR and CUR differ in this case, we consider the current flowing around a unicyclic network of states --- a typical model for biochemical clocks~\cite{Barato2016,Marsland2019} and sensors~\cite{Lang2014,Harvey2023}. To incorporate disorder, we take random transition rates but an overall positive bias for jumps in the clockwise direction; see the caption of Fig.~\ref{fig:bounds} for details. The bias breaks detailed balance and entails a finite rate of entropy production, $\dot{\Sigma}$. We focus on the thermodynamic current flowing between the first two sites which, due to the constrained network geometry, equals the current flowing between any other pair of adjacent sites. Figure~\ref{fig:bounds}(b) shows the corresponding precision-cost ratio $\mathcal{S}/\mathcal{C}$ for the three different bounds in Table~\ref{tab:bounds}. At low bias, the TUR is clearly the tightest bound, while the CUR and KUR are similarly loose. However, for high bias the precision tends to $\mathcal{S}\to \mathcal{T}^{-1}$, demonstrating that the CUR is the most informative bound. 

These findings are consistent with previous studies showing that precision is bounded tightly by dissipation close to equilibrium and by activity far away from it~\cite{Terlizzi2018,Hiura2021}. Yet, in addition, Fig.~\ref{fig:bounds}(b) reveals that the relevant measure of activity far from equilibrium is the inverse mean residual time, $\mathcal{T}^{-1}$, as this generically yields the tightest bound on current fluctuations within this regime. In Appendix~\ref{SM:traffic_flow}, we also compare the CUR with two entropic bounds on counting observables derived recently in Ref.~\cite{Pietzonka2024}, finding again that the CUR is the tightest bound far from equilibrium.

\subsection{Relation between the KUR and CUR}
\label{sec:CUR_vs_KUR}

While the kinetic (KUR) and clock (CUR) uncertainty relations are quantitatively different bounds, they are related both conceptually and mathematically. In Ref.~\cite{Terlizzi2018}, the KUR was derived using the fluctuation-response inequality~\cite{Dechant2020} but it can also be cast in terms of a Cram\'er-Rao bound (CRB)~\cite{Shiraishi2021,Vu2022}, as can other uncertainty relations in stochastic thermodynamics~\cite{Dechant2018,Hasegawa2019,Hiura2021,Vo2022}. However, the nature of the parameter estimation problem is different in each case. 

To obtain the KUR from the CRB, one considers a perturbed generator $\mathsf{L}\to(1+\lambda)\mathsf{L}$, where $\lambda$ is a small parameter that will ultimately be taken to zero. The corresponding trajectory probability distribution, including the jump times, is obtained from Eq.~\eqref{path_probability_with_times} as
\begin{equation}
    \label{KUR_perturbed_path_prob}
P(\bm{\sigma},\mathbf{t}|t,\lambda) = (1+\lambda)^N \ee^{-\lambda \sum_{j=0}^N \Gamma_{\sigma_j} \tau_{j}} P(\bm{\sigma},\mathbf{t}|t),
\end{equation}
 where $\tau_j = t_{j+1}-t_j$ is the waiting time after jump $j$ and $N$ denotes the total number of jumps. For clarity, in this section we write expectation values with respect to the perturbed distribution~\eqref{KUR_perturbed_path_prob} as
\begin{equation}
    \label{perturbed_expectation}
\EE[\Theta|t,\lambda] = \sum_{\bm{\sigma}} \int_0^t\dd t_N \cdots \int_0^{t_2}\dd t_1\, P(\bm{\sigma},\mathbf{t}|t,\lambda) \Theta(\bm{\sigma},\bm{\tau}),
\end{equation}
where $\Theta(\bm{\sigma},\bm{\tau})$ is an arbitrary function of the sequence $\bm{\sigma}$ and the waiting times $\bm{\tau}$, while expectations with respect to the unperturbed distribution ($\lambda=0$) are written as $\EE[\Theta|t]$. Treating the observable $\Theta(\bm{\sigma},\bm{\tau})$ as a (biased) estimator of $\lambda$, its variance obeys the CRB
\begin{equation}
    \label{CRB_KUR}
	\Var[\Theta|t,\lambda]\geq\frac{ \left (\partial_\lambda \EE[\Theta|t,\lambda]\right )^2}{\mathcal{F}_\lambda[P(\bm{\sigma},\mathbf{t}|t,\lambda)]},	
\end{equation}
where the relevant Fisher information here is~\cite{Kay2013}
\begin{equation}
	\label{Fisher_info_KUR}
	\mathcal{F}_\lambda[P(\bm{\sigma},\mathbf{t}|t,\lambda)] = -\EE\left [\partial_\lambda^2\ln P(\bm{\sigma},\mathbf{t}|t,\lambda)\big|t,\lambda\right] = \frac{\EE[N|t,\lambda]}{(1+\lambda)^2}.
\end{equation}
The second equality above follows easily from Eq.~\eqref{KUR_perturbed_path_prob}.

Now, the KUR is obtained from Ineq.~\eqref{CRB_KUR} in the limit $\nolinebreak{\lambda\to 0}$. For a generic observable, one has 
\begin{equation}
   	\label{KUR_expect_expand}
	\frac{\partial}{\partial \lambda}\EE\left [\Theta|t,\lambda\right ] = t \frac{\partial}{\partial t} \EE[\Theta|t] - \EE\left [\sum_{j=0}^N \tau_j\frac{\partial \Theta}{\partial \tau_j}\Bigg|t\right ] +\mathcal{O}(\lambda).
\end{equation}
This can be shown by introducing new integration variables $t_j' = (1+\lambda)t_j$ in Eq.~\eqref{perturbed_expectation}, as detailed in Appendix~\ref{SM:KUR}. For counting observables, or any other observable $\Theta(\bm{\sigma})$ that does not depend explicitly on the waiting times $\bm{\tau}$, the second term on the right-hand side of Eq.~\eqref{KUR_expect_expand} vanishes. Therefore, combining Eqs.~\eqref{CRB_KUR}--\eqref{KUR_expect_expand}, setting $\lambda =0$, and rearranging, we obtain the KUR (valid for all $t>0$)~\cite{Terlizzi2018}
\begin{equation}
    \label{KUR}
	\mathcal{S} = \frac{(\partial_t \EE[\Theta|t])^2}{\Var[\Theta|t]/t} \leq \mathcal{A},
\end{equation}
because the expected number of jumps is $\EE[N|t]= \mathcal{A}t$.

However, Ineq.~\eqref{KUR} holds only for observables that are independent of the waiting times $\bm{\tau}$. These observables are governed by the distribution $P(\bm{\sigma}|t)$, obtained in Eq.~\eqref{path_probability} by marginalising over the jump times $\mathbf{t}$. We now show that the KUR cannot generally be saturated for such observables, due to the existence of the tighter CUR. To see this, note that any observable $\Theta(\bm{\sigma})$ depending only on the sequence $\bm{\sigma}$ obeys the refined CRB
\begin{equation}
    	\label{sequence_CRB}
		 \Var[\Theta|t,\lambda]\geq \frac{(\partial_\lambda \EE[\Theta|t,\lambda])^2}{\mathcal{F}_\lambda[P(\bm{\sigma}|t,\lambda)]},
\end{equation}
where $\mathcal{F}_\lambda[P(\bm{\sigma}|t,\lambda)]$ is the Fisher information for the distribution $P(\bm{\sigma}|t,\lambda)$, obtained by marginalising Eq.~\eqref{KUR_perturbed_path_prob} over the jump times $\mathbf{t}$ (or, equivalently, by shifting $\mathsf{L} \to (1+\lambda)\mathsf{L}$ in Eq.~\eqref{path_probability}). This yields a tighter bound because of the data refinement inequality for the Fisher information~\cite{Zamir1998}
\begin{equation}
	\label{data_refinement_inequality}
	\mathcal{F}_\lambda[P(\bm{\sigma}|t,\lambda)] \leq  \mathcal{F}_\lambda[P(\bm{\sigma},\mathbf{t}|t,\lambda)].
\end{equation}
Therefore, the KUR~\eqref{KUR} cannot be saturated unless Ineq.~\eqref{data_refinement_inequality} is saturated in the limit $\lambda \to 0$. 

In fact, we show in Appendix~\ref{SM:KUR} that
\begin{equation}
	\label{P_KUR_speed_up}
	P(\bm{\sigma}|t,\lambda) = P\big(\bm{\sigma}|(1+\lambda)t\big),
\end{equation}
i.e.~for observables that do not depend on the jump times $\mathbf{t}$, shifting the rates $\mathsf{L} \to (1+\lambda )\mathsf{L}$ is equivalent to rescaling time as $t\to (1+\lambda)t$. It follows that 
\begin{equation}
\label{Fisher_info_t_from_lambda}
\lim_{\lambda\to 0}\mathcal{F}_\lambda[P(\bm{\sigma}|t,\lambda)] = t^2 \mathcal{F}_{t},
\end{equation}
where $\mathcal{F}_{t}\equiv  \mathcal{F}_t[P(\bm{\sigma}|t)]$ is the Fisher information for time estimation, defined in Eq.~\eqref{Fisher_info}. As a consequence, for $\lambda\to 0$, Ineq.~\eqref{sequence_CRB} reduces to the CRB~\eqref{CRB} for time estimation, which in turn implies the CUR. The fact that the CUR is tighter than the KUR can thus be seen as a consequence of the data refinement inequality~\eqref{data_refinement_inequality}. Furthermore, combining Eqs.~\eqref{Fisher_info_KUR},~\eqref{data_refinement_inequality}, and~\eqref{Fisher_info_t_from_lambda}, we deduce an upper bound on the Fisher information for stochastic time estimation
\begin{equation}
	\label{Fisher_upper_bound}
	t\mathcal{F}_{t} \leq \mathcal{A},
\end{equation}
which holds for all $t> 0$. This bound is generally loose but it is saturated at short times, $ \lim_{t\to 0} t\mathcal{F}_t = \mathcal{A},$ as discussed in Sec.~\ref{sec:asymptotic_fisher}. 

\begin{figure}
\includegraphics[width=\linewidth]{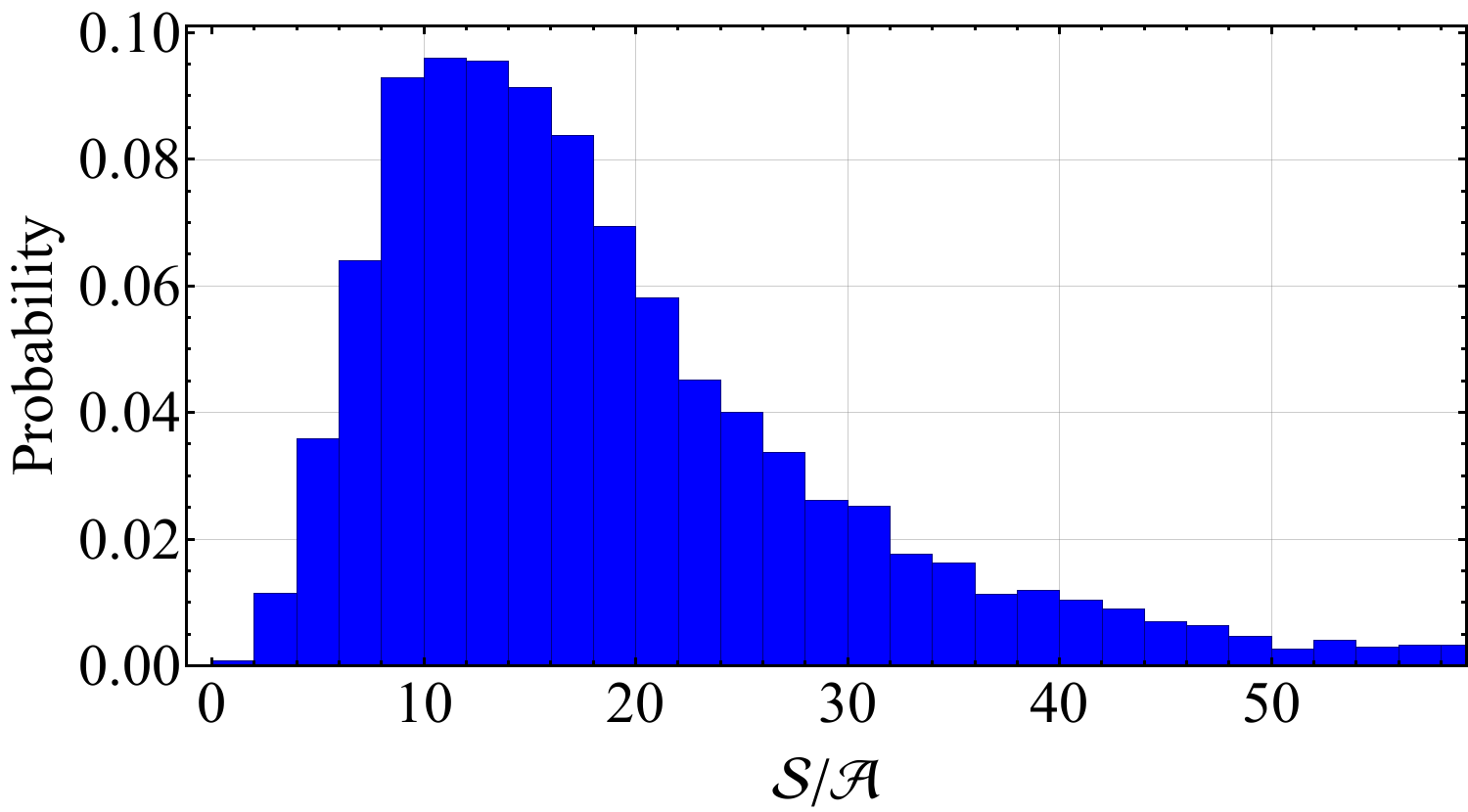}
\caption{Asymptotic signal-to-noise ratio $\mathcal{S}$ of the observable $\Theta(\bm{\sigma},\bm{\tau}) = \sum_{j=0}^N \Gamma_{\sigma_j}\tau_j$ computed for $10^4$ randomly generated, fully connected stochastic networks (see Appendix~\ref{SM:KUR} for details of the calculation). We use the sampling procedure described in the caption of Fig.~\ref{fig:bounds}, with connectivity $c=1$ and dimension $d=10$. \label{fig:KUR_violation}}
\end{figure}

Conversely, the KUR does not hold whenever the second term on the right-hand side of Eq.~\eqref{KUR_expect_expand} is non-zero. This is the case for generic observables that depend explicitly on the waiting times, which are measurable only for observers who have access to an external clock. A simple example of such an observable is the sum of waiting times, $\Theta(\bm{\tau}) = \sum_{j=0}^N \tau_j = t$, equal to the total time $t$. Its variance therefore vanishes, which is nonetheless consistent with the CRB~\eqref{CRB_KUR} because, from Eq.~\eqref{KUR_expect_expand}, we have $\partial_\lambda \EE[\Theta|t,\lambda]_{\lambda=0} = t - t =0$. A non-trivial observable whose fluctuations are not bounded by the KUR is $\Theta(\bm{\sigma},\bm{\tau}) = \sum_{j=0}^N \Gamma_{\sigma_j}\tau_j$. We demonstrate this by computing its SNR for a sample of randomly generated stochastic networks, as shown in Fig.~\ref{fig:KUR_violation}. In the vast majority of cases, we see that $\mathcal{S}\gg \mathcal{A}$, strongly violating the KUR.

Finally, one may wonder whether it is possible to formulate the time estimation problem for this more general class of observables that depend explicitly on the waiting times. Interestingly, the CRB for time estimation does not exist in this case: the support of the distribution $P(\bm{\sigma},\mathbf{t}|t)$ depends on $t$ and so the regularity condition $\EE[\partial_t\ln P(\bm{\sigma},\mathbf{t}|t)]=0$, which is necessary to derive the CRB~\cite{Kay2013}, does not hold.  Naturally, therefore, the problem of stochastic time estimation is only well posed if a clock is not already available to record the jump times $\mathbf{t}$, so that only the information in the marginal distribution $P(\bm{\sigma}|t)$ is accessible.

\subsection{Testing the CUR to detect hidden transitions}
\label{sec:testing_CUR}

\begin{figure}
	\includegraphics[width=\linewidth]{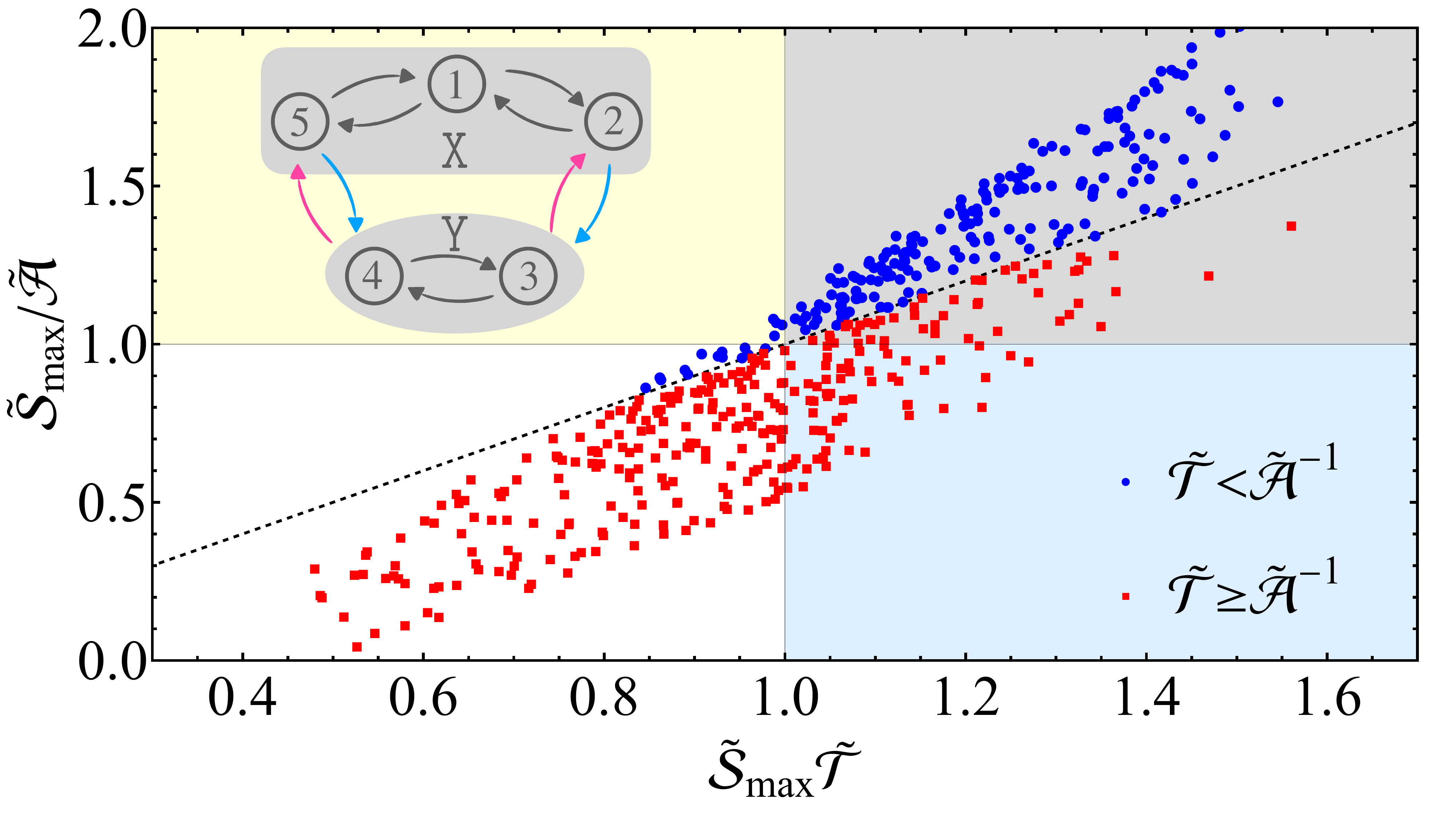}
	\caption{Detecting hidden transitions.  We consider the hidden Markov model shown in the inset, with (hidden) transition rates $R_{\mu\sigma} = r_\sigma \delta_{\mu,\sigma+1} + \ee^{-\Delta } r_\mu \delta_{\mu,\sigma-1}$ and periodic boundary conditions. Each data point corresponds to one random realisation generated by independently sampling the rates $r_\sigma \in [0,1]$ and bias $\Delta \in [0,5]$ from uniform distributions. We compute the optimal SNR $\tilde{\mathcal{S}}_{\rm max}$ and compare to the mean residual time $\tilde{\mathcal{T}}$ and waiting time $\tilde{\mathcal{A}}^{-1}$ for observable transitions $\mathtt{X} \leftrightarrow \mathtt{Y}$; see Appendix~\ref{SM:hidden_markov} for details of the calculation. Hidden transitions are detected whenever $\tilde{\mathcal{S}}_{\rm max} \tilde{\mathcal{T}}\neq 1$.  Hidden transitions are also witnessed if $\tilde{\mathcal{T}}  < \tilde{\mathcal{A}}^{-1}$ (blue circles), or by violations of the KUR (top-left yellow quadrant), CUR (bottom-right blue quadrant), or both (top-right grey quadrant). 
		\label{fig:hidden_markov}}
\end{figure}

We have shown in Sec.~\ref{sec:counting_observables} that the CUR~\eqref{CUR} is saturated by a class of optimal counting observables with maximal SNR. 
For such observables, the CUR becomes an equality
\begin{equation}
\label{CUR_equality}
    \mathcal{S}_{\rm max} = \mathcal{T}^{-1}.
    \end{equation}
If all transitions are observable, this relation can be verified experimentally from the measured jump statistics, as explained below. Conversely, a violation of Eq.~\eqref{CUR_equality} would indicate that the observed dynamics fails to satisfy one of the conditions for the validity of the CUR. To recap, these conditions are that the system is statistically stationary and undergoes classical, autonomous, Markovian dynamics as described by a time-homogeneous master equation $\partial_t \mathbf{p}=\mathsf{L}\mathbf{p}$. While these assumptions hold microscopically in a broad range of biological, chemical, and mesoscopic scenarios, often only a subset of the system's states or transitions can be observed directly. The observable degrees of freedom are coupled to the remaining, hidden states, resulting in effectively non-Markovian dynamics (a so-called hidden Markov model). The presence and influence of such hidden states can thus be detected through violations of the CUR. Crucially, this approach works in the steady state, thereby complementing the recent proposal of Ref.~\cite{Ito2020} that is restricted to non-stationary dynamics.

To illustrate the general idea, consider a system switching between two observable ``meso-states'' $\mathtt{X}$ and $\mathtt{Y}$, each of which may comprise numerous microscopic states~\cite{Esposito2012}; an example is illustrated in the inset of Fig.~\ref{fig:hidden_markov}. This scenario could describe a cellular chemoreceptor with ``bound'' and ``unbound'' states~\cite{Mora2010,Harvey2023}, a gene flipping between ``on'' and ``off'' transcription states~\cite{Suter2011}, or a molecular motor switching between distinct motional states~\cite{Asbury2003,Bai2010}. Switching between the meso-states manifests as a telegraph process, which appears non-Markovian since the waiting times between jumps depend on the inaccessible micro-state before and after each jump. 

To test the relation~\eqref{CUR_equality}, an experimenter must use their observations to construct the maximum SNR and the mean residual time. The latter can be inferred in various ways from the empirical waiting-time statistics (in this section only, we assume the experimentalist has a reliable external clock). For example, if $\tau$ denotes the waiting time between any pair of consecutive transitions, then~\cite{McFadden1962,Cox1982} (see also Appendix~\ref{SM:inspection_paradox})
\begin{equation}
\label{mean_residual_time_hmm}
    \tilde{\mathcal{T}} = \frac{\EE[\tau^2]}{2\EE[\tau]},
\end{equation}
which holds for an arbitrary (non-Markovian) stationary process. Here, we use a tilde to distinguish an observable quantity $\tilde{\mathcal{T}}$ from its ``true'' value $\mathcal{T}$ pertaining to the hidden Markov network. Clearly, $\tilde{\mathcal{T}}\geq \mathcal{T}$ because some transitions are missed, and so the inequality $\mathcal{S} \leq \tilde{\mathcal{T}}^{-1}$ can be violated if the SNR is large enough.

To increase the likelihood of detecting such a violation, our experimenter should find the maximum SNR attainable from the empirical jump statistics, $\tilde{\mathcal{S}}_{\rm max}$. Let $N_{\mathtt{XY}}$ denote the number of transitions $\mathtt{Y}\to \mathtt{X}$ (shown as pink arrows in the inset of Fig.~\ref{fig:hidden_markov}) and $N_{\mathtt{Y}\mathtt{X}}$ is the corresponding number of jumps $\mathtt{X}\to\mathtt{Y}$ (light blue arrows) observed over a duration $t$. Assuming stationary statistics and a large observation time, one constructs the mean currents $\tilde{\vec{J}}$ and diffusion tensor $\tilde{\mathbb{D}}$ as [c.f.~Eq.~\eqref{current_noise_def}]
\begin{align}
    & \tilde{\vec{J}} = \lim_{t\to\infty}\frac{1}{t} \begin{pmatrix} 
    \EE[N_{\mathtt{XY}}] \\ \EE[N_{\mathtt{Y} \mathtt{X}}] 
    \end{pmatrix}, \\ 
    \label{hmm_diffusion}
    & \tilde{\mathbb{D}} = \lim_{t\to\infty} \frac{1}{t}  \begin{pmatrix} 
    \Var[N_{\mathtt{XY}}] & \Cov[N_{\mathtt{XY}},N_{\mathtt{YX}}] \\
     \Cov[N_{\mathtt{YX}},N_{\mathtt{XY}}] & \Var[N_{\mathtt{YX}}]
     \end{pmatrix}.
    \end{align}
Following the same logic as Sec.~\ref{sec:counting_observables} and Appendix~\ref{SM:current_statistics}, the maximal SNR is then given by [c.f.~Eq.~\eqref{SNR_max}]
\begin{equation}
\label{SNR_max_hmm}
    \tilde{\mathcal{S}}_{\rm max} = \tilde{\vec{J}}\cdot \tilde{\mathbb{D}}^{+}\cdot \tilde{\vec{J}}.
\end{equation}
This procedure directly generalises to an arbitrary number of observable meso-states. However, in the present case of two meso-states, we show in Appendix~\ref{SM:hidden_markov} that $N_{\mathtt{X}\mathtt{Y}}$ and $N_{\mathtt{Y}\mathtt{X}}$ are perfectly correlated at long times. Without loss of generality, therefore, it suffices to measure only one of them, say $N_{\mathtt{X}\mathtt{Y}}$, to obtain $\tilde{\mathcal{S}}_{\rm max} = \lim_{t\to\infty} \EE[N_{\mathtt{XY}}]^2/(t \Var[N_{\mathtt{XY}}])$. 

Figure~\ref{fig:hidden_markov} displays a numerical example inspired by chemoreception models of Refs.~\cite{Mora2010,Skinner2021,Harvey2023}. We randomly sample transition rates between the microscopic states of the hidden Markov model, including a random bias favouring jumps in the clockwise direction to break detailed balance. The maximal SNR is reduced by the omission of unobserved jumps, while the mean residual time is increased. However, they are generally changed by different amounts, so that $\tilde{\mathcal{S}}_{\rm max}$ may be either larger or smaller than $\tilde{\mathcal{\mathcal{T}}}^{-1}$. Whenever $\tilde{\mathcal{S}}_{\rm max} \tilde{\mathcal{\mathcal{T}}} \neq 1$, therefore, the influence of hidden transitions is detected. We observe in Fig.~\ref{fig:hidden_markov} that this is generically the case, except for in scenarios of high symmetry that arise when the bias is small so that detailed balance is approximately restored. 

In principle, non-Markovian dynamics could also be witnessed by a violation of the KUR bound $\tilde{\mathcal{S}}_{\rm max} \leq \tilde{\mathcal{A}}$, where $\tilde{\mathcal{A}}$ is the mean rate of observed transitions. This is an imperfect witness, however, because the KUR bound is generally not achievable even for perfectly Markovian networks, as shown in Sec.~\ref{sec:CUR_vs_KUR}. To illustrate this, we also show the ratio $\tilde{\mathcal{S}}_{\rm max} /\tilde{\mathcal{A}}$ in Fig.~\ref{fig:hidden_markov}. Only for those points in the upper-left (yellow) quadrant of Fig.~\ref{fig:hidden_markov} is the KUR apparently violated while the CUR is not. In such cases, however, one could also detect non-Markovianity by simply noting that $\tilde{\mathcal{A}}\tilde{\mathcal{T}} < 1$. This can be tested directly from the mean waiting time, $\tilde{\mathcal{A}}^{-1}$, and its variance: from Eq.~\eqref{mean_residual_time_hmm} we thus have 
\begin{equation}
    \tilde{\mathcal{A}} \tilde{\mathcal{T}}  = \frac{1}{2}\left(1 + \frac{\Var[\tau]}{\EE[\tau]^2}\right).
\end{equation}
A Markov jump process has $\Var[\tau]\geq \EE[\tau]^2$ (see Appendix~\ref{SM:inspection_paradox}), while this condition is generally violated for a non-Markovian process, as indicated by the blue points in Fig.~\ref{fig:hidden_markov}. By comparison, the CUR saturation condition~\eqref{CUR_equality} provides the most stringent test of Markovian dynamics, as it is violated by all points away from the centre of Fig.~\ref{fig:hidden_markov}. Observation of such violations could also be used to detect the presence of quantum-coherent or non-autonomous dynamics, as these are also inconsistent with the CUR.

One may envision other tests of classical Markovianity based on the waiting times, e.g.~by confirming that they are (conditionally) exponentially distributed. This would require reconstruction and hypothesis testing of conditional waiting-time distributions such as $W(\tau|\mathtt{X})$ and $W(\tau|\mathtt{Y})$, the distribution of waiting time $\tau$ after a jump is observed into state $\mathtt{X}$ or $\mathtt{Y}$, respectively. On the contrary, testing the CUR requires only the mean and variance of the \textit{unconditional} waiting times to evaluate Eq.~\eqref{mean_residual_time_hmm}. Our approach is thus comparable, yet provides complementary information, to recent proposals for inferring entropy production from conditional waiting-time statistics~\cite{Skinner2021,Harunari2022, vanderMeer2022}. In general, entropy is produced both by observable transitions and by unobserved dynamics within the meso-states~\cite{Esposito2012}. Identifying the presence of hidden states can therefore help to elucidate these different contributions to the dissipation inferred from partial observations of a non-equilibrium system. 

\section{Clock performance beyond precision}
\label{sec:beyond_precision}

So far we have formulated timekeeping as a local parameter estimation problem, where performance is quantified by the signal-to-noise ratio (SNR) and bounded by the Fisher information. In this section, we discuss other measures of clock performance and demonstrate their connection to the local estimation framework used thus far. Specifically, we consider the resolution (Sec.~\ref{sec:resolution}) and accuracy (Sec.~\ref{sec:accuracy}) parameters introduced in Ref.~\cite{Erker2017} and studied widely in the recent literature on quantum clocks~\cite{Pearson2021, Woods2021,Schwarzhans2021,Meier2023,Xuereb2023,Silva2023,Woods2022}. Finally, we turn to the Allan variance~\cite{Allan1966} (Sec.~\ref{sec:allan_variance}), a standard measure of frequency stability used to assess atomic clocks at the precision frontier~\cite{Bothwell2022}. In the following, we define each of these performance measures for generic counting estimators, using the specific estimators summarised in Table~\ref{tab:estimators} as examples.

\subsection{Resolution}
\label{sec:resolution}

\begin{table}
	\begin{ruledtabular}
	\begin{tabular}{lllll}
		Estimator & Weights & SNR & Resolution & Accuracy \\
		\toprule
		BLUE & $w_{\mu\sigma} = \Gamma_\sigma^{-1} $ & $\mathcal{S} = \mathcal{T}^{-1}$ & $\nu=\mathcal{A}$ & $\mathcal{N} = 1/\mathcal{A}\mathcal{T} $\\
		Erlang & $w_{1d}=J_{1d}^{-1}$ & $\mathcal{S} = \mathcal{T}^{-1}$ & $\nu = \mathcal{A}/d$ & $\mathcal{N} = d/\mathcal{A}\mathcal{T}$ \\ 
		Uniform & $w_{\mu\sigma} = \mathcal{A}^{-1}$ & $\mathcal{S} \leq \mathcal{T}^{-1}$ & $\nu = \mathcal{A}$ & $\mathcal{N} = \mathcal{S}/\mathcal{A}$ 
	\end{tabular}
\end{ruledtabular}
	\caption{Three unbiased time estimators that are counting observables of the form $\nolinebreak{\Theta = \vec{w}\cdot\vec{N}}$. The properties of the Erlang estimator quoted above hold only for ring clocks, with arbitrary escape rates but only clockwise jumps allowed (see Appendix~\ref{SM:Erlang}). \label{tab:estimators}}
\end{table}

A clock establishes a temporal order by associating a time estimate to each event. Two events at the same location can be discriminated only if they are assigned different time estimates. High-resolution clocks are those that can discriminate events that are closely separated in time. Returning to our castaway example, counting waves lapping on a beach yields a higher time resolution than counting sunrises: while the former could feasibly time the cooking of a fish, the latter could not. Intuitively, therefore, a high resolution means that the time estimator changes in small increments.

To formalise this, consider an unbiased linear counting estimator in the form of Eq.~\eqref{counting_estimator}, which counts only those transitions $\sigma\to\mu$ within the alphabet $\aleph$, i.e.~$(\sigma\to\mu) \in \aleph$ if and only if $w_{\mu\sigma}\neq 0$. We define the resolution as $\nu = \bar{w}^{-1}$, where $\bar{w}$ is the average increment of the estimator when a jump is registered:
\begin{equation}
	\label{tick_time}
	\bar{w} = \EE[w_{\mu\sigma}|\aleph] =  \sum_{(\sigma\to\mu)\in \aleph} w_{\mu\sigma} P(\sigma\to\mu|\aleph),
\end{equation}
with $P(\sigma\to\mu|\aleph)$ the probability for a given monitored jump to be $\sigma\to\mu$. Eq.~\eqref{tick_time} describes the mean duration between recorded jumps, as defined by the estimator itself. Following the same logic used to obtain the jump steady state in Sec.~\ref{sec:residual_vs_waiting} and Appendix~\ref{SM:jump_ss}, it is straightforward to show that $P(\sigma\to\mu|\aleph) = J_{\mu\sigma}/\sum_{(\sigma\to\mu)\in \aleph} J_{\mu\sigma}$. Then, using the unbiased constraint, $\vec{w}\cdot\vec{J}=1$, we obtain
\begin{equation}
	\label{resolution}
	\nu = \mathcal{A}_\aleph \equiv \sum_{(\sigma\to\mu)\in \aleph} J_{\mu\sigma},
\end{equation}
where $\mathcal{A}_\aleph$ is the dynamical activity for transitions within the subset $\aleph$. Note that for any other alphabet $\aleph' \subseteq \aleph$, we have $\mathcal{A}_{\aleph^\prime} \leq  \mathcal{A}_{\aleph}$. It follows that all unbiased estimators using the same alphabet have the same resolution, and the highest resolution is obtained for estimators that account for all transitions, i.e.~$\max_\aleph \mathcal{A}_\aleph = \mathcal{A}$.

\begin{figure}
	\includegraphics[width=\linewidth]{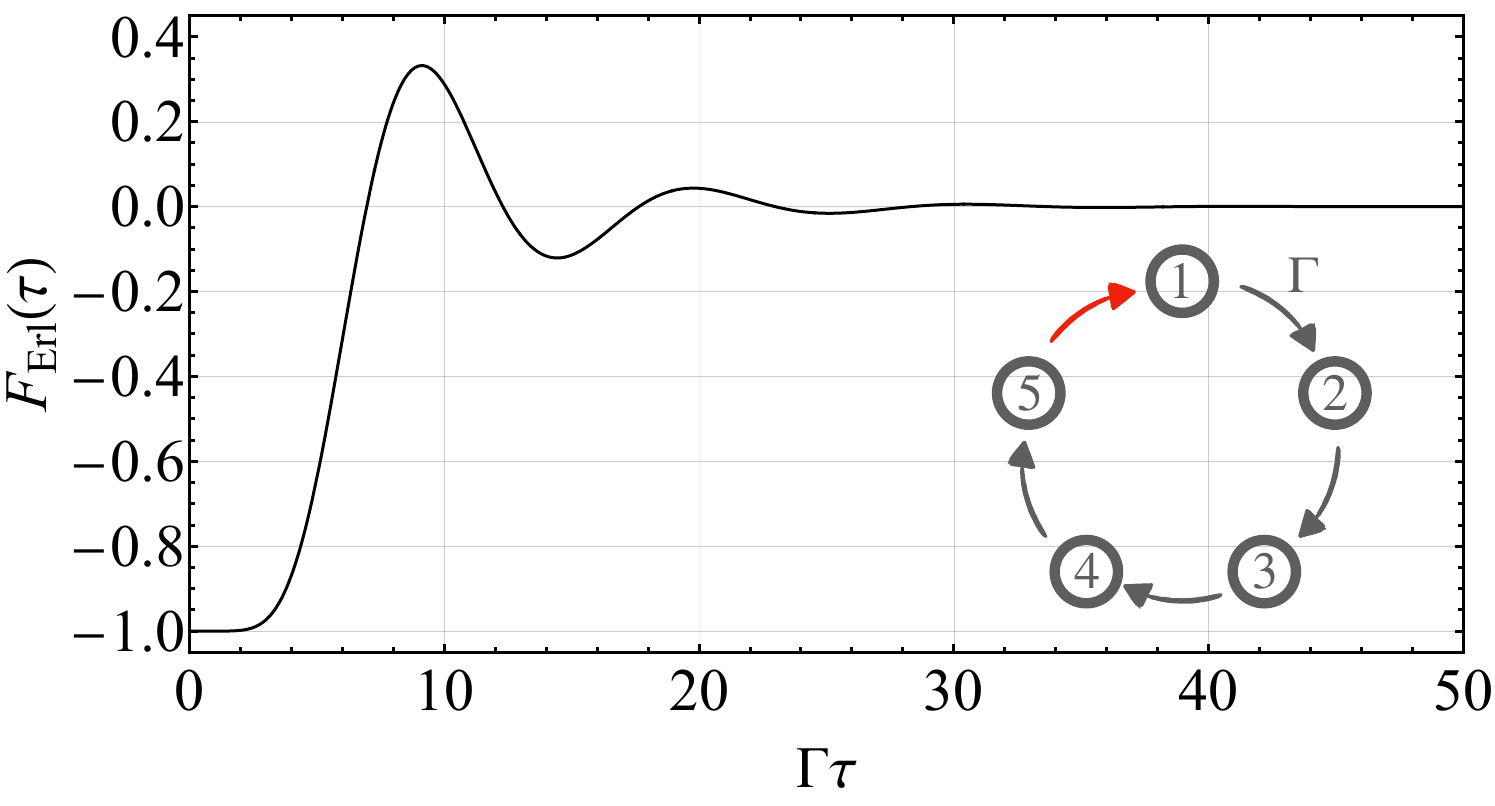}
	\caption{Autocorrelation function for the Erlang estimator on the homogeneous Erlang clock: a ring network with only clockwise transitions allowed, $R_{\mu\sigma} = \Gamma \delta_{\mu,\sigma+1}$ (see inset for a sketch with $d=5$). The Erlang estimator is obtained by counting only the jump $d\to 1$ (highlighted red in the inset), and weighting this by $w_{1d} = d/\Gamma$. Results are plotted for $d=10$; the singular contribution at $\tau=0$ is omitted.
		\label{fig:Erlang}}
\end{figure}

To demonstrate how the choice of estimator affects resolution, we consider the so-called Erlang clock~\cite{Erlang1917,Woods2022, Meier2024}. This is a ring network with $d$ states and fully biased transition rates, $R_{\mu\sigma} = \Gamma_\sigma \delta_{\mu,\sigma+1}$, such that only jumps in the clockwise direction are allowed (Fig.~\ref{fig:Erlang} inset). This model is exactly solvable in the homogeneous case, $\Gamma_\sigma=\Gamma$, as detailed in Appendix~\ref{SM:Erlang}. While the BLUE~\eqref{BLUE_estimator} weights every transition equally, $w_{\mu\sigma}=1/\Gamma$, one may also consider the Erlang estimator, where only the transition $d\to 1$ is counted, i.e.~$w_{1d}=d/\Gamma$ and all other $w_{\mu\sigma}=0$. This is analogous to measuring time by counting seconds (single jumps) or minutes (complete revolutions around the ring). Both estimators are unbiased, $\EE[\Theta] = t$, and have asymptotic variance scaling as $\Var[\Theta]/t \to \Gamma^{-1} = \mathcal{T}$, so they both saturate the CUR (see Appendix~\ref{SM:Erlang}).

However, the two estimators yield significantly different tick statistics, as evidenced by their current autocorrelation functions, $F(\tau) = \Cov[I(t+\tau),I(t)]$. A surprising property of the BLUE is that its current fluctuations are delta-correlated in time (see Eq.~\eqref{autocorrelation_function}). This property is characteristic of white or Poissonian noise, and it contradicts the intuitive expectation that the ticks of a reliable clock should be anti-bunched in time, which would imply that $F(\tau) > \lim_{\epsilon\to 0}F(\epsilon)$ for all $\tau>0$~\cite{Landi2024}. By contrast, the autocorrelation function of the Erlang estimator does show this anti-bunched character, as shown in Fig.~\ref{fig:Erlang}. This apparent regularity comes at the cost of decreased resolution: the resolution of the Erlang estimator is $\nu_{\rm Erl}=\Gamma/d$ while the BLUE has $\nu_{\rm BLUE}=\Gamma$ (see Appendix~\ref{SM:Erlang}). Indeed, the BLUE ticks after every jump, whereas each tick of the Erlang estimator requires the clock to complete $d$ jumps around the ring. Nevertheless, sacrificing resolution does not improve precision as measured by the SNR. In fact, at short times the Erlang estimator has a significantly larger variance than the BLUE. In  Appendix~\ref{SM:Erlang} we show that $\lim_{t\to 0}\Var[\Theta_{\rm Erl}]/t = d/\Gamma$, while for the BLUE we have $\Var[\Theta_{\rm BLUE}]/t =  \Gamma^{-1}$ at all times.

To illustrate how the variance of the Erlang estimator scales at intermediate times, we return to the two-state system with unbalanced rates in Fig.~\ref{fig:estimators}. In the case of inhomogeneous rates $\Gamma_\sigma$, the Erlang estimator is defined by weighting the transition $d\to 1$ as $w_{1d} = J_{1d}^{-1}$, setting all other $w_{\mu\sigma}=0$. This yields an unbiased estimator with minimal variance at long times, $\lim_{t\to \infty}\Var[\Theta_{\rm Erl}]/t = \mathcal{T}$, and resolution $\nu_{\rm Erl} = \mathcal{A}/d$ that is $d$ times smaller than the BLUE. These expressions hold for any $d$, so long as only clockwise transitions are allowed; see Table~\ref{tab:estimators} for a summary and Appendix~\ref{SM:Erlang} for details. Figure~\ref{fig:estimators} shows the scaled variance of this estimator for $d=2$ as a function of time (black squares), demonstrating that the expected error is persistently larger than the variance of the BLUE (upper dashed line) while tending to the same value as $t\to\infty$. Overall, we conclude that maximising both resolution and precision requires the estimator to count all observable events.

\subsection{Accuracy}
\label{sec:accuracy}

The accuracy parameter was introduced in Ref.~\cite{Erker2017} to quantify the stability of an autonomous clock's time estimate. Closely related measures of temporal regularity have been discussed in the context of molecular motors and oscillators~\cite{Moffitt2014, Wierenga2018,Marsland2019}. Unlike the SNR defined by Eqs.~\eqref{general_bound} and~\eqref{SNR_counting}, the accuracy is dimensionless. Its informal definition is the average number of ticks, $\mathcal{N}$, before the clock's time estimate is off by one tick. 

To define this precisely within our framework, where the time estimate $\Theta$ is continuous rather than discrete, we divide $\Theta$ by the mean size of the time increment $\bar{w}$ [see Eq.~\eqref{tick_time}] to obtain a dimensionless number of ticks. We thus solve the equation $\Var[\Theta|t_\mathcal{N}]^{1/2} = \bar{w}$ to find the time $t_\mathcal{N}$ at which the expected error in $\Theta$ equals the mean duration of one tick, $\bar{w}$. We then express $t_\mathcal{N}$ in units of $\bar{w}$ to find the corresponding accuracy, $\mathcal{N} =t_\mathcal{N}/\bar{w} = \nu t_\mathcal{N}$, where $\nu = \bar{w}^{-1}$ is the resolution defined in Eq.~\eqref{resolution}. Since $\Var[\Theta|t]\to t/\mathcal{S}$ for unbiased estimators, this yields the asymptotic accuracy
\begin{equation}
	\label{accuracy}
	\mathcal{N} = \frac{\mathcal{S}}{\nu}.
\end{equation}
This can also be expressed as the inverse of the asymptotic Fano factor~\cite{Landi2024} of the integrated ``tick current'' $\Theta/\bar{w}$, i.e. 
\begin{equation}
	\label{Fano_factor}
	\mathcal{N} = \lim_{t\to\infty} \frac{\EE[\Theta/\bar{w}]}{\Var[\Theta/\bar{w}]}.
\end{equation}
In the special case where $\Theta/\bar{w}$ is an integer, this agrees with the definition recently put forward in Ref.~\cite{Silva2023}.

Returning to Eq.~\eqref{accuracy}, the CUR now imposes the accuracy-resolution trade-off on arbitrary counting estimators given in Ineq.~\eqref{accuracy_resolution_CUR} and restated here for convenience
\begin{equation}
	\label{accuracy-resolution-trade-off}
	\nu  \mathcal{N} \leq \mathcal{T}^{-1}.
\end{equation}
Increased accuracy therefore comes at the cost of reduced resolution, such that at best the accuracy scales with the inverse resolution, $\mathcal{N}\sim \nu^{-1}$. This is a tighter special case of the general trade-off found in Ref.~\cite{Meier2023}, which states that any (classical or quantum) clock's accuracy scales at most quadratically with the inverse resolution. That is, $\mathcal N \leq \mathcal{A}^2_{\rm max}/\nu^{2}$, where $\mathcal{A}_{\rm max}$ is the mean rate at which monitored transitions can occur maximised over all states of the clock, e.g.~in our setup $\mathcal{A}_{\rm max} = \max_{\mathbf{p}} \sum_{(\sigma\to\mu)\in \aleph} R_{\mu\sigma}p_{\sigma}$, where the maximum is performed over all    probability distributions $\mathbf{p}$.

The accuracy parameter captures the intuitive notion that a good clock should tick at regular intervals. This is seen clearly for the Erlang estimator, which shows a $d$-fold improvement in accuracy relative to the BLUE. This follows immediately from Eq.~\eqref{accuracy}, because the Erlang estimator has resolution $\nu_{\rm Erl}=\mathcal{A}/d$ as compared with $\nu_{\rm BLUE} = \mathcal{A}$ for the BLUE, while the SNR $\mathcal{S}=\mathcal{T}^{-1}$ for both estimators. As a result, the BLUE must have an accuracy parameter less than or equal to unity, since $\mathcal{N}_{\rm BLUE} = \mathcal{T}^{-1}/\mathcal{A} \leq 1$ by properties of the arithmetic and harmonic means. Since the BLUE has maximal SNR, any other estimator that counts all transitions also has $\mathcal{N}\leq 1$. However, the accuracy of the Erlang estimator is $\mathcal{N}_{\rm Erl} = d\mathcal{N}_{\rm BLUE}$, which scales linearly with the number of states $d$ and thus can saturate known bounds on classical clocks~\cite{Woods2022}. This embodies the trade-off~\eqref{accuracy-resolution-trade-off}, which states that accuracy can be increased by reducing resolution while holding the SNR fixed.

\subsection{Allan variance}
\label{sec:allan_variance}

The Allan variance is used to assess the stability of frequency standards, such as atomic clocks, as a function of the integration time over which a measurement is performed. For a general unbiased estimator $\Theta$, the time error is $x(t) = \Theta - t$ and the corresponding frequency error is $y(t) = \dd x/\dd t$. The frequency error averaged over a time interval $T$ is
\begin{equation}
	\label{mean_fractional_frequency}
	\bar{y}_T(t) = \frac{1}{T}\int_t^{t+T}\dd t'\,y(t')  = \frac{1}{T}\int_{t}^{t+T} \dd t'\,\left [I(t') - 1\right ],
\end{equation} 
where the current $I(t) = \dd \Theta/\dd t$ plays the role of frequency for a counting estimator. Note that $\EE[\bar{y}_T(t)] = 0$ for an unbiased estimator. The Allan variance is then defined as
\begin{equation}
	\label{Allan_variance}
	\sigma_y^2(T) = \frac{1}{2}\Var\left [\bar{y}_T(t+T) - \bar{y}_T(t) \right ],
\end{equation}
which measures deviations of the average frequency between two consecutive averaging periods of duration $T$. 

Using the techniques of Ref.~\cite{Landi2024}, an explicit expression for the Allan variance can be found in terms of the current autocorrelation function $F(\tau) = \Cov[I(t+\tau),I(t)]$. Full details are given in Appendix~\ref{SM:Allan_variance}, while here we simply note the short- and long-time limits
\begin{equation}
	\label{allan_limits}
	\sigma_y^2(T) \sim \begin{cases}
		\dfrac{K}{T} & (T\to 0),\\ \vspace{-3mm}\\
		\dfrac{D}{T} & (T\to \infty),
	\end{cases}
\end{equation}
where $D$ is the diffusion coefficient and $K=\sum_{\mu,\sigma}w_{\mu\sigma}^2 J_{\mu\sigma}$ is a weighted dynamical activity, which both depend on the choice of weights $\vec{w}$. For long times, minimising the Allan variance is equivalent to minimising $D$ and therefore the BLUE is optimal, yielding the long-time scaling $\sigma_y^2(T)\sim \mathcal{T}/T$. Any other solution to Eq.~\eqref{BLUE} would similarly minimise the long-time Allan variance. For short times, however, the Allan variance is minimised by choosing weights that extremise $K$ instead. Taking into account the unbiased constraint, $\vec{w}\cdot\vec{J}=1$, this extremum is achieved by the uniform weighting $w_{\mu\sigma} = \mathcal{A}^{-1}$, i.e.~$\Theta = N/\mathcal{A}$ where $N$ is the total number of jumps and $\mathcal{A}$ is the dynamical activity (see Appendix~\ref{SM:Allan_variance}). This yields the minimal Allan variance $\sigma_y^2(T) \sim 1/(\mathcal{A}T)$ at short times. Notice that the same estimator saturates the CRB at short times, as discussed in Sec.~\ref{sec:asymptotic_fisher}.

\begin{figure}
	\includegraphics[width=\linewidth]{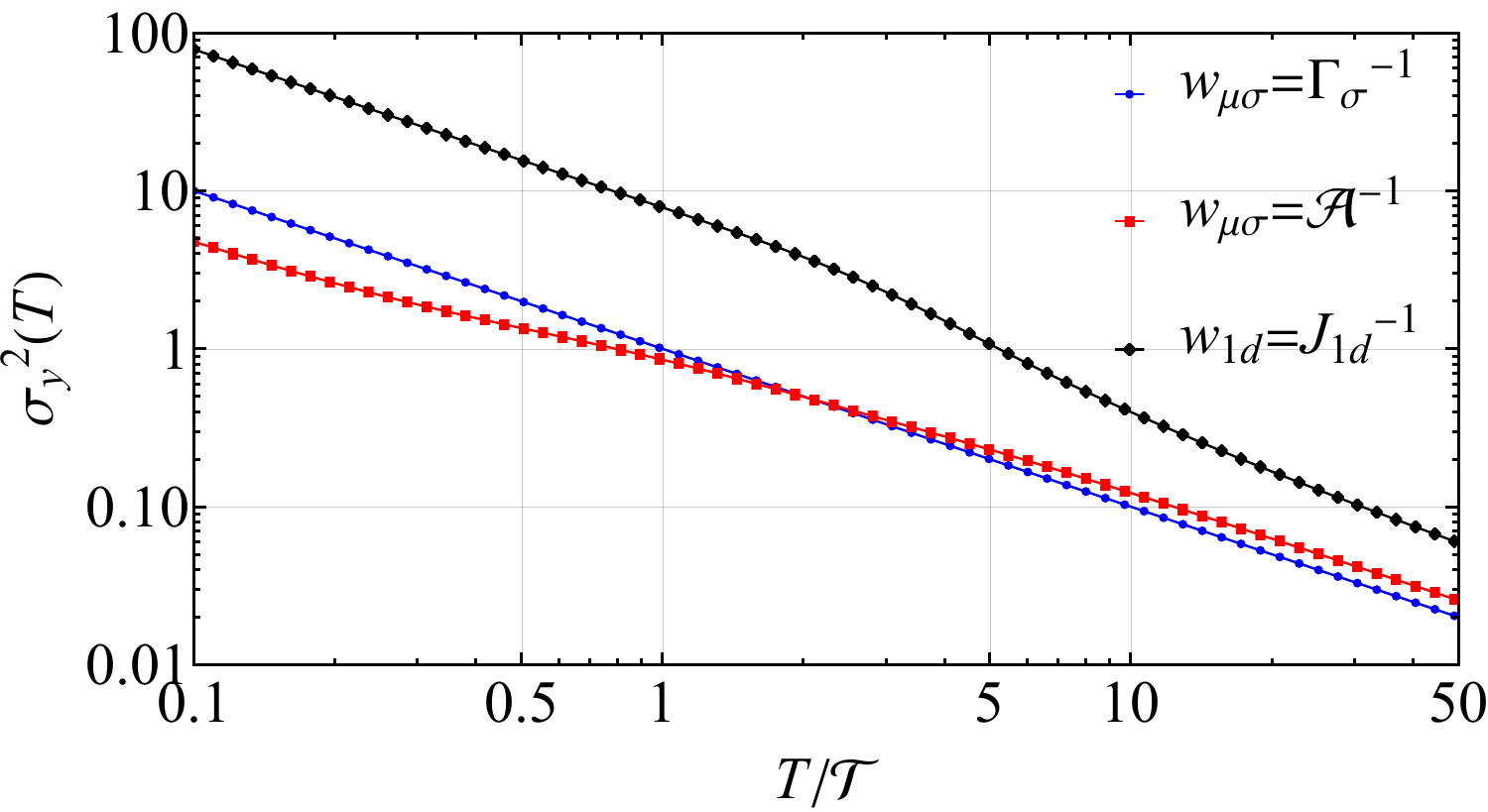}
	\caption{Allan variance of time estimators for a ring clock with a similar geometry to that of Fig.~\ref{fig:bounds}(b). We take a disordered but non-random set of rates given by $R_{\mu\sigma} = r_\sigma\delta_{\mu,\sigma+1} +  \ee^{-\Delta } r_\mu \delta_{\mu,\sigma-1}$, with clockwise bias $\Delta =2$ and $r_\sigma = \sin^2(2\sigma/3)$ (other choices for $r_\sigma$ give qualitatively similar results). Results are shown for $d=10$ and three counting estimators: the BLUE with weights $w_{\mu\sigma} = \Gamma_\sigma^{-1}$ (blue circles), the uniform weighting $w_{\mu\sigma} = \mathcal{A}^{-1}$, and counting a single transition, with $w_{1d}=J_{1d}^{-1}$ and all other $w_{\mu\sigma}=0$. \label{fig:allan}}
\end{figure}

To illustrate the crossover from this short- to the long-time regime, we plot results for a unicyclic clock model in Fig.~\ref{fig:allan}. We allow both clockwise and anticlockwise jumps with disordered rates to ensure that differences between $\mathcal{A}$ and $\mathcal{T}^{-1}$ are visible, but to ensure reproducibility we generate these rates from a deterministic sequence; see the caption of Fig.~\ref{fig:allan} for details. We see that the uniformly weighted estimator has the smallest Allan variance at short times (red squares in Fig.~\ref{fig:allan}). However, after a timescale comparable to a single jump, $T\sim \mathcal{T}$, this uniform estimator transitions to a suboptimal scaling that is worse than that of the BLUE. Notably, for the BLUE one can show that $\sigma_y^2(T) = \mathcal{T}/T$ holds exactly at all times (blue circles in Fig.~\ref{fig:allan}). This can be understood due to the delta-correlated nature of the current (cf.~Eq.~\eqref{autocorrelation_function}), so that correlations across different averaging periods vanish (see Appendix~\ref{SM:Allan_variance})

For comparison, we also plot $\sigma_y^2(T)$ for an unbiased ``Erlang'' estimator that counts only the transition $d\to 1$ (black diamonds in Fig.~\ref{fig:allan}). The Allan variance for this estimator is initially higher than the BLUE but begins to decrease more quickly around the timescale $T\sim d\mathcal{T}$, after which the integration period is long enough to average over several ticks. However, unlike for the Erlang clock introduced in Sec.~\ref{sec:resolution}, the Erlang estimator is not asymptotically efficient for this model. This is because jumps are allowed both clockwise and anti-clockwise around the ring in this case, whereas the Erlang estimator only accounts for completed clockwise rotations. The long-time scaling therefore remains suboptimal compared with the BLUE, which accounts for all transitions.

\section{Discussion} 
\label{sec:discussion}
Stochastic uncertainty relations are often discussed using terminology borrowed from metrology (e.g.~precision, SNR, etc.) and they have been derived in many cases using the mathematics of parameter estimation, with the Cram\'er-Rao bound (CRB) playing a prominent role~\cite{Dechant2018,Hasegawa2019,Shiraishi2021,Hiura2021,Vu2022,Vo2022}. Information theory applied to Markov jump processes has also been used to bound the precision of measurements under resource constraints~\cite{Mora2010,Lang2014,Harvey2023,Nagase2024}. The clock uncertainty relation (CUR) presented here weaves yet another thread in the tapestry of connections between metrology and stochastic thermodynamics. The CUR states that steady-state fluctuations are bounded by the system's capacity to measure time precisely. In simple terms, the less time before a system makes its first jump, the more precisely it can function as a clock at long times, and the optimal time estimators coincide with the observables that fluctuate least. We emphasise that the CUR is a tight bound because it quantifies freneticity via the inverse mean residual time. Crucially, this feature can be exploited to detect the presence of hidden states using the observable jump and waiting-time statistics. This contrasts with the kinetic uncertainty relation, which cannot generally be saturated except for in the short-time limit, where it coincides with the CRB for time estimation.

Central to our approach is the Fisher information with respect to time, which also features explicitly in several classical~\cite{Ito2018,Ito2020,Nicholson2020} and quantum~\cite{GarciaPintos2022} speed limits, where it can be interpreted geometrically as a metric. The notion of information geometry and the Fisher metric has been extended recently to the space of trajectories in Ref.~\cite{Zheng2024}, where the relevant parameters in that case are the entries of the rate matrix, $R_{\mu\sigma}$. Unlike these previous works, however, our Fisher information is not associated with the instantaneous state $p_\sigma(t)$ nor the time-tagged trajectory distribution $P(\bm{\sigma},\mathbf{t}|t)$, but with the marginal sequence distribution $P(\bm{\sigma}|t)$. The latter is the natural distribution to consider for time estimation, but its geometric properties have received comparatively little attention.

It is notable that the mean residual time $\mathcal{T}$ emerges as a key quantity from our analysis. This captures information on both the mean and variance of the waiting times in a single number.  The distribution of waiting times is itself an object of interest for stochastic thermodynamics, since it can be used to infer entropy production from partial information, as shown in recent works~\cite{Skinner2021,vanderMeer2022,Harunari2022}. One may speculate that a similar role might be found for the residual-time distribution, and that its mean $\mathcal{T}$ could be used to tighten bounds involving the entropy production, in the same manner as the dynamical activity appears in thermo-kinetic tradeoff relations~\cite{Pietzonka2016,Maes2017, Shiraishi2018,Vo2020,Vo2022,Vu2023}.
 
While we remain hopeful that no unfortunate readers will need to use our results on stochastic time estimation to survive a shipwreck, some remarks on their practical implementation are still called for. In particular, it is not necessary to already know the waiting times $\Gamma_\sigma^{-1}$ precisely in order to construct an asymptotically optimal time estimator. Such an estimator can be determined experimentally by observing the long-time current statistics to infer $\mathbb{D}$ and $\vec{J}$ from Eqs.~\eqref{current_noise_def}. An optimal counting observable can then be constructed using any pseudoinverse of $\mathbb{D}$ to solve Eq.~\eqref{BLUE}. Moreover, while application of Eqs.~\eqref{current_noise_def} would seem to require \textit{a priori} knowledge of $t$, this is not the case: one can use a given process, say $N_{12}$, as a ``time standard'' by setting $J_{12}=1$ (tantamount to a choice of units), and then define all other quantities in terms of ratios $N_{\mu\sigma}/N_{12}$. It is intriguing that any estimator thus obtained is generally not a thermodynamic current in the sense of being odd under time reversal. Indeed, the optimal estimator does not depend on the order of the sequence $\bm{\sigma}$ at all, so it would assign the same estimates to a trajectory and its time reverse. Counter-intuitively, therefore, the most precise estimators of time cannot distinguish the direction of time's arrow. 

The BLUE defined in Eq.~\eqref{BLUE_estimator} has other surprising properties: its current fluctuations are delta-correlated [Eq.~\eqref{autocorrelation_function}] and its accuracy parameter $\mathcal{N}\leq 1$. That is, the BLUE is already wrong --- as compared to the timescale of its own ticks --- after just a single tick on average. The underlying reason is that accuracy always comes at the cost of resolution: any estimator that counts all transitions has high resolution but poor accuracy [Ineq.~\eqref{accuracy-resolution-trade-off}]. This also suggests that the accuracy parameter $\mathcal{N}$ is better suited to quantify clock precision when only raw counting of events is considered, without any post-processing of the data, e.g.~by weighting events according to their type. In fact, just post-processing alone can improve accuracy (while, of course, sacrificing resolution). The Erlang clock example shows that coarse-graining several jump events into a single tick can lead to a $d$-fold increase in $\mathcal{N}$. If, in addition, the observer stores a record of past events, coarse-graining over multiple revolutions of the Erlang clock would yield accuracy gains bounded only by the size of the observer's memory~\cite{Woods2022,Meier2023}. By contrast, the asymptotic signal-to-noise ratio $\mathcal{S}$ is invariant under such trivial rescaling and is well suited for quantifying how the precision of a post-processed estimator varies on some externally defined timescale, e.g.~the time required to safely cook a fish. According to the CUR, this time must be longer than the internal timescale of the clock's physical dynamics, $\mathcal{T}$, for the estimate to be trustworthy.

From a more foundational perspective, our work takes a significant step forward in the quest to establish the ultimate limits on the measurement of time. Precise clocks are crucial for gravitational sensing~\cite{Bothwell2022,Stray2022}, large-scale quantum computation~\cite{Xuereb2023, Meier2024}, and the search for new physics~\cite{Rosenband2008,Derevianko2014,Roussy2023}, so understanding their limits is of both fundamental and practical importance. All clocks must operate irreversibly, which implies a natural link between entropy production and clock performance that has been established in many examples~\cite{Barato2016,Erker2017,Marsland2019,Milburn2020,Pearson2021} (albeit challenged in some underdamped systems such as the pendulum clock~\cite{Pietzonka2022}). Conversely, the importance of non-equilibrium activity or freneticity in timekeeping has been less studied, with recent work mainly focussing on the role of dynamical activity in quantum clocks~\cite{Meier2023,He2023}. The clock uncertainty relation shows that activity, as quantified by the inverse mean residual time, imposes a tight bound on time estimation for a large class of classical clocks. This serves as a natural starting point to search for tighter precision bounds in the fully quantum regime. 

\section*{Note added}
Soon after the completion of our work, Macieszczak~\cite{Macieszczak2024} reported similar results obtained using large-deviation theory. This complementary method allowed CUR-like bounds (referred to in Ref.~\cite{Macieszczak2024} as ``ultimate KUR'') to be derived for counting observables and first-passage times, not only for classical Markov dynamics but also semi-Markov processes, a class which includes many examples of quantum reset clocks studied in the recent literature~\cite{Erker2017, Schwarzhans2021, Woods2021, Woods2022,Meier2023,Dost2023,Manikandan2023}.

\section*{Data availability}

The data and code needed to reproduce all plots can be found in a Mathematica notebook at Ref.~\cite{mmitchison2024}.

\begin{acknowledgments}
We thank Michael J.~Kewming for helpful discussions on the Allan variance of counting estimators, and Juan P.~Garrahan for insightful correspondence on the manuscript. This research was supported in part by the National Science Foundation under Grants No.~NSF PHY-1748958 and PHY-2309135. We specifically acknowledge the KITP program New Directions in Quantum Metrology and the conference Time in Quantum Theory (TiQT) 2023, where some of this work was performed. This project is co-funded by the European Union (Quantum Flagship project ASPECTS, Grant Agreement No.~101080167). Views and opinions expressed are however those of the authors only and do not necessarily reflect those of the European Union, Research Executive Agency or UKRI. Neither the European Union nor UKRI can be held responsible for them. KP and PPP acknowledge funding from the Swiss National Science Foundation (Eccellenza Professorial Fellowship PCEFP2\_194268). RS acknowledges funding from the Swiss National Science Foundation via an Ambizione grant PZ00P2\_185986. MTM is supported by a Royal Society University Research Fellowship.
\end{acknowledgments}

\section*{Author contributions}
MTM and GTL conceived of the project and developed the mathematical framework. NN and RS established the degeneracy of the diffusion tensor and conjectured the form of the optimal time estimator and the clock uncertainty relation. PPP proved the conjecture using the saddle-point approximation, and KP performed analytical and numerical calculations for the two-state model to verify the results and numerical simulations to detect a hidden Markov model. FM developed the concept of resolution and introduced and solved the Erlang clock example. MTM drafted the manuscript and produced the plots with input from all authors, and NN created the illustrations in Figs.~\ref{fig:fishing} and~\ref{fig:residual_time}. All authors contributed substantially to the discussion and interpretation of the results and ideas underpinning this work.
\appendix 

\section{Path probabilities and waiting-time distributions}
\label{SM:t-ensemble}

In this appendix, we state some basic properties of the path probabilities for the classical master equation $\dot{\mathbf{p}} = \mathsf{L}\mathbf{p}$. The generator can be written as $\mathsf{L} = \mathsf{R}- \mathsf{\Gamma}$, where $[\mathsf{R}]_{\mu\sigma} = R_{\mu\sigma}$ is the matrix of transition rates and $[\mathsf{\Gamma}]_{\mu\sigma} = \Gamma_\sigma \delta_{\mu\sigma}$ is a diagonal matrix of escape rates. The generator is assumed to have a unique stationary state $\mathbf{p}^{\ss}$, which is a null right eigenvector of the generator satisfying $\mathsf{L}\mathbf{p}^{\ss} = 0$. The normalisation is given by $\mathbf{u}^\T \mathbf{p}^{\ss} = 1$, where  $\mathbf{u}^\T$, where $\mathbf{u}^\T = (1,1,\ldots,1)$ is a constant row vector. Probability conservation implies that $\mathbf{u}^\T \mathsf{L}=0$.

The general solution of the master equation is
\begin{equation}
	\label{master_equation_solution}
	\mathbf{p}(t) = \mathsf{G}(t)\mathbf{p}(0),
\end{equation}
where $\mathsf{G}(t) = \ee^{\mathsf{L}t}$ is the propagator. Defining $\mathsf{\tilde{R}}(t) = \ee^{\mathsf{\Gamma }t} \mathsf{R} \ee^{-\mathsf{\Gamma}t}$, we can rewrite the propagator as a Dyson series,
\begin{align}
	\label{Dyson_series}
	\mathsf{G}(t) &  = \ee^{-\mathsf{\Gamma}t} \,  \Texp\left [\int_0^t \dd t' \, \mathsf{\tilde{R}}(t')\right ]  \\ 
	&  =  \ee^{-\mathsf{\Gamma}t} + \int_0^{t} \dd t_1\, \ee^{-\mathsf{\Gamma}(t-t_1)} \mathsf{R}\ee^{-\mathsf{\Gamma}t_1}  + \cdots \notag \\ 
& =	\sum_{N=0}^\infty \int_0^t\dd t_N \cdots \int_0^{t_2} \dd t_1\, \ee^{-\mathsf{\Gamma}(t-t_N)} \mathsf{R} \cdots \ee^{-\mathsf{\Gamma} (t_2-t_1)} \mathsf{R} \ee^{-\mathsf{\Gamma} t_1}.\notag 
\end{align}
Here, $\overleftarrow{\rm T}$ denotes the time-ordering symbol, which reorders all subsequent time-dependent operators such that their time arguments increase from right to left, i.e. 
\begin{equation}
    \label{time-ordering}
    \overleftarrow{\rm T} \mathsf{A}(t_1) \mathsf{B}(t_2) = \begin{cases}
        \mathsf{A}(t_1) \mathsf{B}(t_2), & (t_1 \geq t_2) \\
        \mathsf{B}(t_2) \mathsf{A}(t_1), & (t_1 < t_2)
    \end{cases}.
\end{equation}

Taking the distribution to be stationary, $\mathbf{p}(0) = \mathbf{p}(t) = \mathbf{p}^{\ss}$, and writing out all index contractions explicitly, we have
\begin{align}
	\label{prob_path_integral}
	&p_{\mu}^{\ss} = \sum_{N=0}^\infty \sum_{\sigma_0=1}^d \cdots \sum_{\sigma_{N}=1}^d \delta_{\mu\sigma_N}\int_0^t\dd t_N \cdots \int_0^{t_2} \dd t_1 P(\bm{\sigma},\mathbf{t}|t),\\
	\label{path_probability_sm}
	&P(\bm{\sigma},\mathbf{t}|t) = \ee^{-\Gamma_{\sigma_N} (t-t_N)} R_{\sigma_N\sigma_{N-1}} \cdots \ee^{-\Gamma_{\sigma_1}(t_2-t_1)} R_{\sigma_1\sigma_0} \ee^{-\Gamma_{\sigma_0}t_1} p_{\sigma_0}^{\ss}.
\end{align}
This can be interpreted as a sum over all possible trajectories connecting the initial and final states $\sigma_0$ and $\mu$, where $P(\bm{\sigma},\mathbf{t}|t)$ is the probability density to observe the sequence of $N$ jumps $\bm{\sigma} = \sigma_0\to\sigma_1\to\cdots\to\sigma_N$ at the times $\mathbf{t} = (t_1,t_2,\cdots,t_N)$, where $0\leq t_1 \leq\cdots \leq t_N\leq t$. Normalisation of the steady-state distribution implies the following normalisation of the path probabilities:
\begin{align}
	\label{path_prob_normalisation}
1 & = 	\sum_{N = 0}^\infty \sum_{\sigma_0=1}^d\cdots \sum_{\sigma_N=1}^d \int_0^t\dd t_N \cdots \int_0^{t_2}\dd t_1\, P(\bm{\sigma},\mathbf{t}|t) \notag \\ 
&\equiv \sum_{\bm{\sigma}}  \int_0^t\dd \mathbf{t}\,  P(\bm{\sigma},\mathbf{t}|t).
\end{align}
The total probability for $N$ jumps to have occurred at time $t$ is given by
\begin{equation}
    \label{P_N_of_t}
P(N|t) = \sum_{\sigma_0=1}^d \cdots \sum_{\sigma_N=1}^d \int_0^t\dd\mathbf{t}\, P(\bm{\sigma},\mathbf{t}|t) = \sum_{\sigma_0} \cdots \sum_{\sigma_N}P(\bm{\sigma}|t),
\end{equation}
whose normalisation $\sum_{N=0}^\infty P(N|t)=1$ follows from Eq.~\eqref{path_prob_normalisation}.

The interpretation of Eq.~\eqref{path_probability_sm} as the probability for the trajectory ($\bm{\sigma},\mathbf{t})$ is corroborated by the following observation. Assuming that the system starts in state $\sigma$, the joint probability density for a jump to occur to state $\mu$ after a waiting time $\tau\geq 0$ is given by~\cite{Landi2024} 
\begin{equation}
	\label{joint_waiting_time}
	W(\tau, \mu|\sigma) =  R_{\mu\sigma} \ee^{-\Gamma_\sigma \tau}.
\end{equation}
The distribution of waiting times for any jump, conditioned on the initial state $\sigma$, is therefore
\begin{equation}
	\label{total_waiting_time}
	W(\tau|\sigma) = \sum_{\mu=1}^d W(\tau,\mu|\sigma) =\Gamma_\sigma \ee^{-\Gamma_\sigma \tau}.
\end{equation}
This is consistent with the definition of the waiting-time distribution~\cite{Landi2024}
\begin{equation}
	\label{survival_prob_waiting}
	W(\tau|\sigma) =-\frac{\dd P_{\rm no}(\tau|\sigma)}{\dd \tau},
\end{equation}
where $P_{\rm no}(\tau|\sigma) = \ee^{-\Gamma_\sigma \tau}$ is the survival (no-jump) probability (for $\tau\geq 0$). Therefore, the probability of observing precisely $N$ jumps $\bm{\sigma}$ at times $\mathbf{t}$ is simply the product
\begin{equation}
	\label{path_probability_product}
	P(\bm{\sigma},\mathbf{t}|t) = P_{\rm no}(\tau_N|\sigma_N) \prod_{j=0}^{N-1} W(\tau_j,\sigma_{j+1}|\sigma_j) p_{\sigma_0}^\ss,
\end{equation}
where $\tau_j = t_{j+1}-t_j$ with $t_0\equiv 0$ and $t_{N+1}\equiv t$. This is identical to Eq.~\eqref{path_probability_sm}.

The total probability to observe the jump $\sigma\to\mu$ is found by marginalising Eq.~\eqref{joint_waiting_time} over $\tau$, yielding
\begin{equation}
	\label{transition_prob}
	\pi(\mu|\sigma) = \int_0^\infty \dd\tau\, W(\tau,\mu|\sigma) = \frac{R_{\mu\sigma}}{\Gamma_\sigma}.
\end{equation}
Therefore, the joint probability~\eqref{joint_waiting_time} factorises as
\begin{equation}
	\label{joint_prob_factorise}
	W(\tau,\mu|\sigma) = W(\tau|\sigma) \pi(\mu|\sigma).
\end{equation}
From this it follows that the waiting time $\tau$ and the final state $\mu$ of any jump are independent random variables: they depend on the initial state $\sigma$ but not on each other. In other words, the conditional waiting time for a jump $\sigma\to\mu$ is independent of $\mu$. This can also be shown using Bayes' rule,
\begin{equation}
	\label{conditional_waiting_time_dist}
	W(\tau|\sigma\to\mu) = \frac{W(\tau,\mu|\sigma)}{\pi(\mu|\sigma)} = W(\tau|\sigma).
\end{equation}
Using Eq.~\eqref{joint_prob_factorise} in Eq.~\eqref{path_probability_product} and noting that $P_{\rm no}(\tau|\sigma) =  W(\tau|\sigma)/\Gamma_\sigma$, we recover Eq.~\eqref{path_probability_with_times} in the main text. 

As discussed in the main text, we are interested in the distribution $P(\bm{\sigma}|t)$, which is obtained by marginalizing $P(\bm{\sigma},\mathbf{t}|t)$ [see~Eq.~\eqref{path_probability}]. Here we prove Eq.~\eqref{Fisher_Neyman_factorise}, i.e. that 
\begin{equation}
\label{eq:supp_pfact}
    P(\bm{\sigma}|t) =\Pi(\bm{\sigma}) g(\mathbf{n},t),
\end{equation}
with $\Pi(\bm{\sigma}) = \Gamma_{\sigma_N}^{-1} \prod_{j=0}^{N-1}\pi(\sigma_{j+1}|\sigma_j) p^{\ss}_{\sigma_0}$ and $g(\mathbf{n},t)$ given in Eq.~\eqref{Fisher_Neyman_factor}. To this end, we write
\begin{align}
\label{eq:supp_gnt}
        \frac{P(\bm{\sigma}|t)}{\Pi(\bm{\sigma})}=&\int_0^t  \!\dd t_N \cdots \! \int_0^{t_2} \dd t_1  \prod_{j=0}^{N} W(\tau_j|\sigma_j) \notag\\
        =&\int_0^\infty  \!\dd\tau_N \cdots \! \int_0^\infty \! \dd\tau_0 \,\delta\left (t-\sum_{j=0}^N\tau_j\right ) \prod_{j=0}^{N} W(\tau_j|\sigma_j) \notag\\
        =&\int_{-\infty}^\infty \frac{\dd\chi}{2\pi} \ee^{-\ii\chi t}\prod_{j=0}^N \int_0^\infty \dd\tau_j \Gamma_{\sigma_j}\ee^{-\tau_j(\Gamma_{\sigma_j}-\ii\chi)} \\
        =& \int_{-\infty}^\infty \frac{\dd \chi}{2\pi}\, \ee^{-\ii \chi t} \prod_{j = 0}^N \frac{\Gamma_{\sigma_j}}{\Gamma_{\sigma_j} - \ii \chi} \notag\\
        =& \int_{-\infty}^\infty \frac{\dd \chi}{2\pi}\, \ee^{-\ii \chi t} \prod_{\mu = 1}^d \left( \frac{\Gamma_\mu}{\Gamma_\mu - \ii \chi}\right )^{n_\mu}\equiv g(\mathbf{n},t),\notag
\end{align}
where on the third line we write the delta function in its Fourier representation, $\delta(u)=\int\dd \chi\, \ee^{-\ii\chi u}/2\pi$, and the last line is obtained by recognising that the product involves precisely $n_\mu(\bm{\sigma}) =\sum_{j=0}^N \delta_{\mu\sigma_j}$ identical factors for each state $\mu$.

\section{Jump steady state and jump-conditioned transition probabilities}
\label{SM:jump_ss}

In this appendix we provide more details on the jump steady state discussed in Sec.~\ref{sec:residual_vs_waiting}, and the conditional transition probabilities used to define resolution in Sec.~\ref{sec:resolution}. Suppose that a jump is observed in a given small time interval $\dd t \to 0$. Under steady-state conditions, the probability that the observed jump was $\sigma\to\mu$ is given by
\begin{align}
    \label{prob_mu_sigma}
	P(\sigma\to\mu|\, {\rm jump}) = \frac{R_{\mu\sigma} p_\sigma^\ss}{\mathcal{A}},
\end{align}
where $\mathcal{A} = \sum_{\mu,\sigma} R_{\mu\sigma} p_\sigma^\ss$ is the dynamical activity. This follows from the master equation, which defines the probability to observe a jump $\sigma\to\mu$ in the time interval $\dd t$ as $P(\sigma\to\mu,\, {\rm jump}) = R_{\mu\sigma} p^\ss_\sigma \dd t$, such that the total probability for any jump to be observed is $P({\rm jump}) = \mathcal{A}\dd t$. Putting these together gives $P(\sigma\to\mu|\, {\rm jump})  = P(\sigma\to\mu,\, {\rm jump})/P({\rm jump}).$

Marginalising Eq.~\eqref{prob_mu_sigma} over the final state after the jump, $\mu$, then yields 
\begin{equation}
    \label{jump_ss_derivation}
 	\sum_{\mu} \frac{R_{\mu\sigma} p_\sigma^\ss}{\mathcal{A}} = \frac{\Gamma_\sigma p_\sigma^\ss}{\mathcal{A}} \equiv  p^\jss_\sigma,
\end{equation}
where the jump steady state, $p^\jss_\sigma$, was introduced in Eq.~\eqref{jump_steady_state}. Marginalising Eq.~\eqref{prob_mu_sigma} over the initial state $\sigma$ yields the same distribution by stationarity, i.e.~$\sum_\sigma R_{\mu\sigma}p_\sigma^\ss/\mathcal{A}=p_\mu^\jss$. Therefore, $p_\mu^\jss$ is both the fraction of jumps into and out of state $\mu$. Equivalently, $p_\mu^\jss$ is the frequency of the state $\mu$ within a long sequence $\bm{\sigma}$. The jump steady state is also the stationary distribution of the transition probabilities $\pi(\mu|\sigma) = R_{\mu\sigma}/\Gamma_\sigma$, as defined below Eq.~\eqref{path_probability_with_times} in the main text. This means that 
\begin{equation}
    \label{jss_stationary}
    \sum_{\sigma=1}^d\pi(\mu|\sigma) p^\jss_\sigma = p^\jss_\mu.
\end{equation}

\begin{figure}
    \centering
    \includegraphics[width=\linewidth]{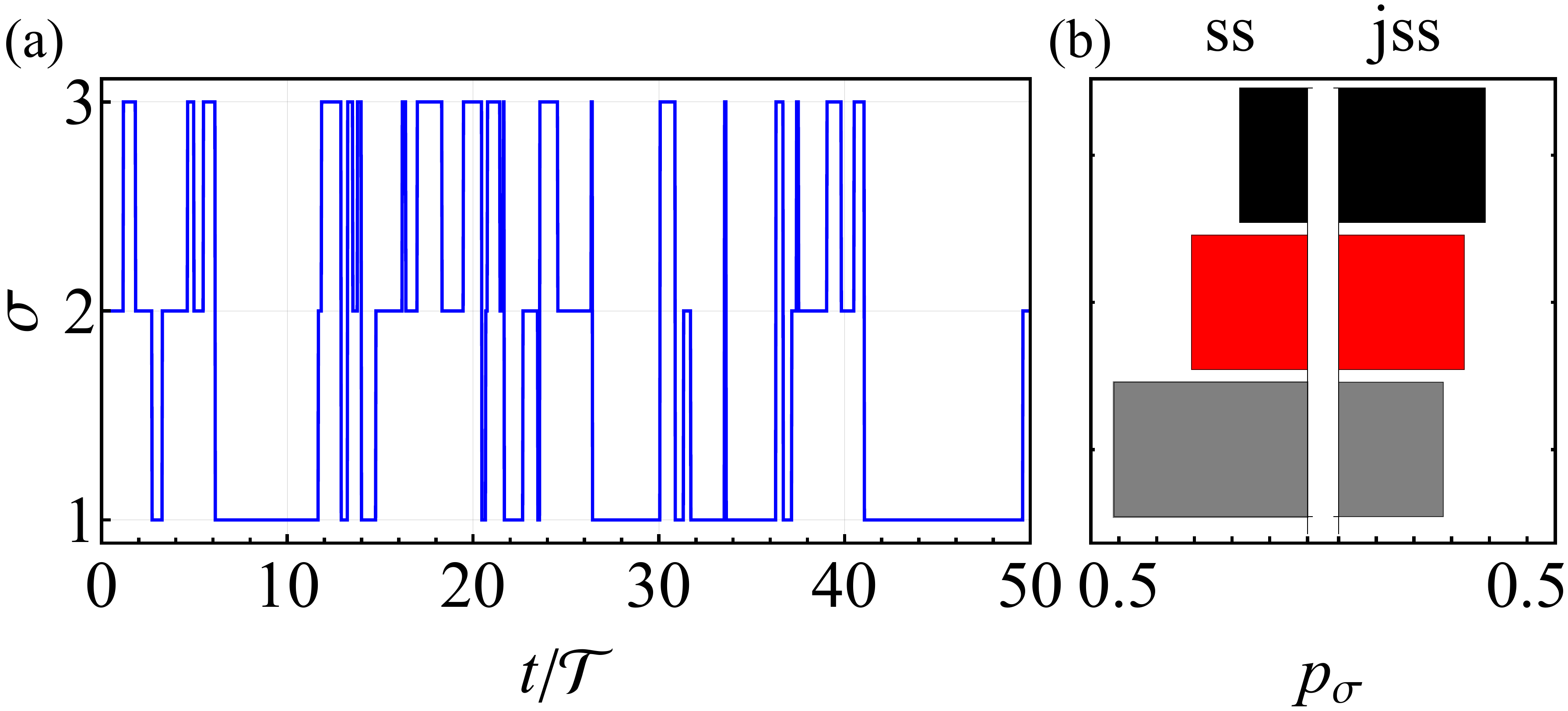}
    \caption{(a) Example trajectory of a three-state system with the generator defined by Eq.~\eqref{jss_L}. (b) Comparison of the steady state distribution $p^{\ss}_\sigma$ (left) and jump steady state $p_\sigma^\jss$ (right) for this system. The most time is spent in the long-lived state $\sigma=1$ but the short-lived state $\sigma=3$ is visited most frequently.}
    \label{fig:jss}
\end{figure}

To illustrate how the jump steady state differs from the steady-state distribution $p_\sigma^\ss$, Fig.~\ref{fig:jss} shows a random trajectory for a three-state system with generator
\begin{equation}
\label{jss_L}
    \mathsf{L}=\begin{pmatrix}
        -2 & 1 & 4 \\
        1 & -4 & 4 \\
        1 & 3 & -8
    \end{pmatrix},
\end{equation}
chosen to highlight the differences between $p_\sigma^\ss$ and $p_\sigma^\jss$. As Fig.~\ref{fig:jss} makes clear, $p_\sigma^\ss$ represents the fraction of the total time spent by the system in each state. For a given random trajectory, this fraction is known as the empirical measure in large-deviation theory~\cite{Garrahan2017}, and its limiting value for long observation times is $p_\sigma^\ss$. Conversely, $p_\sigma^\jss$ represents the frequency of visits to each state: it is therefore the limiting value of the normalised empirical distribution, i.e. $ \lim_{t\to\infty} n_\sigma/(N+1) = p_\sigma^\jss$. See Ref.~\cite{Landi2023a} for further details on the jump steady state in general Markovian open quantum dynamics. 

Now suppose that we monitor only those jumps $\sigma\to\mu$ within an alphabet $\aleph$, as discussed above Eq.~\eqref{resolution}, which defines the resolution. Given that some jump within $\aleph$ is observed, the probability that it was the specific jump $\sigma\to\mu$ is
\begin{align}
    \label{prob_mu_sigma_aleph}
	P(\sigma\to\mu|\aleph) =\frac{R_{\mu\sigma} p_\sigma^\ss}{\mathcal{A}_\aleph},
\end{align}
by a straightforward adaptation of Eq.~\eqref{prob_mu_sigma}. Above, the partial dynamical activity for the alphabet $\aleph$ is defined as $\mathcal{A}_{\aleph}=\sum_{(\sigma\to\mu)\in\aleph} R_{\mu\sigma} p_\sigma^\ss$. For a given generator $\mathsf{L}$ and its associated steady state $p_\sigma^\ss$, $\mathcal{A}_\aleph$ depends only on the alphabet $\aleph$. Moreover, since it is a sum of positive terms, decreasing the size of the alphabet can only decrease $\mathcal{A}_\aleph$. Therefore, for any other alphabet $\aleph' \subseteq \aleph$, we have $\mathcal{A}_{\aleph^\prime} \leq  \mathcal{A}_{\aleph}$.

\section{Distribution of residual time}
\label{SM:inspection_paradox}

In this appendix, we explain the inspection paradox~\cite{Pal2022} by deriving the distribution of the residual time (first recurrence time), following the discussion in Refs.~\cite{McFadden1962,Cox1982}. Consider the diagram in Fig.~\ref{fig:residual_time}, showing jump events distributed randomly on the time axis. The unconditional distribution of waiting times in the bulk of the sequence, $\tau_j$ with $1\leq j<N$, is denoted by
\begin{equation}
    \label{bulk_waiting_time_distribution}
    W(\tau) = \sum_\sigma p_\sigma^\jss W(\tau|\sigma),
\end{equation}
where the jump steady state is defined in Eq.~\eqref{jump_ss_derivation} and the conditional waiting-time distribution is $W(\tau|\sigma) = \Gamma_\sigma \ee^{-\Gamma_\sigma \tau}$. 

Note that, by stationarity, each waiting time is identically distributed but generally not independent. To see this, consider the joint distribution of unconditional waiting times for two consecutive jumps, 
\begin{equation}
\label{joint_waiting_times}
    W(\tau_1,\tau_2) = \sum_{\sigma_1}\sum_{\sigma_2} W(\tau_2|\sigma_2)\pi(\sigma_{2}|\sigma_1) W(\tau_1|\sigma_1) p_{\sigma_1},
\end{equation}
given some initial distribution $p_{\sigma_1}$ and after marginalising over the initial and final states of both jumps. In Eq.~\eqref{joint_waiting_times} we are conditioning on precisely $N=2$ jumps having occurred but the total time is unconstrained. This is known as the $N$-ensemble, as opposed to the $t$-ensemble used in the majority of this work. (See Sec.~VI.B of Ref.~\cite{Landi2024} for a detailed discussion of the $t$- and $N$-ensembles; see also Refs.~\cite{Budini2014,Kiukas2015} where they are referred to as the $s$- and $x$-ensembles, respectively.) Stationarity means that the marginal distributions of $\tau_1$ and $\tau_2$ are identical. This is the case only if we set $p_{\sigma_1} = p_{\sigma_1}^\jss$ equal to the jump steady state, whence
\begin{equation}
    W(\tau_2) = \int_0^\infty \dd \tau_1 \, W(\tau_1,\tau_2) = \sum_{\sigma_2} p_{\sigma_2}^\jss W(\tau_2|\sigma_2),
\end{equation}
where we have used Eq.~\eqref{jss_stationary}, and the same holds for $W(\tau_1)$. Clearly, however, $W(\tau_1,\tau_2)\neq W(\tau_1)W(\tau_2)$, so the waiting times are correlated in general.

To find the properties of the residual time, $\tau_0$, we consider a long time interval $T$ with $M\gg 1$ jumps, and randomly select an initial time $t_0$ as shown in Fig.~\ref{fig:residual_time}. The time $t_0$ will fall within a given interval between two jumps, indexed by $i$. Assuming $t_0$ is distributed uniformly over its entire range $T$, the probability that the interval $i$ is selected is proportional to its length $\tau_i$, as $P(i|\tau_i=\tau) = \tau/T$. The joint probability of selecting interval $i$ with length $\tau$ is therefore $P(i,\tau) = P(i|\tau_i=\tau) W(\tau)$. Marginalising over $i$, we find the probability that an interval with length $\tau$ is selected, as
\begin{equation}
\label{length-bias}
    P(\tau) = \sum_{i=1}^M P(i,\tau) = \frac{M \tau W(\tau)}{T} = \mathcal{A}\tau W(\tau) + \mathcal{O}(T^{-1/2}),
\end{equation}
where we used the fact that $M/T= \mathcal{A} + \mathcal{O}(T^{-1/2})$ for large $T$. We assume $T\to \infty$ from here on. 

Given that the selected interval has duration $\tau$, the conditional distribution of the residual time $\tau_0$ is uniform, i.e.
\begin{equation}
    \label{P_tau_0_given_tau}
    P(\tau_0|\tau) = \begin{cases}
       \tau^{-1} & (0\leq \tau_0 < \tau) \\
        0 & (\tau_0 \geq \tau),
    \end{cases}
    \end{equation}
which follows from stationarity and the uniformly random selection of $t_0$. We thus finally arrive at the distribution of residual time
\begin{equation}
    \label{residual_time_distribution}
    W_0(\tau_0) = \int_{0}^\infty \dd\tau \, P(\tau_0|\tau)P(\tau)  = \int_{\tau_0}^\infty \dd\tau \, \mathcal{A}W(\tau).
\end{equation}
This result is valid for any discrete, stationary process where intervals between events are characterised by a distribution $W(\tau)$ with mean $\mathcal{A}^{-1}$~\cite{McFadden1962,Cox1982}. In particular, it holds for arbitrary non-Markovian dynamics so long as the process is stationary. For the specific case of Markovian jump processes considered here, we find simply
\begin{equation}
    \label{residual_time_dist_specific}
    W_0(\tau) = \sum_{\sigma}p_\sigma^\ss W(\tau|\sigma),
\end{equation}
as expected from first principles. 

In full generality, the mean of the residual time can be found from Eq.~\eqref{residual_time_distribution}, after changing the order of integration, to be
\begin{equation}
    \label{mean_tau0}
    \EE[\tau_0] = \int_0^\infty \dd\tau_0   W_0(\tau_0) \tau_0 = \frac{\EE[\tau^2]}{2\EE[\tau]},
\end{equation}
where expectations on the right-hand side are taken with respect to the waiting-time distribution~\eqref{bulk_waiting_time_distribution}. This can be rearranged suggestively as
\begin{equation}
    \label{mean_residual_var_mean}
    \EE[\tau_0] = \frac{\EE[\tau]}{2}\left(1 + \frac{\Var[\tau]}{\EE[\tau]^2}\right).
\end{equation}
A strong form of the inspection paradox arises if the residual time exceeds the waiting time on average, i.e.~$\EE[\tau_0] > \EE[\tau]$. According to Eq.~\eqref{mean_residual_var_mean}, this occurs whenever $\Var[\tau]> \EE[\tau]^2$. This is characteristic of super-Poissonian statistics, where jumps cluster together in bunches, interspersed by long periods without any jumps. According to Eq.~\eqref{length-bias}, uniformly selecting the initial time $t_0$ generates a bias in favour of these longer intervals. 

The specific waiting-time distribution in Eq.~\eqref{bulk_waiting_time_distribution} is hyper-exponential, i.e.~a mixture of exponential distributions. This satisfies $\Var[\tau]>\EE[\tau]^2$ unless all escape rates $\Gamma_\sigma$ are equal, in which case it reduces to a simple exponential distribution. Specifically, we have $\EE[\tau] = \mathcal{A}^{-1}$ and
\begin{equation}
    \label{var_tau}
    \frac{\Var[\tau]}{\EE[\tau]^2} = 2\mathcal{T}\mathcal{A}-1 \geq 1,
\end{equation}
using the relation between arithmetic and harmonic means, $\mathcal{A}\mathcal{T}\geq 1.$ Plugging this into Eq.~\eqref{mean_residual_var_mean} recovers $\EE[\tau_0]=\mathcal{T}$, as expected.

\section{Limiting behaviour of the Fisher information}
\label{SM:Fisher_limits}

In this appendix, we derive the long- and short-time limits of the Fisher information quoted in Sec.~\ref{sec:optimal_estimator}.

\subsection{Saddle-point approximation}
\label{SM:saddle_point}

We first derive Eq.~\eqref{score_saddle_point} in the main text, from which the asymptotic Fisher information given in Eq.~\eqref{asymptotic_Fisher} follows. To this end, we use Eqs.~\eqref{eq:supp_pfact}--\eqref{eq:supp_gnt} to write
\begin{equation}
    \label{eq:probsaddle1}
    P(\bm\sigma|t) =\Pi(\bm\sigma)\int_{-\ii \infty}^{\ii \infty} \frac{\dd z}{2\pi \ii} \ee^{-z t}\ee^{-\sum_\mu n_\mu\ln(1-z/\Gamma_\mu)}.
\end{equation}
The idea of the saddle-point approximation is to deform this integral in the complex plane such that it passes a saddle point of the exponent along the direction of steepest descent, for details see Sec.~V.~E.~of Ref.~\cite{Landi2024}. The saddle point can be found from the saddle-point equation
\begin{equation}
    \label{eq:supp_spe}
    \sum_\mu\frac{n_\mu}{\Gamma_\mu-z_{\rm sp}} = t,
\end{equation}
which is simply obtained by setting the derivative of the exponent in Eq.~\eqref{eq:probsaddle1} with respect to $z$ to zero. The integral in Eq.~\eqref{eq:probsaddle1} may then be approximated as a Gaussian integral around the saddle point and in particular
\begin{equation}
\label{eq:supp_logpsadd}
    \ln P(\bm\sigma|t) = -z_{\rm sp}t -\sum_\mu n_\mu \ln(1-z_{\rm sp}/\Gamma_\mu) +\mathcal{O}(\ln t),
\end{equation}
which is just the exponent of the integral evaluated at the saddle-point $z_{\rm sp}$.
From Eqs.~\eqref{eq:supp_spe} and \eqref{eq:supp_logpsadd}, one may show that
\begin{equation}
    \label{eq:supp_derlogprob}
    \partial_t \ln P(\bm\sigma|t) = -z_{\rm sp} +\mathcal{O}(t^{-1}).
\end{equation}

To make progress, we anticipate that $z_{\rm sp}$ is small at long times in the sense that it is a random variable with zero mean ($\EE[\partial_t\ln P(\bm{\sigma}|t)]= \sum_{\bm{\sigma}}\partial_t P(\bm{\sigma}|t)=0$ by normalisation) and a variance that scales as $1/t$. The same behavior holds for $\delta n_\mu/t=(n_\mu-\EE[n_\mu])/t.$ We therefore expand Eq.~\eqref{eq:supp_spe} in both $z_{\rm sp}$ and $\delta n_\mu/t$ which results in
\begin{equation}
    \label{eq:supp_saddlepoint}
    z_{\rm sp} \simeq -\left(\sum_\mu\frac{\EE[n_\mu]}{\Gamma_\mu^2}\right)^{-1}\left(\sum_\mu\frac{n_\mu}{\Gamma_\mu}-t\right).
\end{equation}
Together with Eq.~\eqref{eq:supp_derlogprob}, this recovers Eq.~\eqref{score_saddle_point} in the main text.

\subsection{Short-time Fisher information}
\label{SM:short_time_Fisher}

We now derive Eq.~\eqref{Fisher_info_short_time} for the Fisher information in the short-time limit. As $t\to 0$ only trajectories with $N=1$ jump are significant for time estimation. Indeed, from Eqs.~\eqref{path_probability_with_times}--\eqref{path_probability} we have
\begin{align}
    \label{short_time_probs}
    P(\bm{\sigma}|t) = \begin{cases} (1-\Gamma_{\sigma_0}t)p_{\sigma_0}^\ss + \mathcal{O}(t^2) &  (N=0) \\
    R_{\sigma_1\sigma_0} t p^\ss_{\sigma_0} + \mathcal{O}(t^2) & (N=1) \\
    \mathcal{O}(t^2) & (N\geq 2).
    \end{cases}
\end{align}
The score is therefore given to leading order by 
\begin{align}
    \label{short_time_score}
    \partial_t \ln P(\bm{\sigma}|t) = \begin{cases} -\Gamma_{\sigma_0} + \mathcal{O}(t) &  (N=0) \\
    t^{-1} + \mathcal{O}(t^0) & (N=1) \\
    \mathcal{O}(t^{-1}) & (N\geq 2).
    \end{cases}
\end{align}
Combining Eqs.~\eqref{short_time_probs} and~\eqref{short_time_score}, we have 
\begin{align}
    \label{mean_score_sqd}
    \mathcal{F}_t & = \sum_{\bm{\sigma}} P(\bm{\sigma}|t) \left[\partial_t \ln P(\bm{\sigma}|t)\right]^2 \notag \\
& = \sum_{\sigma_0 = 1}^d \Gamma_{\sigma_0}^2 p_{\sigma_0}^\ss + \sum_{\sigma_0,\sigma_1 = 1}^d \frac{R_{\sigma_1\sigma_0} p_{\sigma_0}^\ss}{t} + \mathcal{O}(t^0),
\end{align}
from which Eq.~\eqref{Fisher_info_short_time} follows because $\mathcal{A} = \sum_{\mu,\sigma}R_{\mu\sigma}p^\ss_\sigma$. As a consistency check, note that  Eqs.~\eqref{short_time_probs} and~\eqref{short_time_score} also predict that the average score vanishes to the same order of approximation, $\EE[\partial_t\ln P(\bm{\sigma}|t)] = \mathcal{O}(t)$. This agrees with the exact result $\EE[\partial_t\ln P(\bm{\sigma}|t)] = 0$, thus validating our approximate treatment of the short-time limit.

Finally, we prove that the uniform estimator $\Theta(N) = N/\mathcal{A}$ saturates the CRB for time estimation in the $t\to 0$ limit. The probability $P(N|t)$ for $N$ jumps to occur after a short time $t$ is given by
\begin{equation}
    \label{P_1_short_time}
    P(1|t) = \mathcal{A} t + \mathcal{O}(t^2) = 1 - P(0|t),
\end{equation}
while $P(N|t) = \mathcal{O}(t^2)$ is negligible for $N\geq 2$, as can be shown using Eq.~\eqref{short_time_probs} in Eq.~\eqref{P_N_of_t}. Therefore, $N\approx N^2$ behaves like a Poisson increment with mean and variance
\begin{align}
    \label{N_mean_short_time}
    \EE[N] = \mathcal{A}t + \mathcal{O}(t^2) = \Var[N].
\end{align}
It follows that $\Theta(N)= N/\mathcal{A}$ is an unbiased time estimator in this limit, with variance $\Var[\Theta]=t/\mathcal{A} + \mathcal{O}(t^2)$, thus saturating the CRB~\eqref{CRB} with the Fisher information given by Eq.~\eqref{Fisher_info_short_time}.

\section{Statistics of the asymptotically efficient estimator}
\label{SM:small_n_estimator}

In this appendix derive exact expressions for the mean and variance of the estimator~\eqref{efficient_estimator}, i.e.
\begin{equation}
	\label{linear_estimator_sm}
	\Theta(\mathbf{n}) = \sum_{\mu} \frac{n_\mu}{\Gamma_\mu},
\end{equation}
as well as the final expression for the Fisher information in Eq.~\eqref{asymptotic_Fisher}. The derivations in this appendix exploit several results from Appendix~\ref{SM:current_statistics}: specifically, the statistics of the elementary counting observables $N_{\mu\sigma}(\bm{\sigma})$, i.e.~the number of jumps $\sigma\to\mu$ in a given sequence $\bm{\sigma}$. The reader may thus find it helpful to consult Appendix~\ref{SM:current_statistics} before proceeding.

Our starting point is the representation
\begin{equation}
	n_\mu(\bm{\sigma}) = \sum_{\sigma=1}^d N_{\mu\sigma}(\bm{\sigma}) + \delta_{\mu\sigma_0},
\end{equation}
i.e.~$n_\mu(\bm{\sigma})$ is the total number of jumps entering state $\mu$, plus a contribution if the initial state $\sigma_0=\mu$. Using Eqs.~\eqref{Dyson_series} and~\eqref{prob_path_integral}, we have that 
\begin{align}
	\label{initial_state_fixed}
	\EE[\delta_{\mu\sigma_0}] & = \sum_{N=0}^\infty  \sum_{\sigma_0=1}^d \delta_{\mu\sigma_0}\sum_{\sigma_1=1}^d \cdots \sum_{\sigma_{N}=1}^d \int_0^t \dd\mathbf{t}\, P(\bm{\sigma},\mathbf{t}|t) \\
	& = \sum_{\sigma_N=1}^d [\mathsf{G}(t)]_{\sigma_N\mu} p_{\mu}^\ss =  p_{\mu}^\ss,
\end{align}
using the fact that $\mathbf{u}^\T\mathsf{G}(t)=1$, i.e.~$\mathsf{G}(t)$ is a stochastic matrix. Combining this with Eq.~\eqref{mean_current_av}, we find
\begin{equation}
    \label{SM_emp_dist_mean}
    \EE[n_\mu] = \sum_{\sigma=1}^d R_{\mu\sigma}p^\ss_\sigma t  + p_{\mu}^\ss = \left(1+\Gamma_\mu t\right)p_\mu^\ss,
\end{equation}
where we invoked stationarity, $\mathsf{R}\mathbf{p}^\ss = \mathsf{\Gamma}\mathbf{p}^\ss$. The above expression is used to obtain the final result for the Fisher information in Eq.~\eqref{asymptotic_Fisher}.

We also have 
\begin{align}
	\label{N_initial_state_fixed}
	\EE[N_{\alpha\sigma} \delta_{\mu\sigma_0}] = \sum_{N=0}^\infty  \sum_{\sigma_1=1}^d \cdots \sum_{\sigma_{N}=1}^d \int \dd\mathbf{t}\,  N_{\alpha\sigma} P(\sigma_1\cdots\sigma_N,\mathbf{t}|\mu,t) p_{\mu}^\ss,
\end{align}
where $	P(\sigma_1\cdots\sigma_N,\mathbf{t}|\sigma_0,t) = 	P(\bm{\sigma},\mathbf{t}|t)/p_{\sigma_0}^\ss$ is the probability for a trajectory, conditioned on the system starting in the state $\sigma_0$. Using Eqs.~\eqref{elementary_stochastic_increment} and \eqref{total_jump_count}, we therefore have 
\begin{equation}
	\label{N_initial_state_fixed_solution}
	\EE[N_{\alpha\sigma} \delta_{\mu\sigma_0}] = \int_0^t \dd t' R_{\alpha\sigma} G_{\sigma\mu}(t')p_\mu^\ss,
\end{equation}
since $G_{\sigma\mu}(t) = p_{\sigma}(t)$ is the solution of the master equation with initial condition $p_{\sigma}(0) = \delta_{\sigma\mu}.$ Adding and subtracting a term, and then summing over $\sigma$, we obtain
\begin{align}
	\label{var_calculation}
	\sum_\sigma \EE[N_{\alpha\sigma} \delta_{\mu\sigma_0}] & = \sum_\sigma\left ( \int_0^t \dd t' R_{\alpha\sigma} \left [G_{\sigma\mu}(t') - p_\sigma^\ss\right ]  + J_{\alpha\sigma} t \right )p_\mu^\ss\notag 
	\\
	& =\left (\int_0^t\dd t'\, \left [\mathsf{R}\left (  \mathsf{G}(t') - \mathsf{
	\Pi}_0\right )\right ]_{\alpha\mu} + \Gamma_\alpha p_\alpha^\ss t\right ) p_\mu^\ss,
\end{align}
where $\mathsf{\Pi}_0 = \mathbf{p}^\ss\mathbf{u}^\T$ is the steady-state projector. Putting everything together, we find the mean 
\begin{equation}
	\label{linear_bias}
	\EE[\Theta] = \sum_{\mu} \frac{\left (1+\Gamma_\mu t\right )p_\mu^\ss}{\Gamma_\mu} = \mathcal{T} + t,
\end{equation}
and, following some simple manipulations, the variance
\begin{align}
	\label{linear_var}
	\Var[\Theta] & = \mathcal{T}t + \Var[\Gamma_\sigma^{-1}] \notag \\ 
	& \quad + 2\int_0^t\dd t'\,\mathbf{u}^\T\mathsf{\Gamma}^{-1}\mathsf{L}\left [\mathsf{G}(t') - \mathsf{\Pi}_0\right ]\mathsf{\Gamma}^{-1}\mathbf{p}^\ss.
\end{align}
Here, we define
\begin{equation}
	\label{variance_waiting_time}
	\Var[\Gamma_\sigma^{-1}] = \sum_{\sigma} \frac{p^\ss_\sigma}{\Gamma_\sigma^2} - \mathcal{T}^2.
\end{equation}

These expressions are exact for all $t$, but simple approximations are possible in the limit of times that are very short, $\Gamma_\sigma t \ll 1$, or very long, $\Gamma_\sigma t \gg 1$. For short times, the integral in Eq.~\eqref{linear_var} is negligible, while for long times we can extend the integration limit to infinity and use Eq.~\eqref{Drazin_inverse_integral}. We thus obtain
\begin{equation}
	\label{linear_var_approx}
	\Var[\Theta] \approx  \begin{cases}
		\mathcal{T}t + \Var[\Gamma_\sigma^{-1}] & (\Gamma_\sigma t \ll 1) \\
			\mathcal{T}t - \Var[\Gamma_\sigma^{-1}] & (\Gamma_\sigma t \gg 1) 
	\end{cases}.
\end{equation}

It is interesting to compare these results with the best unbiased linear estimator (BLUE): a minimal-variance counting observable that is defined in Sec.~\ref{sec:counting_observables}. As discussed there, and shown in detail in Appendix~\ref{SM:hyperaccurate_current}, the variance of the BLUE is exactly equal to $\mathcal{T}t$ for all times. For short times, therefore, the uncertainty in Eq.~\eqref{linear_var_approx} is increased relative to the BLUE because $\Theta(\mathbf{n})=\Gamma_{\sigma_0}^{-1}\neq 0$ even though no jump has yet occurred. For large times, conversely, the uncertainty is decreased by the same amount since all states along the trajectory are accounted for. Note that this decrease is of order $\mathcal{O}(t^0)$ and is therefore negligible under the same approximations that lead to Eq.~\eqref{asymptotic_Fisher}. Hence, there is no inconsistency with the CRB within the saddle-point approximation, which neglects subleading corrections in time.

\section{Exact solution for the two-state system}
\label{SM:exact_solution_d2}

To illustrate the estimator $\Theta(\mathbf{n})$ given in Eq.~\eqref{efficient_estimator} and the validity of the saddle-point approximation in obtaining Eqs.~\eqref{score_saddle_point} and~\eqref{asymptotic_Fisher}, which was presented in Appendix~\ref{SM:saddle_point}, we investigate a two-state system. Let $\Gamma_\mu$ be the escape rate from the state $\mu = 1, 2$, such that the transition rates are given by $R_{21} = \Gamma_1$ and $R_{12} = \Gamma_2$. The steady state of the system reads $p_1^{\rm ss} = \Gamma_2/(\Gamma_1 + \Gamma_2) $, $p_2^{\rm ss} = \Gamma_1/(\Gamma_1 + \Gamma_2) $. For the two-state system, each trajectory $\bm{\sigma}$ is unambiguously determined by the number of visits of each state, $\mathbf{n} = (n_1, n_2)$, together with the initial state $\sigma_0$.

Starting from Eqs.~\eqref{eq:supp_pfact} and~\eqref{eq:supp_gnt}, we may write
\begin{equation}
    g(\mathbf{n},t) = \Gamma_1^{n_1} \Gamma_2^{n_2} \int_{-\infty}^\infty \frac{\dd \chi}{2\pi}\, \ee^{-\ii \chi t}  \left( \frac{\ii }{\chi + \ii\Gamma_1}\right )^{n_1} \left( \frac{\ii }{\chi + \ii\Gamma_2}\right )^{n_2}.
\end{equation}
To make progress in analytically solving the integral, we move to the complex plane and express $g(\mathbf{n},t)$ as a contour integral around a clockwise-oriented closed loop $\mathcal{C}$ that crosses the imaginary axis at $- \ii \infty$:
\begin{equation}
    g(\mathbf{n},t) = \Gamma_1^{n_1} \Gamma_2^{n_2} \oint_{\mathcal{C}} \frac{\dd \chi}{2\pi}\, \ee^{-\ii \chi t}  \left( \frac{\ii }{\chi + \ii\Gamma_1}\right )^{n_1} \left( \frac{\ii }{\chi + \ii\Gamma_2}\right )^{n_2}.
\end{equation}
This is justified, because the integrand vanishes when the imaginary part of $\chi$ equals $- \ii \infty$. The integrand has two poles, $\chi_\mu = - \ii \Gamma_\mu$, which both lie inside $\mathcal{C}$. Using the residue theorem, we find
\begin{equation} \label{eq: g res}
    g(\mathbf{n},t) = - 2 \pi \ii \Gamma_1^{n_1} \Gamma_2^{n_2} \left( \mathcal{R}_1 + \mathcal{R}_2 \right),
\end{equation}
where
\begin{equation}
    \mathcal{R}_\mu = \frac{1}{2 \pi ({n_\mu}-1)!} \lim_{\chi \to \chi_\mu} \frac{\dd^{{n_\mu}-1}}{\dd \chi^{{n_\mu}-1}} \ee^{-\ii \chi t} \frac{ \ii^{n_{\mu}+n_{\bar{\mu}}} }{(\chi - \chi_{\bar{\mu}})^{n_{\bar{\mu}}}}
\end{equation}
is the residue of the pole $\chi_\mu$, and $\bar{\mu} \neq \mu$. This leads to a formal solution
\begin{align}
\label{residues_exact}
    \mathcal{R}_\mu =  &\frac{\ii (-1)^{n_{\bar{\mu}}} \ee^{-\Gamma_\mu t}}{2 \pi (\Gamma_\mu - \Gamma_{\bar{\mu}})^{n_\mu+n_{\bar{\mu}}-1}}  \notag \\
    & \times \sum_k^{n_{\mu}-1} \binom{n_{\mu}-1}{k} \frac{(n_{\bar{\mu}}-1+k)!}{(n_{\mu}-1)!(n_{\bar{\mu}}-1)!} \left [(\Gamma_\mu - \Gamma_{\bar{\mu}}) t\right ]^{n_{\mu}-1-k},
\end{align}
which allows us to exactly compute $P(\bm{\sigma}|t)$ using Eqs.~\eqref{eq:supp_pfact} and~\eqref{eq:supp_gnt}, the Fisher information $\mathcal{F}_t$ given in Eq.~\eqref{Fisher_info}, and the average and variance of the estimator $\Theta(\mathbf{n})$ [Eq.~\eqref{efficient_estimator}].
These quantities are illustrated in Fig.~\ref{fig:estimators}.

In the reminder of this section, we illustrate the validity of the saddle-point approximation used to analytically obtain Eqs.~\eqref{score_saddle_point} and~\eqref{asymptotic_Fisher} in the asymptotic limit, which is shown in Appendix~\ref{SM:saddle_point}. In the asymptotic limit, we have $n_\mu \approx n_\mu +1$, and thus we set $n_1 = n_2= n$. The sum of residues in Eq.~\eqref{residues_exact} may be expressed as
\begin{align}
    \mathcal{R}_1 + \mathcal{R}_2 = &\frac{ \ee^{-t(\Gamma_1 + \Gamma_2)/2}}{2 \sqrt{\pi} t (n-1)! (\Gamma_1 - \Gamma_2)^{2n}}  \left [(\Gamma_1 - \Gamma_2)t\right ]^{n+\frac{1}{2}} \notag \\
    & \times \mathbb{I}_{n-\frac{1}{2}} \left(t \frac{\Gamma_1 - \Gamma_2}{2}\right),
\end{align}
where $\mathbb{I}_a(x)$ is a modified Bessel function of the first kind, and we assume without loss of generality $\Gamma_1 > \Gamma_2$. Taking the time derivative results in
\begin{equation}
    \partial_t \ln P(\bm{\sigma}|t) = -\frac{\Gamma_1 + \Gamma_2}{2} + \frac{(n-\frac{1}{2})}{t} + \partial_t \ln{  \mathbb{I}_{n-\frac{1}{2}} \left(t \frac{\Gamma_1 - \Gamma_2}{2}\right) } ,
\end{equation}
where we may substitute $n - 1/2 \approx n$. To tackle the last term, we use the approximation of $\mathbb{I}_a(x)$ for large arguments:
	\begin{equation}
    \mathbb{I}_a(x) \approx \frac{1}{\sqrt{2 \pi} (a^2 + x^2)^{\frac{1}{4}}} \exp\left(  -a \text{ arcsinh}(a/x) + \sqrt{a^2 + x^2}  \right),
\end{equation}
resulting in
\begin{align}
   &  \partial_t \ln \mathbb{I}_{n-\frac{1}{2}} \left(t \frac{\Gamma_1 - \Gamma_2}{2}\right)  \notag \\
   &\approx \frac{1}{2} \sqrt{ \frac{4n^2}{t^2} + (\Gamma_1 - \Gamma_2)^2 } - \frac{1}{2 } \frac{(\Gamma_1 - \Gamma_2)^2 t}{4n^2 + (\Gamma_1 - \Gamma_2)^2 t^2} .
\end{align}
The last term asymptotically vanishes, leading to the expression
\begin{equation}
\label{eq: SP approx}
    \partial_t \ln P(\bm{ \sigma}|t) = -\frac{\Gamma_1 + \Gamma_2}{2} + \frac{n}{t} + \frac{1}{2} \sqrt{ \frac{4n^2}{t^2} + \left (\Gamma_1 - \Gamma_2\right )^2 }.
\end{equation}
In Appendix~\ref{SM:saddle_point}, the saddle-point approximation was employed to show that in the asymptotic regime $\partial_t \ln P(\bm{\sigma}|t)=-z_{\rm sp}$, where the saddle point, $z_{\rm sp}$, is determined by Eq.~\eqref{eq:supp_spe}, which may equivalently be expressed as a quadratic equation:
\begin{equation}
    z_{\rm sp}^2 + z_{\rm sp} \left( \frac{2n}{t} -(\Gamma_1 + \Gamma_2) \right) + \Gamma_1 \Gamma_2  - \frac{n}{t} (\Gamma_1 + \Gamma_2) = 0 .
\end{equation}
The right hand side of Eq.~\eqref{eq: SP approx} is a solution to this equation, which shows that applying the saddle-point method to find $\partial_t \ln P(\bm{ \sigma}|t)$ is justified.

\section{Current statistics}
\label{SM:current_statistics}

\subsection{Elementary stochastic currents}
\label{SM:elementary_statistics}

In this appendix, we recap the formalism for computing current statistics, adapting the approach of Ref.~\cite{Landi2024} to the problem of interest in this work. We define the elementary stochastic increments $\dd N_{\mu\sigma}(t)$: binary random variables such that $\dd N_{\mu\sigma}(t) = 1$ when a jump $\sigma\to\mu$ occurs and $\dd N_{\mu\sigma}(t) =0$ otherwise. They are completely specified by the properties
\begin{align}
		\label{elementary_stochastic_increment}
	& 	\EE[\dd N_{\mu\sigma}(t)] = R_{\mu\sigma} p_\sigma(t)\dd t, \notag \\
	& \dd N_{\alpha\beta}(t) \dd N_{\mu\sigma}(t)= \delta_{\alpha\mu}  \delta_{\beta\sigma}\dd N_{\mu\sigma}(t).
\end{align}
The total number of jumps $\sigma\to \mu$ is therefore given by
\begin{equation}
	\label{total_jump_count}
	N_{\mu\sigma}(t) = \int_0^t\dd N_{\mu\sigma}(t')  \equiv   \int_0^t\dd t' I_{\mu\sigma}(t'),
\end{equation}
where $I_{\mu\sigma}(t) = \dd N_{\mu\sigma}/\dd t$ is the corresponding current. 

The asymptotic current statistics are captured by the growth rate of the mean and variance at long times,
\begin{align}
	\label{mean_current_def}
	 & J_{\mu\sigma}  = \lim_{t\to\infty} \frac{\dd}{\dd t} \EE[N_{\mu\sigma}(t)] =  \lim_{t\to\infty} J_{\mu\sigma}(t),\\
	& D_{\alpha\beta\mu\sigma}  = \lim_{t\to\infty} \frac{\dd}{\dd t} \Cov [N_{\alpha\beta}(t) ,N_{\mu\sigma}(t)].
\end{align}
where $J_{\mu\sigma}(t)$ is the average instantaneous current and $D_{\alpha\beta\mu\sigma} $ is the diffusion tensor. Plugging in Eq.~\eqref{total_jump_count} yields~\cite{Landi2024}
\begin{align}
	\label{mean_current_av}
	 & J_{\mu\sigma }(t) = \EE[I_{\mu\sigma}(t)] =  R_{\mu\sigma} p_\sigma(t)\\
	 \label{diffusion_matrix_correlation_function}
	& D_{\alpha\beta\mu\sigma} = \lim_{t\to\infty} \int_{-\infty}^\infty \dd \tau \, F_{\alpha\beta\mu\sigma}(t+\tau,t),
\end{align}
where we defined the autocorrelation function
\begin{equation}
	\label{autocorrelation_function_4index}
	F_{\alpha\beta\mu\sigma}(t+\tau,t) = \EE[I_{\alpha\beta}(t+\tau)I_{\mu\sigma}(t)] -  J_{\alpha\beta}(t+\tau) J_{\mu\sigma}(t).
\end{equation}
For $\tau = 0$, we have, using Eqs.~\eqref{elementary_stochastic_increment},
\begin{align}
	\label{autocorr_zero_time_delay}
	\EE[I_{\alpha\beta}(t)I_{\mu\sigma}(t)] & = \frac{\delta_{\alpha\mu}\delta_{\beta\sigma} }{\dd t} \EE[I_{\mu\sigma}(t)]  \to \delta_{\alpha\mu}\delta_{\beta\sigma} \delta(\tau) J_{\mu\sigma}(t),
\end{align}
where the contribution from $ \dd t^{-1}\to \delta(\tau)$ tends to a Dirac delta function in the continuum limit. For $\tau >0$, we have, again using Eqs.~\eqref{elementary_stochastic_increment},
\begin{align}
\EE[I_{\alpha\beta}(t+\tau)I_{\mu\sigma}(t)] & = \frac{1}{\dd t^2} {\rm Pr} \left (\dd N_{\alpha\beta}(t+\tau) =1\right |\left . \dd N_{\mu\sigma}(t) =1\right ) \notag \\
& \quad \times {\rm Pr}\left ( \dd N_{\mu\sigma}(t) =1\right ) \notag \\
& = R_{\alpha\beta} G_{\beta\mu}(\tau) J_{\mu\sigma}(t), \label{eq:E[I(t+tau)I(t)]_sol}
\end{align}
where the second line follows because $G_{\beta\mu}(\tau)\equiv[\mathsf{G}(\tau)]_{\beta\mu}$ is the probability that the system is found in state $\beta$ after a time delay $\tau$, given that it was previously in state $\mu$, according to Eq.~\eqref{master_equation_solution}. A similar argument can be used for $\tau<0$ by taking $t+\tau$ as the time of the first jump, so we obtain 
	\begin{align}
	\label{autocorr_full}
&F_{\alpha\beta\mu\sigma}(t+\tau,t)  \\ &= \begin{cases}
		\left [\delta_{\alpha\mu}\delta_{\beta\sigma} \delta(\tau) + R_{\alpha\beta} G_{\beta\mu}(\tau) - J_{\alpha\beta}(t+\tau)  \right ] J_{\mu\sigma}(t) & (\tau\geq 0) \\
		\left [\delta_{\alpha\mu}\delta_{\beta\sigma} \delta(\tau) + R_{\mu\sigma} G_{\sigma \alpha}(-\tau) - J_{\mu\sigma}(t) \right ] J_{\alpha\beta}(t+\tau) & (\tau\leq 0). \notag
	\end{cases}
\end{align}

To find the diffusion matrix, we take the limit of large $t$ where $J_{\mu\sigma}(t+\tau) \to J_{\mu\sigma} = R_{\mu\sigma}p^\ss_\sigma$ equals its steady-state value. We also use the fact that 
\begin{align}
	\label{Drazin_inverse_integral}
	\int_0^\infty \dd \tau \left [ G_{\beta\mu}(\tau) - p^{\ss}_\beta \right ] & =  \int_0^\infty \dd \tau \,\left [ \ee^{\mathsf{L}\tau} - \mathbf{p}^{\ss}\mathbf{u}^\T \right ]_{\beta\mu} = - [\mathsf{L}^{+} ]_{\beta\mu},
\end{align}
by the definition of the Drazin inverse, $\mathsf{L}^{+}$, since $\mathbf{p}^{\ss}\mathbf{u}^\T = \mathsf{\Pi}_0$ is the projector onto the null eigenspace of $\mathsf{L}$~\cite{Landi2024}. Using the shorthand $L^{+}_{
\beta\mu} = [\mathsf{L}^{+} ]_{\beta\mu}$, the diffusion tensor is therefore
\begin{equation}
	\label{diffusion_matrix_full}
	D_{\alpha\beta\mu\sigma} = \delta_{\alpha\mu}\delta_{\beta\sigma} J_{\mu\sigma} - R_{\alpha\beta}L^{+}_{\beta\mu}J_{\mu\sigma} -   R_{\mu\sigma}L^{+}_{\sigma\alpha}J_{\alpha\beta}.
\end{equation}

\subsection{Current statistics of a general counting observable}
\label{SM:general_current}

The most general current is formed of a linear combination of elementary currents, as $I(t) = \sum_{\mu,\sigma} w_{\mu\sigma}I_{\mu\sigma}(t) \equiv \vec{w}\cdot \vec{I}(t)$, with real weights $\{w_{\mu\sigma}\}$. The average current $J(t) = \EE[I(t)]$ is
\begin{equation}
	\label{mean_current_jumps}
J(t) = \sum_{\mu,\sigma} w_{\mu\sigma} J_{\mu\sigma }(t) \equiv \vec{w}\cdot \vec{J}(t),
\end{equation}
the autocorrelation function $F(t_1,t_2) = \Cov[I(t_1),I(t_2)] $  is
\begin{equation}
	\label{autocorr_general}
	F(t_1,t_2) = \sum_{\alpha,\beta,\mu,\sigma}w_{\alpha\beta} F_{\alpha\beta\mu\sigma}(t_1,t_2)w_{\mu\sigma} = \vec{w}\cdot \mathbbm{F}(t_1,t_2)\cdot\vec{w},
\end{equation}
and the steady-state diffusion coefficient is
\begin{equation}	
	\label{diffusion_coefficient_jumps}
	D = \lim_{t\to\infty}\frac{\dd}{\dd t }\Var[\Theta(t)] = \sum_{\alpha,\beta,\mu,\sigma}w_{\alpha\beta} D_{\alpha\beta\mu\sigma}w_{\mu\sigma} \equiv \vec{w}\cdot\mathbb{D}\cdot\vec{w},
\end{equation}
where $\Theta(t) = \int_0^t\dd t'\, I(t')$ is the corresponding integrated current. We represent ``vectors'' with $d^2$ elements using the notation $\vec{w}$, $\vec{I}(t)$, etc., while the tensors $\mathbb{D}$ and $\mathbbm{F}$ act as $d^2\times d^2$ matrices, mapping one such vector onto another.\\

The above expressions can be written in a compact form by introducing the ``current operator'' $\tilde{\mathsf{J}}$ with matrix elements $[\tilde{\mathsf{J}}]_{\mu\sigma} = w_{\mu\sigma} R_{\mu\sigma}$. Focussing on the steady state, the average current is then
\begin{equation}
    \label{average_current_superop}
J = \mathbf{u}^\T \tilde{\mathsf{J}}\mathbf{p}^\ss.
\end{equation}
The steady-state autocorrelation function $F(\tau) = F(t+\tau,t)$ is given by
\begin{equation}
\label{autocorr_superop}
	   F(\tau)  = K\delta(\tau) + \mathbf{u}^\T \tilde{\mathsf{J}} \mathsf{G}(|\tau|) \tilde{\mathsf{J}} \mathbf{p}^\ss - J^2,
\end{equation}
where $K=\sum_{\mu,\sigma} w_{\mu\sigma}^2 J_{\mu\sigma}$, which can be interpreted as a weighted dynamical activity. Finally, the steady-state diffusion coefficient is 
\begin{equation}
    \label{diffusion_superop}
	D = K - 2\mathbf{u}^\T \tilde{\mathsf{J}} \mathsf{L}^+ \tilde{\mathsf{J}} \mathbf{p}^\ss.
\end{equation}
These results are directly analogous to the corresponding expressions for open quantum systems quoted in Ref.~\cite{Landi2024}, where $\tilde{\mathsf{J}}$ takes the role of the jump superoperator $\mathcal{J}$ and $\mathsf{L}$ plays the role of the Liouvillian $\mathcal{L}$.

\subsection{Hyperaccurate current and the BLUE}
\label{SM:hyperaccurate_current}

As discussed in Sec.~\ref{sec:counting_observables}, a hyperaccurate current is defined by weights that maximise the SNR, $\mathcal{S} = J^2/D$. Since $\mathcal{S}$ is invariant under uniform rescaling of the weight vector $\vec{w}$, we can fix $J=1$ to ensure an unbiased estimator and minimise the diffusion coefficient $D$. From Eqs.~\eqref{mean_current_jumps} and~\eqref{diffusion_coefficient_jumps}, we thus seek the minimum of the Lagrangian
\begin{equation}
    \label{Lagrangian}
\Lambda  = \vec{w}\cdot \mathbb{D}\cdot\vec{w} + \zeta\left (1- \vec{w}\cdot \vec{J}\right ),
\end{equation}
where $\zeta$ is a Lagrange multiplier. This yields the equation $\mathbb{D}\cdot \vec{w} = \zeta \vec{J}$ and, since $\vec{w}\cdot\vec{J}=1$ we deduce $ \zeta =\vec{w}\cdot \mathbb{D}\cdot \vec{w} = D$, which recovers Eq.~\eqref{BLUE}, i.e.
\begin{equation}
    \label{BLUE_SM}
		\mathbb{D}\cdot \vec{w} = D \vec{J}.
\end{equation}

We now show that the BLUE~\eqref{BLUE_estimator} satisfies this equation and thus corresponds to a hyperaccurate current. For the particular choice of weights defining the BLUE, i.e. 
\begin{equation}
	\label{hyperaccurate_weights}
	w_{\mu\sigma} = \frac{1}{\Gamma_\sigma},
\end{equation}
it is straightforward to show that
\begin{equation}
	\label{hyperaccurate_average}
\EE[I(t)] = \sum_{\mu,\sigma} \frac{R_{\mu\sigma} p_\sigma(t)}{\Gamma_\sigma} = \sum_\sigma p_\sigma(t) = 1,
\end{equation}
which holds for \textit{any} distribution $p_\sigma(t)$. For the autocorrelation function, we obtain for $\tau\geq 0$
\begin{align}
	\label{hyperaccurate_autocorr}
	\EE[I(t+\tau)I(t)]  & = \sum_{\mu,\sigma}\frac{J_{\mu\sigma}(t)}{\Gamma_\sigma^2} \delta(\tau) + \mathbf{u}^\T \mathsf{R}\mathsf{\Gamma}^{-1}\mathsf{G}(\tau)\mathsf{R}\mathsf{\Gamma}^{-1}\mathbf{p}(t) \notag \\
	& = \sum_{\sigma}\frac{p_\sigma(t)}{\Gamma_\sigma} \delta(\tau) +  1,
\end{align}
using $\mathbf{u}^\T \mathsf{R} = \mathbf{u}^\T\mathsf{\Gamma}$ and $\mathbf{u}^\T \mathsf{G}(\tau) = \mathbf{u}^\T$. Identical results are obtained for $\tau\leq 0$. It follows that the current is delta-correlated,
\begin{equation}
	\label{hyperaccurate_delta_correlated}
	F(t+\tau,t) = \Cov[I(t+\tau),I(t)] = \sum_\sigma \frac{p_\sigma(t)}{\Gamma_\sigma} \delta(\tau),
\end{equation}
which again holds for any distribution $\mathbf{p}(t)$. In the steady state, it reduces to $F(\tau) = \mathcal{T}\delta(\tau)$.

To find the steady-state diffusion coefficient, it is sufficient to note from Eq.~\eqref{diffusion_matrix_correlation_function} that 
\begin{equation}
	\label{D_from_F}
	\vec{w}\cdot \mathbb{D}\cdot \vec{w} = \int_{-\infty}^\infty \dd \tau\, F(\tau) = \mathcal{T}.
\end{equation}
However, it is instructive to use Eq.~\eqref{diffusion_matrix_full} to directly compute 
\begin{align}
	\label{D_times_w}
\left [\mathbb{D}\cdot\vec{w}\right ]_{\alpha\beta} &  = \sum_{\mu,\sigma} \frac{D_{\alpha\beta\mu\sigma}}{\Gamma_\sigma} \notag \\
& =  \frac{J_{\alpha\beta}}{\Gamma_\beta}	-R_{\alpha\beta} \left [ \mathsf{L}^+\mathsf{R}\mathsf{\Gamma}^{-1}\mathbf{p}^\ss\right ]_{\beta} -  \left [\mathbf{u}^\T \mathsf{R}\mathsf{\Gamma}^{-1}\mathsf{L}^+\right ]_{\alpha}J_{\alpha\beta} \notag \\
& =  \frac{J_{\alpha\beta}}{\Gamma_\beta} - R_{\alpha\beta}\left [\left (\mathsf{1} - \mathbf{p}^\ss \mathbf{u}^\T\right )\mathsf{\Gamma}^{-1}\mathbf{p}^\ss\right ]_\beta \notag \\
& = \mathcal{T} J_{\alpha\beta},
\end{align}
where the third line is obtained by substituting $\mathsf{R} = \mathsf{L}+\mathsf{\Gamma}$ and using properties of the Drazin inverse, while the final line follows after recognising that $\mathcal{T} = \mathbf{u}^\T \mathsf{\Gamma}^{-1}\mathbf{p}^\ss$. We conclude from Eq.~\eqref{D_times_w} that $\mathbb{D}\cdot \vec{w} = \mathcal{T}\vec{J}$. Since $\vec{w}\cdot\vec{J} = 1$, this further implies that Eq.~\eqref{BLUE_SM} is satisfied with $D=\vec{w}\cdot\mathbb{D}\cdot \vec{w}=\mathcal{T}$, consistent with Eq.~\eqref{D_from_F}. Taken together, these results prove that the weights~\eqref{hyperaccurate_weights} are a valid solution to Eq.~\eqref{BLUE_SM} [Eq.~\eqref{BLUE}] with SNR $\mathcal{S}=\mathcal{T}^{-1}$, thus saturating the CUR bound~\eqref{general_bound}.

\subsection{Degeneracy of the elementary diffusion matrix}
\label{SM:degeneracy}

The diffusion tensor~\eqref{diffusion_matrix_full} possesses a large degeneracy that arises from the combination of probability conservation and stationarity. In particular, stationarity implies $\sum_\mu R_{\sigma\mu} p^\ss_\mu = \Gamma_\sigma p_\sigma^\ss$, while probability conservation implies $\sum_\mu R_{\mu\sigma} p^\ss_\sigma = \Gamma_\sigma p^\ss_\sigma$, which together imply the Kirchoff-like conservation law
\begin{equation}
	\label{current_conservation}
	\sum_{\mu} \left (J_{\mu\sigma} - J_{\sigma\mu}\right ) = 0,
\end{equation}
stating that the total probability current entering and leaving any state $\sigma$ are equal. As a consequence, one can show that
\begin{align}
	\label{zero_mode}
	&\sum_{\mu,\sigma} J_{\mu\sigma}\left (c_\mu - c_\sigma\right )  = 0, \\
	\label{zero_eigenvalue}
	&\sum_{\mu,\sigma} D_{\alpha\beta\mu\sigma} \left (c_\mu - c_\sigma\right ) = 0,
\end{align}
for any collection of $d$ constants $\{c_\sigma\}$. Eq.~\eqref{zero_mode} follows directly from Eq.~\eqref{current_conservation} upon swapping dummy indices in the double sum. To prove Eq.~\eqref{zero_eigenvalue}, we substitute in the explicit matrix elements~\eqref{diffusion_matrix_full} and consider each of the three terms separately. The first term becomes 
\begin{equation}
	\label{diffusion_zero_first}
	\sum_{\mu,\sigma}\delta_{\alpha\mu}\delta_{\beta\sigma} J_{\mu\sigma} \left (c_\mu - c_\sigma\right ) = J_{\alpha\beta}\left (c_\alpha - c_\beta\right ).
\end{equation}
To evaluate the second term in Eq.~\eqref{diffusion_matrix_full}, we define the diagonal matrix $\mathsf{C}$ with elements $[\mathsf{C}]_{\mu\sigma} = c_\mu\delta_{\mu\sigma}$ and then compute
\begin{align}
	\label{diffusion_zero_second}
		&  - \sum_{\mu,\sigma}R_{\alpha\beta}[\mathsf{L}^{+} ]_{\beta\mu}R_{\mu\sigma} p_\sigma^{\ss}\left (c_\mu - c_\sigma\right ) \notag \\
		& = R_{\alpha\beta}\left [\mathsf{L}^{+}\mathsf{R}\mathsf{C}\mathbf{p}^\ss - \mathsf{L}^{+} \mathsf{C}\mathsf{R}\mathbf{p}^{\ss}\right ]_\beta \notag \\
		& = R_{\alpha\beta}\left [\mathsf{L}^{+}\left (\mathsf{R}-\mathsf{\Gamma}\right )\mathsf{C}\mathbf{p}^{\ss}\right ]_\beta\notag  \\
		& = R_{\alpha\beta}\left [\left (\mathsf{1}-\mathbf{p}^\ss \mathbf{u}^\T\right )\mathsf{C}\mathbf{p}^{\ss}\right ]_\beta\notag \\
		& = J_{\alpha\beta} \left ( c_\beta - \mathbf{u}^\T \mathsf{C}\mathbf{p}^\ss\right ).
\end{align}
The first equality follows from the definition of $\mathsf{C}$, the second equality follows from the stationarity condition $\mathsf{R}\mathbf{p}^\ss = \mathsf{\Gamma}\mathbf{p}^\ss$ together with the fact that $\mathsf{\Gamma}$ and $\mathsf{C}$ are diagonal (commuting) matrices, while the third equality follows from the properties of the Drazin inverse. For the final term in Eq.~\eqref{diffusion_matrix_full}, similar manipulations yield 
\begin{align}
		\label{diffusion_zero_third}
& - \sum_{\mu,\sigma}  R_{\mu\sigma}[\mathsf{L}^{+} ]_{\sigma\alpha}J_{\alpha\beta}\left (c_\mu-c_\sigma\right ) \notag \\
& = - \left [\mathbf{u}^\T\mathsf{C}\mathsf{R}\mathsf{L}^{+} -\mathbf{u}^\T\mathsf{R}\mathsf{C}\mathsf{L}^{+} \right ]_\alpha J_{\alpha\beta} \notag \\ 
& = - \left [\mathbf{u}^\T\mathsf{C}\left (\mathsf{R}-\mathsf{\Gamma}\right )\mathsf{L}^{+} \right ]_\alpha J_{\alpha\beta} \notag \\
& = -J_{\alpha\beta}\left (c_\alpha - \mathbf{u}^\T \mathsf{C}\mathbf{p}^\ss\right ),
\end{align}
where here we used probability conservation, $\mathbf{u}^\T\mathsf{R} = \mathbf{u}^\T \mathsf{\Gamma}$, instead of stationarity. Adding together Eqs.~\eqref{diffusion_zero_first}--\eqref{diffusion_zero_third}, we deduce Eq.~\eqref{zero_eigenvalue}. 

If we interpret $D_{\alpha\beta\mu\sigma} = [\mathbb{D}]_{(\alpha\beta)(\mu\sigma)}$ as a $d^2\times d^2$ matrix, then Eq.~\eqref{zero_eigenvalue} is the eigenvalue equation for a null eigenvector. One can always choose precisely $d-1$ orthogonal, non-trivial null eigenvectors of the form $w_{\mu\sigma} = c_{\mu} - c_\sigma$. To see this, note that the diffusion matrix is real and symmetric and therefore its eigenvectors can be chosen to be real and orthogonal. The inner product between two null eigenvectors is
\begin{equation}
	\label{null_inner_product}
	\vec{w}\cdot\vec{w}'=\sum_{\mu,\sigma}\left (c_\mu - c_\sigma\right )\left (c'_\mu-c'_\sigma\right ) = 2\left (d \mathbf{c}^\T\mathbf{c}' \, - \mathbf{u}^\T \mathbf{c}\,  \mathbf{u}^\T\mathbf{c}'\right ).
\end{equation}
This will vanish for any pair of $d$-dimensional vectors $\mathbf{c}$ and $\mathbf{c}'$ if they are orthogonal to each other and to the constant vector $\mathbf{u}$; clearly, there are $d-1$ such linearly independent vectors. The constant vector itself is a trivial solution to Eq.~\eqref{zero_eigenvalue} since $u_\mu - u_\sigma = 0$. We conclude that the null eigenvectors $c_\mu - c_\sigma$ span a $(d-1)$-dimensional subspace in the space of possible weights $w_{\mu\sigma}$.

This large null eigenspace of the diffusion matrix implies that the solution of Eq.~\eqref{BLUE} [Eq.~\eqref{BLUE_SM}] is not unique. Any two sets of weights $w_{\mu\sigma}$ and $w'_{\mu\sigma}$ related by $w'_{\mu\sigma} = w_{\mu\sigma} + c_\mu - c_\sigma$ yield an identical average current~\eqref{mean_current_jumps} and diffusion coefficient~\eqref{diffusion_coefficient_jumps}. The interpretation of this is as follows: adding a fixed weight $c_\mu$ to all currents flowing into state $\mu$ is compensated by removing the same weight from all currents leaving state $\mu$.

Nevertheless, it is possible to use a pseudo-inverse of $\mathbb{D}$ to find a (non-unique) solution to Eq.~\eqref{BLUE_SM} in closed form. A natural choice is the Drazin inverse $\mathbb{D}^+$, which annihilates all vectors in the null eigenspace of $\mathbb{D}$ by construction. Applying $\mathbb{D}^+$ to both sides of Eq.~\eqref{BLUE_SM} yields $\vec{w} = D \mathbb{D}^+\cdot \vec{J}$. Unbiasedness then implies that $1 = \vec{J}\cdot\vec{w} = D \vec{J}\cdot \mathbb{D}^+\cdot \vec{J}$ or, equivalently, that the maximal SNR is
\begin{equation}
    \label{SNR_max}
    \mathcal{S}_{\rm max} = \vec{J}\cdot \mathbb{D}^+\cdot \vec{J}.
\end{equation}

\section{Comparison with thermodynamic bounds on counting observables}
\label{SM:traffic_flow}

In this Appendix, we compare the CUR with the bounds derived in Ref.~\cite{Pietzonka2024}. These bounds are valid for counting observables with binary weights such that $w_{\mu\sigma} \in \{0,1\}$. Following Ref.~\cite{Pietzonka2024}, we consider two kinds of such counting observables: traffic and flow. A traffic observable has $w_{\sigma\mu} = 1$ whenever $w_{\mu\sigma}=1$, a  flow observable has $w_{\sigma\mu} = 0$ whenever $w_{\mu\sigma}=1$. That is, flow observables count transitions in only one direction, whereas traffic is counted in both. In our notation, the bounds derived in Ref.~\cite{Pietzonka2024} can be written as
\begin{equation}
    \mathcal{S} \leq \mathcal{C} = \frac{J}{\Phi(\dot{
    \Sigma}/J)},
\end{equation}
where $J = \partial_t \EE[\Theta]$ is the average current pertaining to the observable $\Theta=\vec{w}\cdot\vec{N}$, $\mathcal{S} = J^2/D$ is the corresponding SNR, and $\dot{
    \Sigma}$ is the entropy production rate defined in Table~\ref{tab:bounds}. Finally, $\Phi(z)$ is a function of the normalised entropy production rate $\dot{\Sigma}/J$ that is different for flow and traffic:
    \begin{align}
        \label{traffic_bound}
    & \Phi_{\rm traffic}(z) = \min_{x } \left[ 1 - x^2 +\frac{x^4}{z/2 - x \,{\rm arctanh}(x) + x^2}\right], \\
        \label{flow_bound}
    & \Phi_{\rm flow}(z) = \min_{y } \left[ 1 +\frac{1}{2 - y + \frac{2y^2}{z + y\ln(1-y)}}\right], 
    \end{align}
where minimisation is performed over the range in which $2x \,{\rm arctanh}(x) \leq  z$ or $-y\ln(1-y) < z$, respectively.

\begin{figure}
    \centering
    \includegraphics[width=\linewidth]{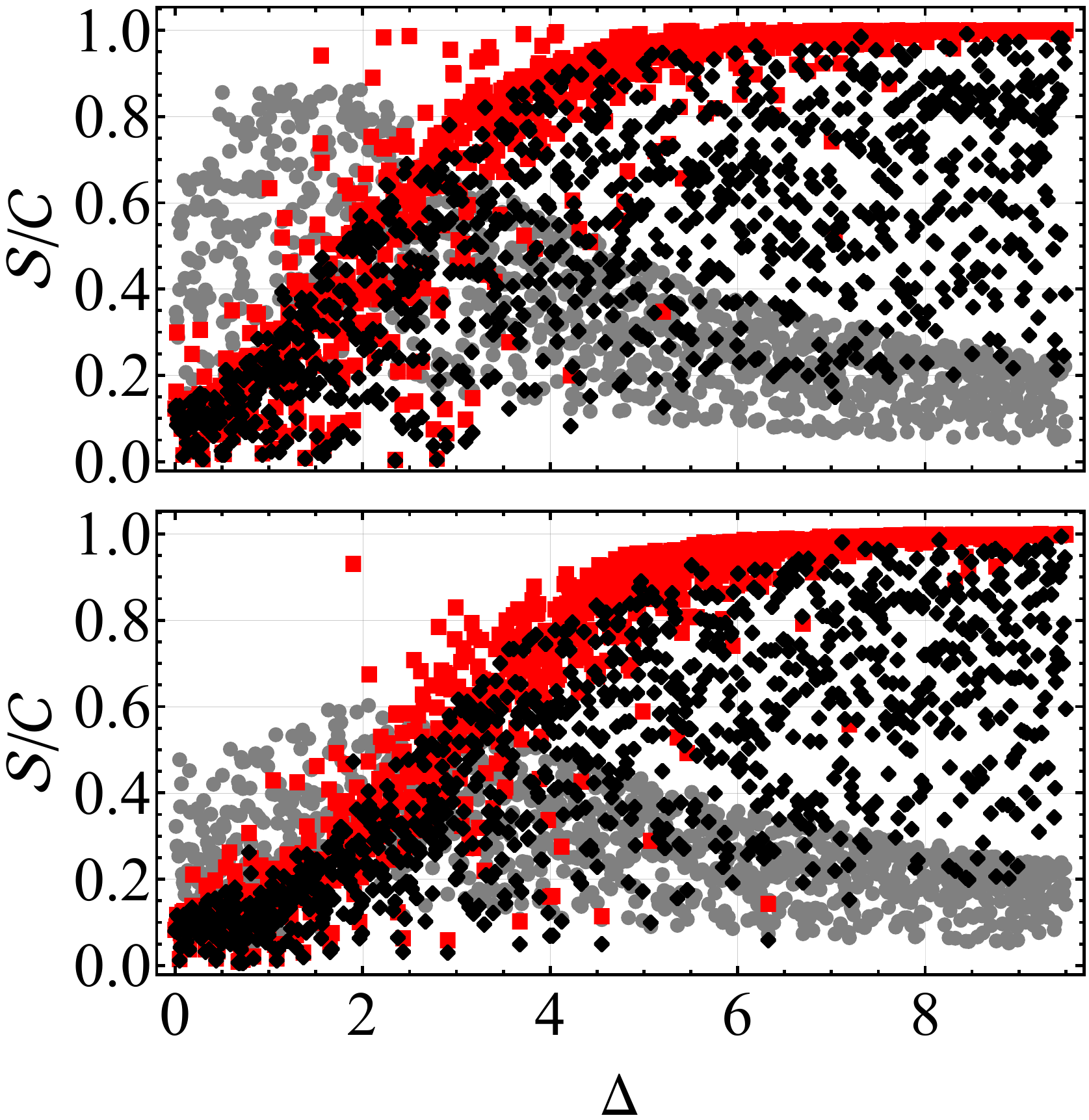}
    \caption{Precision-cost ratio for the CUR (red squares), KUR (black diamonds), and the entropic bounds of Ref.~\cite{Pietzonka2024} (grey circles). In both panels, we randomly generate ring networks with clockwise bias $\Delta$ and $d=5$ states, following the same procedure as detailed in the caption of Fig.~\ref{fig:bounds}(b). The upper panel shows a flow observable with $w_{21}=1$ and all other $w_{\mu\sigma}=0$. The lower panel  shows a traffic observable with $w_{21}= w_{12}=1$ and all other $w_{\mu\sigma}=0$.\label{fig:traffic_flow}}
    
\end{figure}

Fig.~\ref{fig:traffic_flow} shows how these bounds perform for random unicyclic networks, with varying degrees of bias towards jumps in the clockwise direction as in Fig.~\ref{fig:bounds}(b). We observe that the thermodynamic bounds of Ref.~\cite{Pietzonka2024} are the most relevant in a regime of moderate entropy production. Moreover, the SNR for a flow observable is generally bounded more tightly than that of a traffic observable, as shown in the upper and lower panels of Fig.~\ref{fig:traffic_flow}, respectively. However, both bounds become progressively less informative as the bias is increased, whereas the CUR is again the tightest bound in this far-from-equilibrium regime.

\section{Kinetic uncertainty relation and the Cram\'er-Rao bound}
\label{SM:KUR}
In this appendix, we provide further details on the connection between the clock uncertainty relation and the kinetic uncertainty relation (KUR), and show that the latter cannot be saturated by any counting observable in general. We also discuss an observable for which the KUR is violated. 

In Ref.~\cite{Terlizzi2018}, the KUR was derived using the fluctuation-response inequality~\cite{Dechant2020}, which for a small perturbation becomes equivalent to a Cram\'er-Rao bound (CRB). Following Ref.~\cite{Terlizzi2018}, we consider a perturbed generator $\mathsf{L}\to(1+\lambda)\mathsf{L}$, where the limit $\lambda\to 0$ will ultimately be taken. The perturbed trajectory probability distribution is given by Eq.~\eqref{KUR_perturbed_path_prob}, or explicitly
\begin{equation}
	\label{KUR_path_prob}
	P(\bm{\sigma},\mathbf{t}|t,\lambda) =(1+\lambda)^N \ee^{-(1+\lambda) \sum_{j=0}^N \Gamma_{\sigma_j} \tau_j}  \prod_{j=0}^{N-1} R_{\sigma_{j+1}\sigma_j} p_{\sigma_0}^\ss,
\end{equation}
which follows from Eq.~\eqref{path_probability_sm} upon rescaling $R_{\mu\sigma}\to (1+\lambda)R_{\mu\sigma}$ and $\Gamma_\sigma \to(1+\lambda)\Gamma_\sigma$. Consider a general observable $\Theta(\bm{\sigma},\mathbf{t})$, whose expectation value with respect to the perturbed distribution~\eqref{KUR_path_prob} is denoted by
\begin{equation}
	\label{general_expectation}
	\EE[\Theta|t,\lambda] = \sum_{\bm \sigma} 
\int_0^t\dd \mathbf{t}\,P(\bm{\sigma},\mathbf{t}|t,\lambda) \Theta(\bm{\sigma},\mathbf{t}),
\end{equation}
using the notation of Eq.~\eqref{path_prob_normalisation}. Treating this observable as a (biased) estimator of $\lambda$, its variance is bounded by the CRB
\begin{equation}
	\label{CRB_KUR_SM}
\Var[\Theta|t,\lambda]\geq\frac{ \left (\partial_\lambda \EE[\Theta|t,\lambda]\right )^2}{\mathcal{F}_\lambda[P(\bm{\sigma},\mathbf{t}|t,\lambda)]},
\end{equation}
where
\begin{align}
	\label{Fisher_info_KUR_SM}
	\mathcal{F}_\lambda[P(\bm{\sigma},\mathbf{t}|t,\lambda)] &  = \EE\left [\left (\frac{\partial \ln P(\bm{\sigma},\mathbf{t}|t,\lambda)}{\partial\lambda}\right )^2\Bigg|t,\lambda\right ] \notag \\ &= -\EE\left [\frac{\partial^2 \ln P(\bm{\sigma},\mathbf{t}|t,\lambda)}{\partial\lambda^2}\Bigg|t,\lambda\right ]
\end{align}
is the Fisher information of the distribution~\eqref{KUR_path_prob} with respect to $\lambda$. Both forms on the right-hand side of Eq.~\eqref{Fisher_info_KUR_SM} are equivalent~\cite{Kay2013}, but using the second form with Eq.~\eqref{KUR_path_prob} it follows immediately that
\begin{equation}
	\label{KUR_Fisher_deriv}
	\mathcal{F}_\lambda[P(\bm{\sigma},\mathbf{t}|t,\lambda)] = \frac{\EE[N|t,\lambda]}{(1+\lambda)^2}.
\end{equation}

\subsection{Sequence observables}

To recover the KUR, consider an observable $\Theta(\bm{\sigma})$, which depends only on the sequence $\bm{\sigma}$ and not the jump times $\mathbf{t}$. Its expectation value is $\EE[\Theta|t,\lambda] = \sum_{\bm\sigma} P(\bm{\sigma}|t,\lambda)\Theta(\bm{\sigma})$, where
\begin{align}
	\label{KUR_sequence_prob}
	P(\bm{\sigma}|t,\lambda) & = \int_0^t\dd t_N \cdots \int_0^{t_2}\dd t_1 P(\bm{\sigma},\mathbf{t}|t,\lambda) \notag \\
	& = \int_0^{(1+\lambda)t}\dd t'_N \cdots \int_0^{t'_2}\dd t'_1 P\left (\bm{\sigma},\mathbf{t}'|(1+\lambda)t\right ) \notag \\
	& = P\left (\bm{\sigma}|(1+\lambda)t\right ).
\end{align}
On the second line we changed integration variables to $t_j'=(1+\lambda)t_j$, and we omit the explicit conditioning on $\lambda$ when writing quantities with $\lambda=0$, e.g.~$P(\bm{\sigma}|t) = P(\bm{\sigma}|t,\lambda=0)$. Therefore, for such expectation values, rescaling the generator $\mathsf{L}\to (1+\lambda)\mathsf{L}$ is equivalent to rescaling the total time as $t\to(1+\lambda)t$. Explicitly, we can write
\begin{equation}
	\label{KUR_response}
	\EE[\Theta|t,\lambda] = \EE[\Theta|(1+\lambda)t] = \EE[\Theta|t] + \lambda t  \partial_t\EE[\Theta|t]  + \mathcal{O}(\lambda^2).
\end{equation}
Plugging Eqs.~\eqref{KUR_response} and~\eqref{KUR_Fisher_deriv} into the CRB~\eqref{CRB_KUR_SM}, taking the limit $\lambda\to 0$, and rearranging, we get the KUR
\begin{equation}
	\label{KUR_SM}
	\mathcal{S} = \frac{(\partial_t \EE[\Theta|t])^2}{\Var[\Theta|t]/t} \leq \mathcal{A},
\end{equation}
where $\mathcal{A} = \EE[N|t]/t = \sum_{\mu,\sigma}\EE[N_{\mu\sigma}|t] /t = \sum_{\mu,\sigma} R_{\mu\sigma}p_\sigma^{\ss}$ is the dynamical activity. 

\subsection{Waiting-time observables}
\label{SM:KUR_general_observables}

For observables that depend non-trivially on the waiting times $\bm{\tau}$, the perturbed expectation value~\eqref{general_expectation} is given by
\begin{align}
    \label{general_expectation_rescale}	\EE\left [\Theta|t,\lambda\right ] & = \sum_{\bm{\sigma}} \int_0^t\dd\mathbf{t} \,P(\bm{\sigma},\mathbf{t}|t,\lambda)\Theta(\bm{\sigma},\bm{\tau})\notag \\
 & = \sum_{\bm{\sigma}} \int_0^{(1+\lambda)t}\dd\mathbf{t}' \,P(\bm{\sigma},\mathbf{t}'|(1+\lambda)t)\Theta\left (\bm{\sigma},(1+\lambda)^{-1}\bm{\tau}'\right )\notag \\
 & = \EE\left [\Theta\left (\bm{\sigma},(1+\lambda)^{-1}\bm{\tau}\right )\big |(1+\lambda)t\right ],
\end{align}
where on the second line we changed variables to $t'_j = (1+\lambda)t_j$, as in Eq.~\eqref{KUR_sequence_prob}, and the notation $(1+\lambda)^{-1}\bm{\tau}$ means that each waiting time is rescaled as $\tau_j\to (1+\lambda)^{-1}\tau_j$. Expanding for small $\lambda$ yields
\begin{equation}
	\label{expect_expand}
	\partial_\lambda \EE\left [\Theta|t,\lambda\right ] = t \partial_t \EE[\Theta|t] - \EE\left [\sum_{j=0}^N \tau_j\partial_{\tau_j} \Theta(\bm{\sigma},\bm{\tau})\Bigg|t\right ] +\mathcal{O}(\lambda),
\end{equation}
which reproduces Eq.~\eqref{KUR_expect_expand}. As a consequence, the arguments leading to the KUR~\eqref{KUR_SM} no longer hold in general.

\subsection{Observables linear in the waiting times $\tau_j$}

As an example of the general observables discussed in Appendix~\ref{SM:KUR_general_observables} above, consider an observable that is a linear function of the waiting times, $\Theta(\bm{\sigma},\bm{\tau}) = \sum_{j=0}^N a_j(\bm{\sigma}) \tau_j$ with $\{a_j(\bm{\sigma})\}$ arbitrary functions of $\bm{\sigma}$. We then have $\sum_{j=0}^N \tau_j\partial_{\tau_j}\Theta = \Theta$. If the observable is also time-extensive, $\EE[\Theta|t] \sim t$, then for large times $t \partial_t \EE[\Theta|t] = \EE[\Theta|t]$. Under these conditions, the right-hand side of Eq.~\eqref{expect_expand} vanishes up to $\mathcal{O}(\lambda)$ corrections and thus, for $\lambda\to 0$, the CRB~\eqref{CRB_KUR_SM} reduces to the trivial bound $\Var[\Theta]\geq 0$. As discussed in Sec.~\ref{sec:CUR_vs_KUR}, this bound is saturated by the observable $\Theta(\bm{\tau}) = \sum_{j=0}^N \tau_j = t$. Since the total time $t$ is fixed, its variance vanishes.

A non-trivial example of such an observable is given by 
\begin{equation}
    \label{dimensionless_time}
    \Theta(\bm{\sigma},\bm{\tau}) = \sum_{j=0}^N \Gamma_{\sigma_j} \tau_j \equiv Q(\bm{\sigma},\bm{\tau}).
\end{equation}
The statistics of this observable can be found using the following convenient trick. Let us define the moment-generating functions for $Q$ and for the number of jumps, $N$, as 
\begin{align}
    \label{Q_MGF}
M_Q(\xi,t) = \EE\left [\ee^{-\xi Q}\big |t\right ] ,\\
    \label{N_MGF}
M_N(\eta,t) = \EE\left [\ee^{-\eta N}\big |t\right ], 
\end{align}
where $\xi$ and $\eta$ are counting parameters. To derive a relation between these two functions, we write out Eq.~\eqref{Q_MGF} explicitly using Eq.~\eqref{path_probability_sm} and the notation of Eq.~\eqref{path_prob_normalisation}, finding
\begin{align}
\label{MGF_relation_derivation}
    M_Q(\xi,t) & = \sum_{\bm{\sigma}}\int_0^t\dd \mathbf{t} \,\ee^{-(1+\xi)Q(\bm{\sigma},\bm{\tau})} \prod_{j=0}^{N-1} R_{\sigma_{j+1}\sigma_j}p_{\sigma_0}^\ss \notag \\
& = \sum_{N=0}^\infty (1+\xi)^{-N} \!\sum_{\sigma_0=1}^d\cdots  \sum_{\sigma_N=1}^d \int_0^{(1+\xi)t} \!\! \dd \mathbf{t}' P\big (\bm{\sigma},\mathbf{t}'|(1+\xi)t\big ) \notag \\ 
& = \sum_{N=0}^\infty (1+\xi)^{-N} P\big(N|(1+\xi)t\big ) \notag \\
&  = \EE\left [(1+\xi)^{-N}\big |(1+\xi)t\right ].
\end{align}
The second equality above follows after introducing new integration variables $t_j' = (1+\xi)t_j$, and in the third equality we recognise the total probability for $N$ jumps in a time $(1+\xi)t$, defined by Eq.~\eqref{P_N_of_t}. Comparing Eq.~\eqref{MGF_relation_derivation} with Eq.~\eqref{N_MGF}, we deduce the relation
\begin{equation}
    \label{MGF_relation}
	M_Q(\xi,t) = M_N\big(\ln(1+\xi),(1+\xi)t \big).
\end{equation}
This shows that the moments of $Q$ can be expressed in terms of moments of $N$, and vice versa. 

Now, an explicit expression for $M_N(\eta,t)$ can be found by noting that $\ee^{-\eta N} = \prod_{j=0}^{N-1}\ee^{-\eta}$, and therefore
\begin{align}
    \label{M_N_explicit}
M_N(\eta,t) & =  \sum_{\bm{\sigma}}\int_0^t\dd \mathbf{t} \,\ee^{-Q(\bm{\sigma},\bm{\tau})} \prod_{j=0}^{N-1} \ee^{-\eta} R_{\sigma_{j+1}\sigma_j}p_{\sigma_0}^\ss \notag \\
& = \mathbf{u}^\T \mathsf{G}_\eta(t) \mathbf{p}^\ss,
\end{align}
where we resummed the Dyson series (cf.~Eq.~\eqref{Dyson_series}) to obtain the ``tilted'' propagator $\mathsf{G}_\eta(t) = \exp\left [ \left ( \ee^{-
\eta}\mathsf{R} - \mathsf{\Gamma}\right )t\right ]$. This result is well known from the theory of full counting statistics~\cite{Landi2024}. Using $\mathsf{\Gamma} = \mathsf{R} - \mathsf{L}$ and defining $\mathsf{\check{R}}(t) = \ee^{-\mathsf{L}t} \mathsf{R}\ee^{\mathsf{L}t}$, we can write the tilted propagator in terms of a time-ordered exponential as
\begin{equation}
    \label{tilted_propagator}
 \mathsf{G}_\eta(t) = \mathsf{G}(t) \, \Texp \left [ \int_0^t \dd t' \left (\ee^{-\eta}-1\right )\mathsf{\check{R}}(t')\right ].
\end{equation}
Finally, to obtain $M_Q(\xi,t)$ from Eq.~\eqref{MGF_relation}, we simply substitute $\eta \to \ln(1+\xi)$ and $t\to (1+\xi)t$ into Eq.~\eqref{tilted_propagator}, obtaining the relatively simple result 
\begin{equation}
    \label{MGF_explicit}
	M_Q(\xi,t) =\mathbf{u}^\T \Texp\left [-\xi \int_0^{(1+\xi)t} \dd t' \mathsf{\check{R}}(t')\right ] \mathbf{p}^\ss.
\end{equation}

The moments of $Q$ are found from the derivatives of $M_Q(\xi,t)$ at $\xi=0$, as
\begin{equation}
    \label{T_moments}
\EE[Q^n] =(-1)^n\left .\frac{\partial^n M_Q}{\partial \xi^n} \right \vert_{\xi=0}.
\end{equation}
Carefully applying this formula, we find the mean
\begin{equation}
    \label{Q_mean}
\EE[Q] = \int_0^t\dd t'\mathbf{u}^\T \mathsf{\check{R}}(t') \mathbf{p}^\ss = t \mathbf{u}^\T \mathsf{R} \mathbf{p}^\ss = \mathcal{A}t,
\end{equation}
since $\mathbf{u}^\T\mathsf{L} = \mathsf{L}\mathbf{p}^\ss = 0$ and $\mathbf{u}^\T\mathsf{R}\mathbf{p}^\ss = \sum_{\mu,\sigma}R_{\mu\sigma}p_\sigma^\ss = \mathcal{A}$ is the dynamical activity. Similarly, the second moment is
\begin{align}
    \label{Q_second_moment}
\EE[Q^2] & = 2\int_0^t \dd t'\int_0^{t'} \dd t'' \mathbf{u}^\T\mathsf{R} \mathsf{G}(t'-t'') \mathsf{R}\mathbf{p}^\ss,
\end{align}
and thus the variance $\Var[Q] = \EE[Q^2] - (\mathcal{A}t)^2$ is 
\begin{align}
    \label{Q_variance}
\Var[Q] & = 2\int_0^t \dd t'\int_0^{t'} \dd t'' \mathbf{u}^\T\mathsf{R} \left [\mathsf{G}(t'-t'') - \mathbf{p}^\ss \mathbf{u}^\T\right ] \mathsf{R}\mathbf{p}^\ss \notag \\ 
& = 2 \int_0^t\dd\tau (t-\tau) \mathbf{u}^\T \mathsf{R}\left [\mathsf{G}(\tau) - \mathbf{p}^\ss \mathbf{u}^\T\right ] \mathsf{R}\mathbf{p}^\ss\notag \\ 
& = -2 t \mathbf{u}^\T \mathsf{R}\mathsf{L}^+ \mathsf{R}\mathbf{p}^\ss + \mathcal{O}(t^0)
\end{align}
The second line follows after changing integration variables to $\tau = t'-t''$ and $s = (t'+t'')/2$ and carrying out the trivial integral over $s$, while the final line follows from extending the upper integration limit to infinity and using Eq.~\eqref{Drazin_inverse_integral}. 

Notice that $Q$ and $N$ are equal on average, $\EE[N] = \EE[Q]$, but $Q$ has a smaller variance, $\Var[Q]= \Var[N]-\EE[N]$. As a result, the asymptotic signal-to-noise ratio of $Q$, given by $\mathcal{S} = \lim_{t\to\infty}\mathcal{A}^2/(\Var[Q]/t)$, can exceed the dynamical activity. This is illustrated by the results in Fig.~\ref{fig:KUR_violation}, which are generated using Eq.~\eqref{Q_variance}.

\section{Details of the hidden Markov model}
\label{SM:hidden_markov}
In this appendix, we give further details on our calculations for hidden Markov models described in Sec.~\ref{sec:testing_CUR}. We partition the system into two disjoint sets of states $\mathtt{X}$ and $\mathtt{Y}$, i.e. each micro-state indexed by $\sigma$ is assumed to belong to either the set $\mathtt{X}$ or the set $\mathtt{Y}$. By appropriately ordering the indices, the generator $\mathsf{L} = \mathsf{L}_0 + \mathsf{K}$ can be written in block form with
\begin{equation}
    \mathsf{L}_0 = \begin{pmatrix}
        \mathsf{L}_{0\mathtt{X}} & \mathsf{0} \\
        \mathsf{0} & \mathsf{L}_{0\mathtt{Y}}
    \end{pmatrix}, \quad \mathsf{K} = \begin{pmatrix}
        \mathsf{0} & \mathsf{R}_{\mathtt{XY}} \\
        \mathsf{R}_{\mathtt{YX}} & \mathsf{0}
    \end{pmatrix}.
\end{equation}
Here, $\mathsf{L}_{0\mathtt{X}}$ and $\mathsf{L}_{0\mathtt{Y}}$ describe the evolution within meso-states $\mathtt{X}$ and $\mathtt{Y}$ in the absence of an observed jump, while $\mathsf{R}_{\mathtt{XY}}$ and $\mathsf{R}_{\mathtt{YX}}$ contain the off-diagonal elements of the rate matrix $\mathsf{R}$ describing jumps $\mathtt{Y} \to \mathtt{X}$ and $\mathtt{X}\to\mathtt{Y}$, respectively, with $\mathsf{0}$ a zero matrix.

The distribution of waiting time until the first observed jump is~\cite{Landi2024}
\begin{equation}
   \tilde{W}_0(\tau) = \mathbf{u}^\T \mathsf{K} \ee^{\mathsf{L}_0 \tau} \mathbf{p}^\ss = - \mathbf{u}^\T \mathsf{L}_0 \ee^{\mathsf{L}_0 \tau} \mathbf{p}^\ss,
\end{equation}
where the second equality follows from $\mathbf{u}^\T\mathsf{L}=0$. From this, the observed mean residual time is readily obtained as
\begin{equation}
    \tilde{\mathcal{T}} = \int_0^\infty \dd\tau\, \tau \tilde{W}_0(\tau) = -\mathbf{u}^\T \mathsf{L}_0^{-1} \mathbf{p}^\ss.
 \end{equation}
To find the effective dynamical activity, we need the mean rate of observed jumps $\mathtt{Y}\to\mathtt{X}$ and $\mathtt{X}\to\mathtt{Y}$, which can be written respectively  as 
\begin{align}
   &  J_{\mathtt{XY}} = \sum_{\mu\in \mathtt{X}} \sum_{\sigma\in\mathtt{Y}} R_{\mu\sigma} p_{\sigma}^\ss = \vec{J}\cdot \vec{w}_{\mathtt{XY}} , \\
   & J_{\mathtt{YX}} = \sum_{\mu\in \mathtt{Y}} \sum_{\sigma\in\mathtt{X}} R_{\mu\sigma} p_{\sigma}^\ss = \vec{J}\cdot \vec{w}_{\mathtt{YX}}. 
\end{align}
Here, we introduced the weight vector $\vec{w}_{\mathtt{XY}}$, which assigns unit weight to all jumps $\mathtt{Y}\to\mathtt{X}$ and zero weight to all others, while a similar definition holds for $\vec{w}_{\mathtt{YX}}$, i.e.
\begin{equation}
    \label{weight_vectors_XY}
    [\vec{w}_{\mathtt{XY}}]_{\mu\sigma} = \sum_{\alpha \in \mathtt{X}} \sum_{\beta\in \mathtt{Y}} \delta_{\alpha\mu} \delta_{\beta\sigma}, \quad [\vec{w}_{\mathtt{YX}}]_{\mu\sigma} = \sum_{\alpha \in \mathtt{Y}} \sum_{\beta\in \mathtt{X}} \delta_{\alpha\mu} \delta_{\beta\sigma}.
\end{equation}
In fact, we can show that $J_{\mathtt{XY}} = J_\mathtt{YX}$, because the difference between these two vectors can be written in the form $[\vec{w}_{\mathtt{YX}}]_{\mu\sigma} - [\vec{w}_{\mathtt{XY}}]_{\mu\sigma} = c_\mu - c_\sigma$, with
\begin{equation}
\label{c_mu_hmm}
    c_\mu = \sum_{\alpha \in\mathtt{Y}}\delta_{\alpha\mu}.
\end{equation}
To see this, we write 
\begin{align}
\label{w_diff_hmm}
   [\vec{w}_{\mathtt{YX}}]_{\mu\sigma} - [\vec{w}_{\mathtt{XY}}]_{\mu\sigma} & =   \sum_{\alpha \in \mathtt{Y}} \sum_{\beta\in \mathtt{X}} \delta_{\alpha\mu} \delta_{\beta\sigma} -\sum_{\alpha \in \mathtt{X}} \sum_{\beta\in \mathtt{Y}} \delta_{\alpha\mu} \delta_{\beta\sigma} \notag \\ & = \sum_{\alpha \in \mathtt{Y}} \sum_{\beta=1}^d \delta_{\alpha\mu} \delta_{\beta\sigma} - \sum_{\alpha =1}^d \sum_{\beta\in \mathtt{Y}} \delta_{\alpha\mu} \delta_{\beta\sigma} \notag \\ 
   &= \sum_{\alpha \in \mathtt{Y}} \delta_{\alpha\mu} - \sum_{\beta\in \mathtt{Y}} \delta_{\beta\sigma},
\end{align}
where on the second line we simply added and subtracted the term $\sum_{\alpha \in \mathtt{Y}} \sum_{\beta\in \mathtt{Y}} \delta_{\alpha\mu} \delta_{\beta\sigma}$. It follows that
\begin{equation}
    \label{equal_Js}
    J_{\mathtt{YX}} - J_{\mathtt{XY}} = \sum_{\mu,\sigma} J_{\mu\sigma} \left(c_\mu -c_\sigma\right ) = 0,
\end{equation}
by Eq.~\eqref{zero_mode}, so that the total observed activity is 
\begin{equation}
    \tilde{\mathcal{A}} = J_{\mathtt{XY}} + J_\mathtt{YX} = 2J_{\mathtt{XY}}.
\end{equation}
It also follows from Eq.~\eqref{zero_eigenvalue} that
\begin{equation}
    \lim_{t\to\infty}\frac{\Var[N_{\mathtt{XY}} - N_\mathtt{YX}]}{t} = (\vec{w}_{\mathtt{YX}} - \vec{w}_{\mathtt{XY}})\cdot \mathbb{D} \cdot (\vec{w}_{\mathtt{YX}} - \vec{w}_{\mathtt{XY}}) = 0.
\end{equation}
Therefore, when considering time-extensive quantities in the long-time limit, we can approximate $N_{\mathtt{XY}} - N_\mathtt{YX} \approx 0$ since this difference has mean and variance of order $\mathcal{O}(t^0)$. In other words, $N_{\mathtt{XY}}$ and 
$N_\mathtt{YX}$ become equivalent at the stochastic level and it suffices to consider only $N_{\mathtt{XY}}$, for example. Any other linear combination of $N_{\mathtt{XY}}$ and $N_{\mathtt{YX}}$ (except $N_{\mathtt{XY}} - N_{\mathtt{YX}} \approx 0$) will have the same, maximal SNR. These conclusions, which can be understood as a natural consequence of probability conservation and stationarity, hold for any system with precisely two meso-states, as defined above.

\section{Analytical solution of the Erlang clock}
\label{SM:Erlang}

\subsection{Homogeneous Erlang clock}

Here, we provide an analytical solution to the Erlang clock~\cite{Erlang1917,David1987} shown in Fig.~\ref{fig:Erlang} of the main text.
For readability, we here repeat that this clock for dimension $d$ is given by a ring network with transition rates $R_{\mu\sigma}=\Gamma\delta_{\mu,\sigma+1},$ and periodic boundary conditions, $p_{\mu+d}(t) \equiv p_\mu(t)$. The Erlang estimator counts only every $d$th transition, that is once the system has completed one cycle, the time estimator is increased by one unit.
Formally, this is achieved with the following weights $w_{\mu\sigma}=d \Gamma^{-1} \delta_{\mu,1}\delta_{\sigma,d}.$ Since only a single type of jump is monitored, the alphabet $\aleph = \{d\to 1\}$ has a single element and thus $P(d\to 1|\aleph) = 1$, so it follows immediately from Eq.~\eqref{tick_time} that the resolution is $\nu = \bar{w}^{-1} = \Gamma/d$.

In the following, we show that the Erlang clock is indeed unbiased, $\EE[\Theta]=t,$ and moreover that the asymptotic variance is minimal, $\Var[\Theta]/t\rightarrow \mathcal T^{-1}.$ Finally, we provide an exact expression for the autocorrelation function shown in Fig.~\ref{fig:Erlang} by calculating the propagator matrix $\mathsf G(t) = e^{\mathsf L t}$ analytically.

To simplify the notation, we can assume without loss of generality that $\Gamma=1,$ i.e.~parameter time $t$ is expressed in units of $\Gamma.$
With this, we can directly calculate the expectation value and variance of the Erlang estimator by using results from Sec.~\ref{SM:degeneracy}.
For the given setup the BLUE is defined by the weights $w_{\mu+1,\mu}=1,$ which counts all the jumps with the same weight.
Given the cyclic symmetry of our setup, the long-time variance and expectation value of our estimator are invariant under transformations $w_{\mu\sigma}' = w_{\mu\sigma} + c_\mu-c_\sigma,$ for an arbitrary vector $c_\mu,$ as shown in Sec.~\ref{SM:degeneracy}.
By chosing for example $c_\mu = -\mu,$ this symmetry allows us to obtain the Erlang estimator from the BLUE.
We can explicitly calculate $w'_{\mu+1,\mu}=1 - (\mu+1) + \mu = 0,$ so long as $\mu\leq d-1,$ and for $\mu=d,$ we have $w'_{1d}=1 - 1 + d = d,$ due to cyclicity.
For these indices $\vec w'$ concides with the Erlang estimator, but for some other indices $\mu,\sigma,$ we may have $w'_{\mu\sigma}\neq 0.$
Since there are only jumps $\mu\rightarrow \mu+1$, these terms are irrelevant, and therefore using the results from Sec.~\ref{SM:degeneracy} we find that indeed $\EE[\Theta]=t,$ and $\Var[\Theta]/t\rightarrow \mathcal T^{-1}$ for the Erlang estimator.

For the next step, we want to calculate the current autocorrelation function for the Erlang estimator that is shown in Fig.~\ref{fig:Erlang}.
The autocorrelation function is defined as
\begin{align}
    F(t+\tau,t) &= \mathrm{Cov}[I(t+\tau),I(t)] \\
    &= \EE[I(t+\tau)I(t)]-\EE[I(t+\tau)]\EE[I(t)],
\end{align}
which in case for the Erlang estimator reduces to determining $\EE[I(t+\tau)I(t)],$ because we know that in the steady-state limit we have $\lim_{t\rightarrow\infty}\EE[I(t)]=1$.
For values of $\tau\geq 0,$ we can calculate explicitly
\begin{align}
    \EE[I(t+\tau)I(t)] &= d^2 \EE[I_{1,d}(t+\tau)I_{1,d}(t)] \\
    &=d\left[\delta(\tau) + R_{1,d}G_{d,1}(\tau)J_{1,d}(t)\right],
\end{align}
having used Eq.~\eqref{eq:E[I(t+tau)I(t)]_sol} for arriving at the second line. In the steady-state limit $t\rightarrow\infty$, we find $J_{1,d}(t) = 1,$ and since $R_{1,d}=\Gamma = 1$ in these units, the autocorrelation function is given by
\begin{align}
    \lim_{t\rightarrow\infty} F(t+\tau,t) &= d\left(\delta(\tau) + G_{d,1}(\tau)\right) - 1.
\end{align}

The quantity we still need to determine is the propagator matrix
\begin{align}
    \mathsf{G}(t) = e^{\mathsf{L}t}.
\end{align}
To compute $\mathsf{G}(t)$, we consider the Poisson process described by the infinite-dimensional extension of $\mathsf L$ as $\mathsf L_\infty$. More precisely, we solve a uni-directional hopping process along a semi-infinite chain of sites described by the probability vector $\mathbf{q}(t)$. Then, the rate equation $\dot q_\mu(t)=-q_\mu(t) + q_{\mu-1}(t)$ holds for all values of $\mu\geq 1,$ and we set by definition $q_0(t) \equiv 0.$ To recover the probability vector for the $d$-dimensional Erlang clock, we identify as equivalent all sites differing by a multiple of $d$ and sum over the corresponding probabilities:
\begin{align}
\label{eq:p_mu_q_mu_relation}
    p_\mu(t) = \sum_{\sigma\geq 0} q_{\mu+\sigma d}(t).
\end{align}
We can verify that 
\begin{align}
    \dot p_\mu(t) &= \sum_{\sigma \geq 0}\dot q_{\mu+\sigma d}(t)\\
    &= \sum_{\sigma \geq 0}\left(q_{\mu+\sigma d-1}(t)-q_{\mu+\sigma d}(t)\right)\\
    &=p_{\mu-1}(t) - p_\mu(t),    
\end{align}
which is just what we get from $\dot{\mathbf{p}}(t) = \mathsf L \mathbf{p}(t)$ in the $d$-dimensional case. This proves that the identification in Eq.~\eqref{eq:p_mu_q_mu_relation} is consistent with the two equations of motion for $\mathbf{q}$ and $\mathbf{p}$.

Now, to solve the equation of motion for $\mathbf{q}(t)$ we introduce the moment generating function $\rho(t,\chi) = \sum_{\mu\geq 0} p_\mu(t) \ee^{\ii \mu\chi}$, such that the following two equations are equivalent:
\begin{align}
    \dot{\mathbf{q}}(t) = \mathsf L_\infty \mathbf{q}(t) \Leftrightarrow \dot \rho(t,\chi) = (\ee^{\ii\chi} - 1)\rho(t,\chi).
\end{align}
The one on the RHS we can trivially integrate and we find
\begin{align}
    \rho(t,\chi) &= \ee^{t(\ee^{\ii\chi}-1)}\rho(0,\chi) \\
    &= \sum_{\mu,\sigma\geq 0}\frac{t^\mu}{\mu!}\ee^{\ii\mu\chi}\ee^{-t}q_\sigma(t)\ee^{\ii\sigma\chi}
\end{align}
which allows us to analytically calculate $\mathsf G_\infty(t)=\exp(\mathsf L_\infty t)$.
Given the initial condition $\rho^{(\sigma)}(0,\chi) = \ee^{\ii\sigma \chi},$ we can determine
\begin{align}
    [\mathsf{G}_\infty(t)]_{\mu\sigma} &= \frac{1}{2\pi}\int_0^{2\pi} \dd\chi\, \ee^{-\ii\mu\chi} \rho^{(\sigma)}(t,\chi) \\
    &= \frac{1}{2\pi} \int_0^{2\pi}\dd\chi\, \sum_{\nu\geq 0}\frac{t^\nu}{\nu!}\ee^{-t}\ee^{\ii\chi(\sigma+\nu-\mu)}\\
    &= \frac{t^{\mu-\sigma}}{(\mu-\sigma)!}\ee^{-t}.
\end{align}
Not surprisingly, we recover the Poisson distribution, defined by the rate equation for $\mathbf{q}(t).$

Finally, we can calculate the propagator matrix $\mathsf G(t)$ for the $d$-dimensional system by using Eq.~\eqref{eq:p_mu_q_mu_relation} to get
\begin{align}
    G_{\mu\sigma}(t) &= \sum_{\nu \geq 0} [\mathsf{G}_\infty(t)]_{\mu+\nu d,\sigma} \\
    &=\sum_{\nu \geq0} \frac{t^{\mu-\sigma+\nu d}}{(\mu-\sigma +\nu d)!}\ee^{-t}.
\end{align}
Given this result, we can calculate $G_{d,1}(t)$ which we need for the autocorrelation function,
\begin{align}
    \lim_{t\rightarrow\infty} F(t+\tau,t) = d\left(\delta(\tau) + \sum_{\mu \geq 0}\frac{\tau^{d-1 + \mu d}}{(d-1+\mu d)!}\ee^{-\tau}\right)-1,
\end{align}
the result of which is plotted in Fig.~\ref{fig:Erlang}. Note that from this we can obtain the short-time variance, as 
\begin{equation}
	\label{short_time_variance}
	\Var[\Theta] = \int_0^{t}\dd t_1\int_0^t\dd t_2 F(t_1,t_2) = d t +\mathcal{O}(t^2),
\end{equation}
or, restoring units, we have $\lim_{t\to 0} \Var[\Theta_{\rm Erl}]/t = d/\Gamma$, which is $d$ times larger than the variance of the BLUE, $\Var[\Theta_{\rm BLUE}]/t = \mathcal{T} = \Gamma^{-1}$. 

\subsection{Inhomogeneous Erlang clock}

One can also modify the Erlang clock to accommodate inhomogeneous rates, $R_{\mu\sigma} = \Gamma_\sigma \delta_{\mu,\sigma+1}$. As before, only clockwise motion around the ring is allowed, but now the escape rate $\Gamma_\sigma$ can be different for each site. While a complete solution for the dynamics is cumbersome in this case, we can still obtain simple expressions for the asymptotic precision and resolution of the Erlang estimator, which counts only a single transition.

\subsubsection{Two-state system}

Let us start with the simplest case of $d=2$, whose dynamics was already solved exactly in Appendix~\ref{SM:exact_solution_d2}. Here we recap the steady-state probabilities $p_1^\ss = \Gamma_2/(\Gamma_1 + \Gamma_2) $ and $p_2^\ss = \Gamma_1/(\Gamma_1 + \Gamma_2) $, so that the mean residual time is
\begin{equation}
    \label{T_d2}
\mathcal{T} = \frac{\Gamma_1^2 +\Gamma_2^2}{\Gamma_1\Gamma_2(\Gamma_1+\Gamma_2)}. 
\end{equation}
The Erlang estimator is the unbiased counting estimator that counts only the transition $2\to 1$, i.e. $w_{12}=0$ and 
\begin{equation}
    \label{Erlang_d2}
w_{12} = J_{12}^{-1} = \frac{\Gamma_1+\Gamma_2}{\Gamma_1\Gamma_2}.
\end{equation}
Since only a single transition is counted, its resolution is simply the inverse of this increment:
\begin{equation}
    \label{resolution_Erlang_d2}
	\nu = \frac{\Gamma_1\Gamma_2}{\Gamma_1+\Gamma_2}.
\end{equation}

To show that this estimator has minimal variance in the long-time limit, we use the fact that the ticks of the Erlang estimator are a renewal process~\cite{Cox1982}: the state of the system is reset to the same state, $\sigma=1$, with the same ensuing dynamics after every tick. A renewal process is fully specified by the distribution of waiting times between ticks, 
\begin{equation}
    \label{waiting_time_def}
W_d(t) = -\frac{\dd P_{{\rm no}, d}}{\dd t},
\end{equation}
where $P_{{\rm no}, d}(t)$ is the probability that no tick occurs up to time $t$ given that a tick was detected at $t=0$. For the Erlang estimator with $d=2$, this is equivalent to the probability that at most one jump has occurred after the system was set to state $\sigma=1$, so we find
\begin{equation}
    \label{P_0_Erlang}
P_{{\rm no}, 2}(t) = \ee^{-\Gamma_1 t} + \int_0^t\dd t_1 \ee^{-\Gamma_2(t-t_1)} \Gamma_1 \ee^{-\Gamma_1 t_1},
\end{equation}
where the first term is the probability of $N=0$ jumps and the second term is the probability for $N=1$  (cf. Eq.~\eqref{prob_path_integral}). Plugging this into Eq.~\eqref{waiting_time_def} yields the tick distribution
\begin{align}
    \label{WTD_Erlang_d2}
W_2(t) & = \int_0^t \dd t_1 \Gamma_2 \ee^{-\Gamma_2(t-t_1)} \Gamma_1 \ee^{-\Gamma_1t_1}\\ 
 \label{WTD_Erlang_d2_solution}
& = \frac{\Gamma_1\Gamma_2 \left ( \ee^{-\Gamma_2 t} - \ee^{-\Gamma_1 t}\right )}{\Gamma_1 - \Gamma_2}.
\end{align}

The mean and variance of the tick waiting times are thus found to be
\begin{align}
\label{mean_WT}
 &    \EE[t_2]  =  \int_0^\infty \dd t\, W_2(t) t = \frac{\Gamma_1 + \Gamma_2}{\Gamma_1\Gamma_2} = \nu^{-1}, \\
\label{var_WT}
& \Var[t_2]  =  \int_0^\infty \dd t\, W_2(t) \left (t -\EE[t_2]\right )^2 = \frac{\Gamma_1^2 + \Gamma_2^2}{\Gamma_1^2\Gamma_2^2}.
\end{align}
Now, the Erlang estimator itself is $\Theta_{\rm Erl} = \EE[t_2]N_2$, where $N_2$ is the number of ticks. Since $N_2$ is renewal, its long-time diffusion coefficient can be shown to be (e.g.~see Sec.VI.B.2 of Ref.~\cite{Landi2024})
\begin{equation}
    \label{diffusion_renewal}
\lim_{t\to \infty} \frac{\dd}{\dd t} \Var[N_2] = \frac{\Var[t_2]}{\EE[t_2]^3}.
\end{equation}
It follows that the diffusion coefficient for the Erlang estimator itself is $D=\Var[t_2]/\EE[t_2] = \mathcal{T}$. This proves that the Erlang estimator is an unbiased counting estimator with minimal variance, i.e.~it constitutes a solution to Eq.~\eqref{BLUE}. Its resolution, on the other hand, is given by $\nu= \EE[t_2]^{-1} = \mathcal{A}/2$, where the dynamical activity is $\mathcal{A} = 2\Gamma_1 \Gamma_2/(\Gamma_1+\Gamma_2)$.

\subsubsection{Arbitrary number of states}

To extend this to larger values of $d$, we first find the steady-state probability distribution. Since the equation of motion is $\partial_t p_\sigma = \Gamma_{\sigma-1}p_{\sigma-1} - \Gamma_\sigma p_\sigma$, the steady-state solution obeys the recursion relation
\begin{equation}
	\label{steady_state_recursion}
	p_\sigma^\ss = \frac{\Gamma_{\sigma-1}}{\Gamma_\sigma} p_{\sigma-1}^\ss.
\end{equation}
The solution is $p_\sigma^\ss = {\rm const.}\times \Gamma_\sigma^{-1}$, and taking into account normalisation, we find
\begin{equation}
	\label{Erlang_steady_state}
	p_\sigma^\ss = \frac{\mathcal{A}}{ d \Gamma_\sigma},
\end{equation}
where the dynamical activity is given by
\begin{equation}
    \label{A_Erlang}
\mathcal{A} = \sum_{\sigma} p_\sigma^\ss \Gamma_\sigma = \left ( \sum_{\sigma=1}^d \frac{1}{d \Gamma_\sigma}\right )^{-1}.
\end{equation}
We also find the mean residual time
\begin{equation}
    \label{T_Erlang}
	\mathcal{T} =  \sum_{\sigma} \frac{p_\sigma^\ss}{\Gamma_\sigma} = \sum_{\sigma=1}^d \frac{\mathcal{A}}{d \Gamma_\sigma^2}. \\
\end{equation}

Now, we define the Erlang estimator by setting all weights $w_{\mu\sigma}=0$ except for
\begin{equation}
	\label{Erlang_estimator_general}
	w_{1d} = J_{1d}^{-1}.
\end{equation}
This choice guarantees that the estimator is unbiased, since $\vec{w}\cdot \vec{J} = w_{1d}J_{1d} = 1.$ Only one jump is counted, therefore the resolution is simply the inverse of Eq.~\eqref{Erlang_estimator_general}. More precisely, we have
\begin{equation}
	\label{resolution_Erlang_general}
	\nu =  \frac{\mathcal{A}}{d},
\end{equation}
where we have used $J_{1d} = \Gamma_dp_d^\ss$ together with Eq.~\eqref{Erlang_steady_state}. Therefore, the Erlang estimator has resolution $d$ times smaller than the BLUE, for any set of rates $\{\Gamma_\sigma\}$ and dimension $d$.

As before, we are dealing with a renewal process since the state is reset to $\sigma=1$ after each tick. We therefore need only find the tick waiting-time distribution to obtain the asymptotic statistics of the estimator. For arbitrary $d$, Eq.~\eqref{WTD_Erlang_d2} generalises to 
\begin{equation}
	\label{WTD_Erlang}
	W_d(t) = \int_0^t\dd t_{d-1} \cdots \int_0^{t_2} \dd t_1 \prod_{\sigma=1}^{d} W(t_{\sigma}-t_{\sigma-1}| \sigma),
\end{equation}
where $W(\tau|\sigma) = \Gamma_\sigma \ee^{-\Gamma_\sigma \tau}$ is the distribution of waiting times preceding a jump from state $\sigma$, and we define $t_{d} \equiv t$ and $t_0 \equiv 0$. This expresses the fact that the total time between ticks is given by the sum of $d$ independent waiting times, corresponding to the $d$ sequential jumps $\sigma\to \sigma +1$ that take the system once around the clock. The probability of the total waiting time is thus a convolution of the individual distributions, as in Eq.~\eqref{WTD_Erlang}. 

By the convolution theorem, the Laplace transform of the tick distribution~\eqref{WTD_Erlang} is
\begin{equation}
	\label{Laplace_transform_WTD}
	\tilde{W}_d(z) = \int_0^\infty \dd t\, \ee^{-zt} W_d(t) = \prod_{\sigma=1}^d \frac{\Gamma_\sigma}{z+\Gamma_\sigma},
	\end{equation}
where $\tilde{W}(z|\sigma) = \Gamma_\sigma/(z+\Gamma_\sigma)$ is the Laplace transform of the waiting-time distribution for a single jump. To find the mean and variance of the total waiting time between ticks, it is sufficient to note that $\tilde{W}_d(z)$ is the moment-generating function for the tick waiting times, i.e.
\begin{equation}
    \label{moment_generating_function}
    \EE[t_d^n] = \int_0^\infty \dd t\, W_d(t) t^n = (-1)^n	\left.\frac{\partial^n \tilde{W}_d(z)}{\partial z^n}\right\vert_{z=0}.
\end{equation}
The corresponding cumulants can thus be found by taking derivatives of $\ln \tilde{W}_d(z)$ at  $z=0$, viz.
\begin{align}
	\label{mean_tick_Erlang}
	&	\EE[t_d]  = -\partial_z \ln \tilde{W}_d(z)\vert_{z=0} = \sum_{\sigma=1}^d \Gamma_\sigma^{-1},\\
	\label{var_Erlang}
	& \Var[t_d] = \partial^2_z \ln \tilde{W}_d(z)\vert_{z=0}= \sum_{\sigma=1}^d \Gamma_\sigma^{-2}.
\end{align}
We recognise immediately that the mean time between ticks is $\EE[t_d] = d/\mathcal{A} = \nu^{-1}$ (cf.~Eq.~\eqref{resolution_Erlang_general}), as expected. Using Eqs.~\eqref{A_Erlang}--\eqref{T_Erlang} and Eqs.~\eqref{mean_tick_Erlang}--\eqref{var_Erlang}, we obtain the diffusion coefficient (cf.~the discussion around Eq.~\eqref{diffusion_renewal})
\begin{equation}
	\label{Erlang_diffusion_general}
	D = \frac{\Var[t_d]}{\EE[t_d]} = \mathcal{T}.
\end{equation}
This proves the results for the Erlang estimator summarised in Table~\ref{tab:estimators}. It is notable that these results do not depend on the specific value of $\Gamma_d$, only on aggregate properties of the set $\{\Gamma_\sigma\}$. This shows that there is nothing special about the transition $d\to 1$ even in the inhomogeneous case: an unbiased Erlang estimator can be defined for any other pair of adjacent sites, yielding the same asymptotic precision and resolution. 

\section{Allan variance}
\label{SM:Allan_variance}

In this Appendix, we compute the Allan variance for a stationary counting observable. The Allan variance is defined in Eq.~\eqref{Allan_variance}, which we can also write as
\begin{equation}
	\label{Allan_variance_SM}
	\sigma_y^2(T) = \frac{1}{2}\EE\left [\Big (\bar{y}_T(t+T) - \bar{y}_T(t) \Big )^2\right ],
\end{equation}
where the average fractional frequency error is
\begin{equation}
	\label{av_fractional_frequency}
	\bar{y}_T(t)  = \frac{1}{T}\int_{t}^{t+T} \dd t'\,\left [I(t') - 1\right ].
\end{equation} 
Since $y_T(t)$ is a stationary, zero-mean process, we have
\begin{align}
	\label{Allan_autocorr_derivation}
	\sigma_y^2(T) & = \frac{1}{2}\left \{  \Var[\bar{y}_T(t+T)] +  \Var[\bar{y}_T(t)] \right . \notag\\  & \qquad+  \left . 2\, \Cov[ \bar{y}_T(t+T), \bar{y}_T(t) ]\right \} \notag \\
	& =  \Var[\bar{y}_T(0)] + \Cov[ \bar{y}_T(T), \bar{y}_T(0) ].
\end{align}
Upon plugging in Eq.~\eqref{av_fractional_frequency} for $y_T(t)$, this yields
\begin{align}
	\label{Allan_autocorrelator_SM}
	\sigma_y^2(T) = & \frac{\Var[\Theta |t=T]}{T^2} + \frac{1}{T^2}\int_{T}^{2T}\!\! \dd t_1 \int_0^T \!\!\dd t_2 \, F(t_1-t_2) ,
\end{align}                                     with $F(t_1-t_2)=\Cov[I(t_1),I(t_2)]$.   
The variance in the first term is expressed in terms of the autocorrelation function as
\begin{align}                                   
	\label{finite_T_variance}                         
\Var[\Theta |t=T] & = 	\int_{0}^{T} \!\!\dd t_1 \int_0^T\! \!\dd t_2 \, F(t_1-t_2) \notag \\
& = 2 \int_0^T \!\dd \tau \,(T-\tau) F(\tau),
\end{align}                                     
where we changed integration variables to $\tau = t_1-t_2$ and $s = (t_1+t_2)/2$, and carried out the trivial integral over $s$. Similarly, the second term in Eq.~\eqref{Allan_autocorrelator_SM} can be written as 
\begin{equation}
	\label{cross_correlator_SM}
	\Cov[ \bar{y}_T(T), \bar{y}_T(0) ] = \int_0^T\dd\tau \,\tau F(\tau) + \int_T^{2T} \dd \tau\,(2T-\tau)F(\tau).
\end{equation}                 

Now, for $\tau \geq 0$ we can use Eq.~\eqref{autocorr_superop} to write the autocorrelator as 
\begin{align}
	\label{F_spectral}
F(\tau)  & = K\delta(\tau) + \sum_{k=0}^{d^2}  \ee^{\lambda_k\tau} \mathbf{u}^\T \tilde{\mathsf{J}}  \mathsf{\Pi}_k \tilde{\mathsf{J}} \mathbf{p}^\ss - J^2 \notag\\ 
& = K\delta(\tau) + \sum_{k\neq 0} \ee^{\lambda_k\tau} \mathbf{u}^\T \tilde{\mathsf{J}}  \mathsf{\Pi}_k \tilde{\mathsf{J}} \mathbf{p}^\ss,
\end{align}
where $\mathsf{\Pi}_k = \mathbf{r}_k \mathbf{l}^\T_k $ is the projector onto $\mathbf{r}_k$, the right eigenvector of $\mathsf{L}$ with eigenvalue $\lambda_k$, with $\mathbf{l}_k^\T$ the corresponding left (row) eigenvector~\cite{Landi2024}. The left and right eigenvectors form a biorthonormal set, such that $\mathbf{l}_k^\T \mathbf{r}_{k'} = \delta_{kk'}$. The second line of Eq.~\eqref{F_spectral} follows from the fact that $\lambda_0=0$, with the steady state $\mathbf{p}^\ss =\mathbf{r}_0 $ and  the constant row vector $\mathbf{u}^\T= \mathbf{l}^\T_0$ as the corresponding right and left eigenvectors. We therefore have $\mathbf{u}^\T \tilde{\mathsf{J}}  \mathsf{\Pi}_0 \tilde{\mathsf{J}} \mathbf{p}^\ss = (\mathbf{u}^\T \tilde{\mathsf{J}}\mathbf{p}^\ss)^2 = J^2$ from Eq.~\eqref{average_current_superop}.

Using Eq.~\eqref{F_spectral} in Eq.~\eqref{finite_T_variance} and performing the integral, we have 
\begin{align}                                   
	\label{finite_T_variance_calc}                         
	\Var[\Theta |t=T] & =  K T + 2 \sum_{k\neq 0} \frac{\ee^{\lambda_kT }- 1 - T\lambda_k}{\lambda^2_k} \mathbf{u}^\T \tilde{\mathsf{J}}  \mathsf{\Pi}_k \tilde{\mathsf{J}} \mathbf{p}^\ss.
\end{align}      
Now, using the fact that $\{\mathsf{\Pi}_k\}$ is an orthonormal, complete set of projectors, $\mathsf{\Pi}_k\mathsf{\Pi}_{k'} = \delta_{k,k'}\mathsf{\Pi}_k$ and $\sum_k \mathsf{\Pi}_k=\mathsf{1}$, and that $\sum_{k\neq 0}\mathsf{\Pi}_k/\lambda_k = \mathsf{L}^+$ is the Drazin inverse~\cite{Landi2024}, we can write 
\begin{align}                                   
	\label{finite_T_variance_final}                         
	\Var[\Theta |t=T] & =  D T + 2\mathbf{u}^\T \tilde{\mathsf{J}} \mathsf{L}^+ \left [\mathsf{G}(T) - \mathsf{1}\right ] \mathsf{L}^+\tilde{\mathsf{J}} \mathbf{p}^\ss,
\end{align}   
where we have used Eq.~\eqref{diffusion_superop} to collect terms equal to the diffusion coefficient, $D$. A similar calculation can be performed for Eq.~\eqref{cross_correlator_SM}, yielding a relatively compact expression for the Allan variance,
\begin{equation}
	\label{Allan_final}
	\sigma_y^2(T) = \frac{D}{T} + \frac{1}{T^2} \mathbf{u}^\T \tilde{\mathsf{J}} \mathsf{L}^+ \left \{2\left [\mathsf{G}(T) - \mathsf{1}\right ] -\left [\mathsf{G}(T) - \mathsf{1}\right ]^2\right \} \mathsf{L}^+\tilde{\mathsf{J}}\mathbf{p}^\ss,
\end{equation}
which we use to generate the data in Fig.~\ref{fig:allan}. 

At long times, the second term of Eq.~\eqref{Allan_final} is subleading with respect to the first, and we recover simply 
\begin{equation}
	\label{Allan_long_time}
	\sigma^2_y(T) = \frac{D}{T}  + \mathcal{O}(T^{-2}).
\end{equation}
For short times, a power expansion of $\mathsf{G}(T) = \ee^{\mathsf{L}T}$ gives
\begin{align}
	\label{Allan_short_time}
	\sigma_y^2(T) &  = \frac{D}{T} + \frac{1}{T}  \mathbf{u}^\T \tilde{\mathsf{J}} \mathsf{L}^+ \mathsf{L} \mathsf{L}^+\tilde{\mathsf{J}}\mathbf{p}^\ss + \mathcal{O}(T) \notag \\
	& = \frac{K}{T} + \mathcal{O}(T)
\end{align}
where we again used Eq.~\eqref{diffusion_superop}. Thus, we confirm the long- and short-time limits quoted in Eq.~\eqref{allan_limits}. While the long-time Allan variance~\eqref{Allan_long_time} is clearly minimised by the BLUE, which has minimal $D$, minimising the short-time Allan variance requires extremising $K=\sum_{\mu,\sigma} w_{\mu\sigma}^2 J_{\mu\sigma}$. This is achieved by choosing uniform weights $w_{\mu\sigma}=\rm const.$ which, together with the unbiased constraint $J=\vec{w}\cdot\vec{J}=1$, implies that $w_{\mu\sigma} = \mathcal{A}^{-1}$.

The Allan variance simplifies drastically for the BLUE defined by Eq.~\eqref{BLUE_estimator}, as a consequence of its autocorrelation function $F(\tau) = \mathcal{T}\delta(\tau)$. Plugging this into Eq.~\eqref{Allan_autocorrelator_SM}, the second term vanishes while the variance becomes $\Var[\Theta|t] = Dt$ (cf.~Eq.~\eqref{variance_autocorrelation}). We therefore conclude that, for the BLUE, $\sigma_y^2(T) = D/T$ exactly for all $T$.

\bibliography{biblio}

\end{document}